%% file: odin-1.tex
\documentstyle[twocolumn,ieee]{article}
\input myth.tex

\newcommand{ \Good }[1]{{\mbox{\(\displaystyle{#1}\)}}}
\newcommand{ \good }[1]{{\mbox{$#1$}}}
\long\def\comment#1{}

\newlength{\FRAMEwidth}
\newcommand{ \FBOX }[2]{{\fbox{#2\raise-8pt\hbox{\vbox to #1{}}}}}

\newcommand{\FRAME}[1]
   {\vspace{1ex}\noindent\framebox[\textwidth]
   {\parbox{\FRAMEwidth}{\vspace{1ex}#1\vspace{1ex}}}\vspace{1ex}}
\newcommand{ \HEAD  }[1]
   {\cleardoublepage\markboth{}{}\noindent\framebox[\textwidth]
   {\parbox{\FRAMEwidth}{\chapter{#1}}}%
   \thispagestyle{bookheadings}\vspace{2ex}}
\newcommand{ \HEADa }[1]
   {\clearpage\markboth{}{}\noindent\framebox[\textwidth]
   {\parbox{\FRAMEwidth}{
                                      \section{#1}}}%
   \thispagestyle{artheadings}\vspace{2ex}}

\def\Varrow#1#2{{\hbox{\(\left#1\vbox to #2{}\right.%
               \nulldelimiterspace=0pt\mathsurround=0pt\)}}}

\newcommand{ \thebb }[2]%
 {\if#11\arabic{#2}\else\if#1a\alph{#2}\else\if#1A\Alph{#2}\else%
  \if#1i\roman{#2}\else\if#1I\Roman{#2}\else{#1}\fi\fi\fi\fi\fi}
\newcommand{ \widbb }[1]%
 {\if#11{9}\else\if#1a{a}\else\if#1A{H}\else%
  \if#1i{vi}\else\if#1I{VI}\else{#1}\fi\fi\fi\fi\fi}
\newenvironment{bbize}[1]%
{\begin{list}{{\rm (\thebb{#1}{enumi})}\hfill}%
 {\settowidth\labelwidth{(\widbb{#1})}\leftmargin\labelwidth%
  \advance\leftmargin\labelsep\usecounter{enumi}}}%
{\end{list}}
\newcommand{ \bb }[1]{\begin{bbize}{#1}}
\newcommand{ \ee }   {\end{bbize}}
\newenvironment{bbbize}[2]%
{\begin{list}{{\rm (\thebb{#1}{enumi}\thebb{#2}{enumii})}\hfill}%
 {\settowidth\labelwidth{(\widbb{#1}\widbb{#2})}%
  \leftmargin\labelwidth\advance\leftmargin\labelsep%
  \usecounter{enumii}}}%
{\end{list}}
\newcommand{ \bbb }[2]{\begin{bbbize}{#1}{#2}}
\newcommand{ \eee }   {\end{bbbize}}

\newcommand{ \pH  }{\mbox{$\oplus$-Horn}\ }
\newcommand{ \aH  }{\mbox{{\rm $\&$}-Horn}\ }
\newcommand{ \paH }{\mbox{{\rm ($\oplus$,$\&$)}-Horn}\ }

\newcommand{ \eH  }{\mbox{{\rm !}-Horn}\ }
\newcommand{ \epH }{\mbox{{\rm (!,$\oplus$)}-Horn}\ }

\newcommand{ \OUT  }[1]{\mbox{{\bf Value}$[#1]$}}
\newcommand{ \STACK}[1]{\mbox{{\bf Stack}$[#1]$}}

\newcount\PLv\newcount\PLw\newcount\PLx\newcount\PLy
\newdimen\PLyy\newdimen\PLX \newbox\PLdot \setbox\PLdot\hbox{\tiny.}
\def\scl{.08}
\def\PLot#1{\PLx`#1\advance\PLx-42\PLy\PLx\PLv\PLx\divide\PLy9%
 \PLw\PLy\multiply\PLw9\advance\PLx-\PLw\advance\PLx-4\PLy-\PLy%
 \advance\PLy4\PLX=\the\PLx pt\advance\PLyy\the\PLy pt\wd%
 \PLdot=\scl\PLX\raise\scl\PLyy\copy\PLdot}
\def\draw#1{\ifx#1\end\let\next=\relax\else\PLot#1%
            \let\next=\draw\fi\next}

\newcommand{ \lb }{\Varrow{(}{0.9em}}
\newcommand{ \rb }{\Varrow{)}{0.9em}}
\newcommand{ \ls }{\Varrow{(}{0.8em}}
\newcommand{ \rs }{\Varrow{)}{0.8em}}

\def\gsep{\raise-2pt\hbox{\vrule\vbox{\hrule\kern1em%
                          \hbox to 3.5pt{}\hrule}\vrule}}

\newcommand{ \imply }[2]{\good{({#1} \rightarrow {#2})}}

\newcommand{ \Logic}[2]{\mbox{{\rm{\bf #1}}$(#2)$}}

\def\llto{\mathbin{-\mkern-3mu\circ}}

\def\IA{\hbox{\PLyy=70pt\draw :::;DMV_gqppyyyyyooooxxxnnwvlutkjaWNE%
       =5-./99:::CCCC:::99/..--544=EENWWaajjjkktttttttNNNVVVVVVVV\end
         \hskip7pt}} 
                       \newbox\IAbox\setbox\IAbox\IA
\def\Par{\copy\IAbox}

\newdimen\hauteur\newdimen\largeur
\def\vspec#1{\special{ps:#1}}
\def\rotstart#1{\vspec{gsave currentpoint currentpoint translate
   #1 neg exch neg exch translate}}
\def\rotfinish{\vspec{currentpoint grestore moveto}}
\def\rot#1{\hauteur=\ht#1\advance\hauteur by\dp#1\largeur=\wd#1%
\hbox to\largeur{\kern\largeur\vbox to\hauteur{\kern\hauteur%
\rotstart{180 rotate}\copy#1\vss}\hss}\rotfinish}

\newbox\etnormal\setbox\etnormal=\hbox{\normalsize\rm\&}
\newbox\etscript\setbox\etscript=\hbox{\scriptsize\rm\&}
\newbox\etscriptscript\setbox\etscriptscript=\hbox{\tiny\rm\&}

\newcommand{ \bang}[1]{\good{{\rm\,!}#1}}

\newcommand{ \odin }{\mbox{{\rm 1\hspace{-.5ex}l}}}
\newcommand{ \limply  }[2]{\good{(#1 \llto #2)}}
\newcommand{ \Limply  }[2]{\good{\lb{#1} \llto {#2}\rb}}
\newcommand{ \lembed  }[3]{\Limply{\lb{#1} \llto {#2}\rb}{#3}}
\newcommand{ \lvariant }[3]{\good{\ls #1 \llto (#2 \oplus #3)\rs}}

\newcommand{ \lnondet }[4]%
 {\good{\ls({#1} \llto {#2})\,{\rm\&}\,({#3} \llto {#4})\rs}}
\newcommand{ \lnondets}[6]%
 {\good{\lb\ls{#1} \llto {#2}\rs\,{\rm\&}\,\ls{#3} \llto {#4}\rs\,
     {\rm\&} \cdots {\rm\&}\,\ls{#5} \llto {#6}\rs \rb}}

\newcommand{ \cimply }[2]{\Limply{\lb{#2}\llto p\rb}
                                 {\lb{#1}\llto p\rb}}
\newcommand{ \ang }[1]{\good{\langle #1 \rangle}}

\newcommand{ \Li }[2]{\Good{\lb{#1} \backslash {#2}\rb}} 

\newcommand{ \livariants}[4]%
     {\Li{#1}{\lb{#2} \oplus {#3} \oplus \cdots \oplus {#4}\rb}}

\newcommand{ \RULE}[3]
 {\begin{tabular}[t]{@{}p{#1}}#2\\[1ex]\hspace{2em}#3%
  \end{tabular}}

\newcommand{ \seq }[2]{\good{#1\,\vdash #2}}

\newcommand{ \son }[2]
 {\begin{tabular}[b]{@{}c@{}} #1 \\ \multispan1\hrulefill\\ #2 %
  \end{tabular}}

\newcommand{ \sonN}[3]
{\begin{tabular}[b]{@{}c@{}l@{}} #2 &\\%
  \multispan1\hrulefill&\omit\\%
    #3 &\rlap{\(\smash{\mathop{\phantom{o}}\limits^{\,\mbox{#1}}}\)}%
 \end{tabular}}

\newcommand{ \Sona}[3]
{\begin{tabular}[b]{@{}p{#1}@{}c@{}p{#1}@{}l@{}}%
  \multispan4{#2}\\%
  \multispan3\hrulefill&\omit\\%
  \hfill&{#3}&\hfill&%
 \end{tabular}}

\newcommand{ \SonaN}[4]
{\begin{tabular}[b]{@{}p{#1}@{}c@{}p{#1}@{}l@{}}%
  \multispan4{#3}\\%
  \multispan3\hrulefill&\omit\\%
  \hfill&{#4}&\hfill&%
  \rlap{\(\smash{\mathop{\phantom{o}}\limits^{\,\mbox{#2}}}\)}%
 \end{tabular}}

\newcommand{ \Son }[4]
{\begin{tabular}[b]{@{}p{#1}@{}p{#2}@{}c@{}p{#2}@{}l@{}}%
  \multispan5{#3}\\%
  \omit&\multispan3\hrulefill&\omit\\%
  \hfill&\hfill&{#4}&\hfill&%
 \end{tabular}}

\newcommand{ \SonN}[5]
{\begin{tabular}[b]{@{}p{#1}@{}p{#2}@{}c@{}p{#2}@{}l@{}}%
\multispan5{#4}\\%
 \omit&\multispan3\hrulefill&\omit\\%
 \hfill&\hfill&{#5}&\hfill&%
 \rlap{\(\smash{\mathop{\phantom{o}}\limits^{\,\mbox{#3}}}\)}%
   \end{tabular}}

\newcommand{ \SON }[2]
{\begin{tabular}[b]{@{}c@{}}{#2}\\ \(\Biggm\downarrow\)\\ \fbox{#1}%
 \end{tabular}}

\newcommand{ \SONs }[4]
{\begin{tabular}[b]{@{}p{#1}p{#1}@{}c@{}p{#1}p{#1}@{}}%
  \multicolumn{2}{c}{#3}&&\multicolumn{2}{c}{#4}\\
  &\multicolumn{3}{|c|}{}&\\%
  \cline{2-4}%
 \hfill&\hfill&\(\Biggm\downarrow\)&\hfill&\hfill\\%
  \multicolumn{5}{c}{\fbox{#2}}%
 \end{tabular}}

\newcommand{ \SONsWrap }[4]
{\begin{tabular}[b]{@{}p{#1}p{#1}@{}c@{}p{#1}p{#1}@{}}\\%
  \multicolumn{2}{c}{\(\vdots\)}&&\multicolumn{2}{c}{\(\vdots\)}\\%
  \multicolumn{2}{c}{\(\Biggm\downarrow\)}&%
 &\multicolumn{2}{c}{\(\Biggm\downarrow\)}\\%
  \multicolumn{2}{c}{\fbox{#3}}&&\multicolumn{2}{c}{\fbox{#4}}\\
  &\multicolumn{3}{|c|}{}&\\ \cline{2-4}%
 \hfill&\hfill&\(\Biggm\downarrow\)&\hfill&\hfill\\%
  \multicolumn{5}{c}{\fbox{#2}}\\%
  &&\(\Biggm\downarrow\)&&\\%
  \multicolumn{5}{c}{\(\vdots\)}\\%
 \end{tabular}}

\newcount \WIDTH  
\newcount \HEIGHT 
\newcount \Xcur   
\newcount \Ycur   
\newcount \HALF
\newcount \TreeH  
\newcount \TreeW  

\newcommand{\vPICTURE}[5]%
{\setlength{\unitlength}{1pt}\HEIGHT=#5\multiply\HEIGHT by #2%
 \WIDTH=14\multiply\WIDTH by #3\advance\HEIGHT by \WIDTH%
 \WIDTH=#5\multiply\WIDTH by #4%
\begin{picture}(\WIDTH,\HEIGHT)%
  \HALF=#5\divide\HALF by 2\divide\WIDTH by 2%
  \put(\WIDTH,\HEIGHT){#1{#5}}
\end{picture}}

\def\toLeft{r}
\def\toRight{l}

\newcommand{ \Node }[2]
{\begin{picture}(0,0)
 \put(0,0){\makebox(0,0)[c]{$\bullet$}}
 \put(0,3){\makebox(0,0)[b#2]{\underline{#1}}}
\end{picture}}

\newcommand{ \NodeXIV }[3]
{\begin{picture}(0,0)
 \put(-7,-14){\framebox(14,13){#1}}
 \if#3r \put(-7,-5){\makebox(0,0)[br]{\underline{#2}}}%
 \else  \put( 7,-5){\makebox(0,0)[bl]{\underline{#2}}}\fi
\end{picture}}

\newcommand{ \nodeXIV }[1]
{\begin{picture}(0,0)
 \put(-7,-14){\makebox(14,13){#1}}
\end{picture}}

\newcommand{ \NodeWide }[2]
{\begin{picture}(0,0)
 \put(-13,-14){\framebox(26,13){#1}}
 \put( 13,-5){\makebox(0,0)[bl]{\underline{#2}}}
\end{picture}}

\newcommand{ \NodeWIDE }[1]
{\begin{picture}(0,0)
 \put(-35,-21){\framebox(69,27){#1}}
\end{picture}}

\newcommand{ \SOUTH     }[1]{\vector( 0,-1){#1}}
\newcommand{ \SOUTHWEST }[1]{\vector(-1,-1){#1}}
\newcommand{ \SOUTHEAST }[1]{\vector( 1,-1){#1}}

\newcommand{ \South }[3]
{\begin{picture}(0,0)
 \put(0,0){\SOUTH{#1}}
 \put(0,-\HALF){\makebox(0,0)[b#3]{\underline{#2}}}
\end{picture}}

\newcommand{ \SouthD}[1]
{\begin{picture}(0,0)
 \put(0,0){\line(0,-1){9}}
 \TreeH=#1 \advance \TreeH by -6 \divide \TreeH by 3
 \multiput(0,-6)(0,-3){\TreeH}{\makebox(0,0)[c]{$\cdot$}}
 \TreeH=#1 \advance \TreeH by -12
 \put(0,-\TreeH){\SOUTH{12}}
\end{picture}}

\newcommand{ \SouthWest }[3]
{\begin{picture}(0,0)
 \put(0,0){\SOUTHWEST{#1}}
 \put(-\HALF,-\HALF){\makebox(0,0)[b#3]{\underline{#2}}}
\end{picture}}

\newcommand{ \SouthEast }[3]
{\begin{picture}(0,0)
 \put(0,0){\SOUTHEAST{#1}}
 \put(\HALF,-\HALF){\makebox(0,0)[b#3]{\underline{#2}}}
\end{picture}}

\newcommand{ \SouthEastD}[1]
{\begin{picture}(0,0)
 \put(0,0){\line(1,-1){10}}
 \TreeH=#1 \advance \TreeH by -6 \divide \TreeH by 2
 \multiput(4,-4)(2,-2){\TreeH}{\makebox(0,0)[c]{$\cdot$}}
 \TreeH=#1 \advance \TreeH by -12
 \put(\TreeH,-\TreeH){\SOUTHEAST{12}}
\end{picture}}

\newcommand{ \FiveDotsS }[1]
{\begin{picture}(0,0)%
 \TreeH=#1\multiply\TreeH by 2\divide\TreeH by 11%
 \multiput(0,-\TreeH)(0,-\TreeH){5}{\makebox(0,0)[c]{$\bullet$}}
\end{picture}}
\newcommand{ \FiveDotsSW }[1]
{\begin{picture}(0,0)%
 \TreeH=#1\multiply\TreeH by 2\divide\TreeH by 11%
 \multiput(-\TreeH,-\TreeH)(-\TreeH,-\TreeH){5}
                             {\makebox(0,0)[c]{$\bullet$}}
\end{picture}}
\newcommand{ \FiveDotsSE }[1]
{\begin{picture}(0,0)%
 \TreeH=#1\multiply\TreeH by 2\divide\TreeH by 11%
 \multiput(\TreeH,-\TreeH)(\TreeH,-\TreeH){5}
                              {\makebox(0,0)[c]{$\bullet$}}
\end{picture}}

\newcommand{ \DELTA }[6]
{\begin{picture}(0,0)
 \put(0,0){\line(-1,-1){#1}}
 \put(0,0){\line( 1,-1){#1}}
 \TreeH=#1 \divide \TreeH by 2
 \put(0,-\TreeH){\makebox(0,0)[t]{#2}}
 \put(-#1,-#1){\line(1,0){#1}}
 \put(-#1,-#1){\SOUTH{18}}
 \put( #1,-#1){\line(-1,0){#1}}
 \put( #1,-#1){\SOUTH{18}}
 \TreeH=#1 \advance \TreeH by 4
 \put(0,-\TreeH){\makebox(0,0)[t]{{\bf \ldots}}}
 \advance \TreeH by 14
 \put(-#1,-\TreeH){\Node{#3}{#5}}
 \put( #1,-\TreeH){\Node{#4}{#6}}
\end{picture}}

\newcommand{ \dELTA }[4]
{\begin{picture}(0,0)
 \TreeH=#1 \divide \TreeH by 2
 \put(0,0){\line(-1,-2){\TreeH}}
 \put(0,0){\line( 1,-2){\TreeH}}
 \put(-\TreeH,-#1){\line(1,0){#1}}
 \advance \TreeH by 5
 \put(0,-\TreeH){\makebox(0,0)[t]{#2}}
 \put(0,-#1){\SOUTH{18}}
 \TreeH=#1 \advance \TreeH by 18
 \put(0,-\TreeH){\Node{#3}{#4}}
\end{picture}}

\newcommand{ \deltaD }[2]
{\begin{picture}(0,0)
 \TreeH=#1 \divide \TreeH by 2
 \put(0,0){\line(-1,-2){\TreeH}}
 \put(0,0){\line( 1,-2){\TreeH}}
 \advance \TreeH by 5
 \put(0,-\TreeH){\makebox(0,0)[t]{#2}}
 \divide \TreeH by 3
 \multiput(0,-#1)(-3,0){\TreeH}{\makebox(0,0)[c]{$\cdot$}}
 \multiput(0,-#1)( 3,0){\TreeH}{\makebox(0,0)[c]{$\cdot$}}
\end{picture}}

\newcommand{ \Tree}[3]
{\begin{picture}(0,0)
 \TreeH=#1 \multiply \TreeH by 2
 \TreeW=#1   \divide \TreeW by 2
 \put(0,0){\line(-1,-4){\TreeW}}
 \put(0,0){\line( 1,-4){\TreeW}}
 \put(0,-#1){\makebox(0,0)[t]{#2}}
 \put(-\TreeW,-\TreeH){\line(1,0){#1}}
 \put(-\TreeW,-\TreeH){\SOUTH{18}}                
 \put( \TreeW,-\TreeH){\SOUTH{18}}                
 \advance \TreeH by 4
 \put(0,-\TreeH){\makebox(0,0)[t]{{\bf \ldots}}}
 \advance \TreeH by 14                             
 \put(-\TreeW,-\TreeH){\Node{#3}{\toRight}}
 \put( \TreeW,-\TreeH){\Node{#3}{\toRight}}
\end{picture}}

\newcommand{ \Scheme }[4]
{\begin{picture}(0,0)
 \TreeW=#1 \divide \TreeW by 2
 \put(-\TreeW,0){\Node{#3}{\toLeft}}
 \put(-\TreeW,0){\SOUTH{\TreeW}}
 \put(      0,0){\makebox(0,0)[c]{{\bf \ldots}}}
 \put( \TreeW,0){\Node{#3}{\toRight}}
 \put( \TreeW,0){\SOUTH{\TreeW}}
 \TreeH=#1 \multiply \TreeH by 2
 \put(-#1,-#1)    {\framebox(\TreeH,\TreeW)[c]{#2}}
 \put(-\TreeW,-#1){\SOUTH{\TreeW}}
 \put( \TreeW,-#1){\SOUTH{\TreeW}}
 \advance \TreeH by -\TreeW
 \put(-\TreeW,-\TreeH){\Node{#4}{\toLeft}}
 \put(      0,-\TreeH){\makebox(0,0)[c]{{\bf \ldots}}}
 \put( \TreeW,-\TreeH){\Node{#4}{\toRight}}
\end{picture}}

\newcommand{ \schem }[4]
{\begin{picture}(0,0)
 \TreeW=#1 \divide \TreeW by 2
 \put(0,0){\SOUTH{\TreeW}}
 \TreeH=#1 \multiply \TreeH by 2
 \put(-#1,-#1)    {\framebox(\TreeH,\TreeW)[c]{#2}}
 \put(-\TreeW,-#1){\SOUTH{\TreeW}}
 \put( \TreeW,-#1){\SOUTH{\TreeW}}
 \advance \TreeH by -\TreeW
 \put(-\TreeW,-\TreeH){\Node{#3}{\toLeft}}
 \put(      0,-\TreeH){\makebox(0,0)[c]{{\bf \ldots}}}
 \put( \TreeW,-\TreeH){\Node{#4}{\toRight}}
\end{picture}}

\newcommand{ \PLACE }[2]
{\begin{picture}(0,0)
 \put(-12,-24){\framebox(24,23){#2}}
 \put(-12,-10){\makebox(0,0)[br]{\underline{#1}}}
\end{picture}}

\newcommand{ \PLACEbig }[2]
{\begin{picture}(0,0)
 \put(-80.4,-24){\dashbox{3}(48,24){#1}}
 \put(-32,-24){\framebox(64,23){#2}}
\end{picture}}

\newcommand{ \BAR }[2]
{\begin{picture}(0,0)
 \put(0,0){\makebox(0,0)[c]{\vrule height6pt width36pt depth0pt}}
 \if#2r \put(-18,3){\makebox(0,0)[br]{\underline{#1}}}%
 \else  \put( 18,3){\makebox(0,0)[bl]{\underline{#1}}}\fi
\end{picture}}

\newcommand{ \BARnondet}[2]
{\begin{picture}(0,0)
 \put(0,0){\makebox(0,0)[br]{\vrule height3pt width#1pt depth0pt}}
 \put(0,0){\makebox(0,0)[bl]{\vrule height3pt width#1pt depth0pt}}
 \put(-#1,0){\makebox(0,0)[c]{\vrule height6pt width36pt depth0pt}}
 \put( #1,0){\makebox(0,0)[c]{\vrule height6pt width36pt depth0pt}}
 \put(0,-1){\makebox(0,0)[tc]{\underline{#2}}}
\end{picture}}

\newcommand{ \BARempty}[2]
{\begin{picture}(0,0)
 \put(-20,-3){\framebox(40,5){}}
 \put(0,-1){\makebox(0,0)[c]{\vrule height3pt width40pt depth0pt}}
 \if#2r \put(-20,3){\makebox(0,0)[br]{\underline{#1}}}%
 \else  \put( 20,3){\makebox(0,0)[bl]{\underline{#1}}}\fi
\end{picture}}

\newcommand{\hitEast}
{\begin{picture}(0,0)%
 \put(-8,4){\vector(2,-1){8}}
\end{picture}}

\newcommand{\hitWest}
{\begin{picture}(0,0)%
 \put(8,-4){\vector(-2,1){8}}
\end{picture}}

\newcommand{\dEast}[1]
{\begin{picture}(#1,0)\thicklines%
 \HALF=#1\divide\HALF by 2\TreeW=\HALF\divide\TreeW by 2%
 $\bezier{\TreeW}(0,0)(\HALF,\TreeW)(#1,0)$%
 \put(#1,0){\hitEast}
\end{picture}}

\newcommand{\Darrow}[1]
{\begin{picture}(0,0)%
 $\bezier{\HALF}(-#1,0)(0,\HALF)(#1,0)$%
 \put(#1,0){\hitEast}
\end{picture}}

\newcommand{\Dbarrow}[1]
{\begin{picture}(0,0)%
 $\bezier{\HALF}(-#1,0)(0,-\HALF)(#1,0)$%
 \put(-#1,0){\hitWest}
\end{picture}}

\begin{document}

\title {Simulating Linear Logic\\
                in\\
        \mbox{1-Only} Linear Logic}
\author{Max I. Kanovich\\
        Russian Humanities State University, Moscow, Russia\\
                  and\\
        CNRS, Laboratoire de Math\'{e}matiques Discr\`{e}tes\\[1em]
         {\bf Pr\'{e}tirage n$^\circ$ 94-02}}
\date{January 28, 1994}
\maketitle
\thispagestyle{empty}
\begin{abstract}
 Linear Logic was introduced by Girard as a
 resource-sensitive refinement of classical logic. It turned out
 that full propositional Linear Logic is undecidable
 (Lincoln, Mitchell, Scedrov, and Shankar)
 and, hence, it is more expressive than (modalized)
 classical or intuitionistic logic.
  In this paper we focus on the study of the simplest fragments
 of Linear Logic, such as the one-literal and constant-only
 fragments (the latter contains no literals at all).

 Here we demonstrate that all these extremely simple fragments
 of Linear Logic (one-literal, \mbox{$\bot$-only}, and  even
 unit-only) are exactly of the {\em same} expressive
 power as the corresponding full versions:
\bb{a}
\item On the level of the {\em multiplicatives}
      \mbox{$\{ \otimes, \Par, \llto \}$} we get $NP$-completeness.
\item Enriching this basic set of connectives by {\em additives}
      \mbox{$\{\&,\oplus \}$} yields $PSPACE$-completeness.
\item Using in addition the {\em storage} operator~!, we can
      prove the undecidability of all these three fragments.
\ee

 We present also a {\em complete} computational interpretation (in
 terms of {\em acyclic programs with stack}) for \mbox{$\bot$-free}
 Intuitionistic Linear Logic. Based on this interpretation, we
 prove the {\em fairness} of our encodings and establish the
 foregoing complexity results.
\end{abstract}

\newpage
\tableofcontents
\newpage



\section {Introduction and \mbox{Summary}}
 Linear Logic was introduced by J.-Y.Girard~\cite{Girard} as a
 resource-sensitive refinement of classical logic. It turned out
 that full propositional Linear Logic is undecidable~\cite{LMSS},
 and, hence, it is more expressive than (modalized)
 classical or intuitionistic logic.
 Moreover, an {\em exact correspondence} between natural
 fragments of propositional Linear Logic and natural complexity
 classes can be established~\cite{LMSS,lics}.
  In this paper we focus on the study of the simplest fragments
 of Linear Logic, such as one-literal and constant-only
 fragments (the latter contains no literals at all) and
 demonstrate that these extremely simple fragments are of the
 {\em same} expressive power as the corresponding full versions.

 Formulas of propositional Linear Logic are built up of
 {\em literals} and {\em constants}~\mbox{($\bot,\odin$)} by
 the following {\em connectives:}
 $$ \otimes,\ \Par,\ \llto,\ \&,\ \oplus,\ !,\ \mbox{and}\ ? $$

 According to the well-known approaches, the hierarchy of natural
 fragments of Linear Logic can be developed in the following
 three directions:
\bb{1}
\item We start from the basic set of connectives,
      \mbox{the {\em multiplicatives:} $\otimes, \Par,$}
      and proceed to enrich it either
      \mbox{by {\em additives:} $\&,\oplus,$} or
      \mbox{by {\em exponentials:} $!, ?,$} or
      by both {\em additives} and {\em exponentials.}
\bbb{1}{a}
\item
      Thus we can start with the Multiplicative
      Fragment \Logic{LL}{\otimes,\Par,\bot,\odin}\\
      (which is proved to be \mbox{$NP$-complete~\cite{lics}),}
\item and go either to the Multiplicative-Additive
      Fragment \Logic{LL}{\otimes,\Par,\&,\oplus,\bot,\odin}\\
      (which is \mbox{$PSPACE$-complete~\cite{LMSS}),}
\item or to the Multiplicative-Exponential
      Fragment \Logic{LL}{\otimes,\Par,!,?,\bot,\odin}\\
      (its exact complexity level is unknown),
\item and finish in the full propositional Linear Logic
      \Logic{LL}{\otimes,\Par,\&,\oplus,!,?,\bot,\odin}\\
      (which is undecidable~\cite{LMSS}).
\eee
\item We can confine ourselves to formulas of a certain simple
      structure.
      E.g., it is typical of many logical systems to
      limit the depth of nesting of implications.
      In particular, it leads to the consideration of the
      so-called {\em Horn formulas} having the form~\imply{X}{Y}.

      As a rule, the {\em Horn} fragments are essentially simpler
      than their corresponding full versions.

      Contrary, for Linear Logic we have the following
      results~\cite{lics} demonstrating the maximum possible
      expressive power of Horn fragments:
\bbb{1}{a}
\item the purely Horn fragment \Logic{HLL}{\otimes,\llto}
      consisting of {\em Horn implications} \limply{X}{Y}, 
      is already \mbox{$NP$-complete,}
\item \mbox{the \paH fragment} \Logic{HLL}{\otimes,\llto,\&,\oplus},
      that contains {\em \pH implications}
              \lvariant{X}{Y_1}{Y_2}
      and {\em \aH implications}
              \lnondet{X_1}{Y_1}{X_2}{Y_2},
      is already \mbox{$PSPACE$-complete,}
\item the \eH fragment \Logic{HLL}{\otimes,\llto,\bang{}}
      is still decidable 
      (it is polynomially equivalent to Petri nets),
\item and the \epH fragment \Logic{HLL}{\otimes,\llto,\oplus,\bang{}}
      can simulate many-counter Minsky machines.
\eee
      Theorem~\ref{tNLL} shows the collapse of this hierarchy on
      the next step when we introduce {\em elementary embedded
      implications} \lembed{U}{V}{Y}.

\item Finally, for a given fragment of Linear Logic \Logic{LL}{\sigma}
      (its formulas are built up of literals and constants by
      connectives from the set~$\sigma$, constants are also taken
      from~$\sigma$),
      we can reduce the number of the literals used to a fixed
      number~$k$ and study the corresponding
      fragment~\Logic{LL$^\good{k}$}{\sigma}.
      Following such a {\em bottom-up} approach, we will start
      with the simplest cases when $k$~is small, namely,
      we will study the {\em one-literal}
      fragment~\Logic{LL$^\good{1}$}{\sigma} and {\em constant-only}
      fragment~\Logic{LL$^\good{0}$}{\sigma}.
\ee
 Actually, this approach is also quite traditional.

 E.g., consideration of the one-literal fragment of
 intuitionistic propositional logic allows us to obtain the full
 characterization of this fragment and shed light on the
 true nature of intuitionistic logic as a whole~\cite{Nishimura,Dick}.

 As for the expressive power of constant-only fragments of
 traditional logical systems, it is equal to {\em zero:}
 the entire problem boils down to primitive Boolean calculations
 over constants.

 The intricate story for Linear Logic began with the following
 unexpected results:
\bb{a}
\item The simplest one-literal fragment \Logic{LL$^\good{1}$}{\llto} is
      $NP$-complete~\cite{Kemail}.
\item The simplest constant-only fragment 
      \Logic{LL$^\good{0}$}{\otimes,\Par,\llto,\bot,\odin}
      is $NP$-complete~\cite{LW92}.
\ee

 As for one-literal and constant-only fragments enriched by
 {\em additives} and$/$or {\em exponentials}, {\em a priori} we
 could indicate both {\em pro} and {\em contra} arguments for
 their expressive power to be of high level.

 In particular, we could point out that all known proofs of the
 $PSPACE$-completeness of implicative fragment of intuitionistic
 propositional logic as well as of quantified Boolean propositional
 formulas are essentially based on an {\em unbounded} number of
 variables used.

 Regarding to the expressive power of connectives involved, the
 \mbox{$\bot$-only} case seemed to be easier for consideration,
 because we could use at least the functionally complete set of
 connectives including {\em negation}. The only problem was to wipe
 out the influence of the inference rules specified for~$\bot$ and,
 as a result, cause~$\bot$ to be thought of as an {\em ordinary}
 {\em positive} literal.

 The one-literal and unit-only cases met a problem at this point
 because, in the absence of~$\bot$, the whole system of connectives
        $$ \otimes,\ \Par,\ \llto,\ \&,\ \oplus $$
 is functionally incomplete (even in the Boolean sense).

 The unit-only case was the most complicated one because it is quite
 hard to conceive of the unit~$\odin$ as a literal.

 Nevertheless, Corollary~\ref{cmain} validates the following.
\begin{theorem} \label{tONE}
 {\em For one-literal fragments of Linear Logic,} we prove that
\bb{1}
\item \Logic{LL$^\good{1}$}{\otimes,\Par,\llto} is $NP$-complete.
\item Moreover, the purely implicative one-literal
      fragment \Logic{LL$^\good{1}$}{\llto} is already $NP$-complete.
\item \Logic{LL$^\good{1}$}{\llto,\&} is $PSPACE$-complete.
\item \Logic{LL$^\good{1}$}{\llto,\bang{}} can polynomially simulate
      the whole \mbox{$\bot$-free} Intuitionistic Linear Logic
      \Logic{ILL}{\otimes,\llto,\bang{}}.\footnote%
      {The latter consists of sequents of the form 
                $$ \seq{\Sigma}{A} $$
       where multiset~$\Sigma$ and formula~$A$ belong to the 
       language of \Logic{LL}{\otimes,\llto,\bang{}}
       (containing neither~$\bot$ nor~$\Par$).}
      In particular, the reachability problem for Petri Nets can
       be encoded in this one-literal fragment.
\item \Logic{LL$^\good{1}$}{\llto,\&,\bang{}} can directly simulate
      many-counter Minsky machines, and, hence, it is undecidable.
\ee
\end{theorem}

\begin{theorem} \label{tBOT}
 {\em For $\bot$-only fragments of Linear Logic,} we have
\bb{1}
\item \Logic{LL$^\good{0}$}{\otimes,\Par,\bot} is
      $NP$-complete~\cite{LW92}.
\item Moreover, the purely implicative $\bot$-only fragment
      \Logic{LL$^\good{0}$}{\llto,\bot} is already $NP$-complete.
\item \Logic{LL$^\good{0}$}{\llto,\&,\bot} is $PSPACE$-complete.
\item \Logic{LL$^\good{0}$}{\llto,\bang{},\bot} can polynomially
      simulate the whole \mbox{$\bot$-free} Intuitionistic
      Linear Logic \Logic{ILL}{\otimes,\llto,\bang{}}.
\item \Logic{LL$^\good{0}$}{\llto,\&,\bang{},\bot} can directly
      simulate many-counter Minsky machines, and, hence,
      it is undecidable.
\ee
\end{theorem}

\begin{theorem} \label{tUNIT}
 Finally, {\em for unit-only fragments of Linear Logic,} we prove that
\bb{1}
\item \Logic{LL$^\good{0}$}{\otimes,\llto,\odin} is trivial.
\item Nevertheless, \Logic{LL$^\good{0}$}{\otimes,\Par,\llto,\odin}
      is $NP$-complete~\cite{LW92}.
\item \Logic{LL$^\good{0}$}{\otimes,\Par,\llto,\&,\odin}
      is proved to be $PSPACE$-complete.
\item \Logic{LL$^\good{0}$}{\otimes,\Par,\llto,\bang{},\odin}
      can polynomially simulate~\Logic{ILL}{\otimes,\llto,\bang{}},
      and, hence, the complexity level of this Unit-Only Fragment
      is not less than the level of the whole Multiplicative-Exponential
      Fragment of \mbox{$\bot$-free} Intuitionistic Linear Logic.
\item \Logic{LL$^\good{0}$}{\otimes,\Par,\llto,\&,\bang{},\odin}
      can directly simulate many-counter Minsky machines,
      and, hence, it is undecidable.
\ee
\end{theorem}

 The plan of the paper is as follows:
\bb{a}
\item We present a {\em complete} computational interpretation
      of the {\em Normalized} Intuitionistic Linear Logic
      \Logic{NLL}{\otimes,\llto,\&,\oplus,\bang{}} in terms of
      {\em acyclic programs with stack.}
\item Then we encode all {\em normalized} sequents by one-literal,
      \mbox{$\bot$-only,} and unit-only sequents, and prove the
      {\em fairness} of these encodings.
\item Finally, based on the {\em uniformity} of our encodings,
      we establish the foregoing complexity results for the natural
      fragments of one-literal and constant-only Linear Logic.
\ee

 \begin{table*}[htp]
 \begin{center}
 \begin{tabular}{|llll|}     \hline &&&\\
 {\bf I}          & \son{}{\seq{A}{A}}  &  &           \\[1em]
 {\bf L$\llto$}   &
  \son{\seq{\Sigma_1}{A,\Phi_1} \ \ \ \ \ \seq{B,\Sigma_2}{\Phi_2}}
      {\seq{\Sigma_1,\limply{A}{B},\Sigma_2}{\Phi_1,\Phi_2}} & 
 {\bf R$\llto$}   &
  \son{\seq{\Sigma,A}{B,\Phi}}
      {\seq{\Sigma}{\limply{A}{B},\Phi}}          \\[1em]
 {\bf L$\otimes$} & \son{\seq{\Sigma,A,B}{\Phi}}
			{\seq{\Sigma,(A \otimes B)}{\Phi}} &
 {\bf R$\otimes$} &
  \son{\seq{\Sigma_1}{A,\Phi_1} \ \ \ \ \ \seq{\Sigma_2}{B,\Phi_2}}
      {\seq{\Sigma_1,\Sigma_2}{(A \otimes B),\Phi_1,\Phi_2}}\\[1em]
 {\bf L$\Par$}   &
  \son{\seq{\Sigma_1,A}{\Phi_1} \ \ \ \ \ \seq{\Sigma_2,B}{\Phi_2}}
      {\seq{\Sigma_1,\Sigma_2,(A \Par B)}{\Phi_1,\Phi_2}} & 
 {\bf R$\Par$}   &
  \son{\seq{\Sigma}{A,B,\Phi}}
      {\seq{\Sigma}{(A \Par B),\Phi}}          \\[1em]
 {\bf L$\oplus$}  &
  \son{\seq{\Sigma,A}{\Phi} \ \ \ \ \seq{\Sigma,B}{\Phi}}
      {\seq{\Sigma,(A \oplus B)}{\Phi}}                 &
 {\bf R$\oplus$}   & 
  \son{\seq{\Sigma}{A,\Phi}}
      {\seq{\Sigma}{(A \oplus B),\Phi}}      \\[1em]
&&
& \son{\seq{\Sigma}{B,\Phi}}
      {\seq{\Sigma}{(A \oplus B),\Phi}}       \\[1em]
 {\bf L$\&$}      &
  \son{\seq{\Sigma,A}{\Phi}}
      {\seq{\Sigma,(A \& B)}{\Phi}}   & 
 {\bf R$\&$}      &
  \son{\seq{\Sigma}{A,\Phi} \ \ \ \ \ \seq{\Sigma}{B,\Phi}}
      {\seq{\Sigma}{(A \& B),\Phi}}              \\[1em]
& \son{\seq{\Sigma,B}{\Phi}}
      {\seq{\Sigma,(A \& B)}{\Phi}}      &&  \\[1em] 
 {\bf L!}         &
  \son{\seq{\Sigma, A}{\Phi}}
      {\seq{\Sigma,\bang{A}}{\Phi}}                        &
 {\bf R!}         &
  \son{\seq{\bang{\Sigma}}{C}}
      {\seq{\bang{\Sigma}}{\bang{C}}}                 \\[1em]
 {\bf W!}         &
  \son{\seq{\Sigma}{\Phi}}
      {\seq{\Sigma,\bang{A}}{\Phi}}                        &
 {\bf C!}         &
  \son{\seq{\Sigma,\bang{A},\bang{A}}{\Phi}}
      {\seq{\Sigma,\bang{A}}{\Phi}}                       \\[1em]
 {\bf L$\bot$}         & \son{}{\seq{\bot}{ }}       &
 {\bf R$\bot$} &
  \son{\seq{\Sigma}{\Phi}}
      {\seq{\Sigma}{\Phi,\bot}}             \\[1em]
 {\bf L\odin}         &
  \son{\seq{\Sigma}{\Phi}}
      {\seq{\Sigma,\odin}{\Phi}}                        &
 {\bf R\odin} & \son{}{\seq{}{\odin}}   \\[1em]
 \hline
 \end{tabular}
 \end{center}
 \caption {The Inference Rules of Linear~Logic.}
 \label{tLL}
 \end{table*}

\section {Normalized Sequents}

 Here we consider formulas of propositional Linear Logic that are
 built up of {\em positive literals}
     $$ p_1,\ p_2,\ \ldots,\ p_m,\ \ldots,\ p_{m},\ \ldots $$
 and {\em constants}
       $$  \bot,\ \odin $$
 by the following {\em connectives:}
 $$ \otimes,\ \Par,\ \llto,\ \&,\ \oplus,\ \mbox{and}\ \bang{} $$
 The inference rules for these {\em connectives} are given
 in Table~\ref{tLL}.

 Without loss of generality, we can confine ourselves to
 {\em normalized sequents,} i.e.~sequents of the form%
 \footnote{ Where \bang{\Gamma} stands for the multiset
            resulting from putting the {\em modal storage}
            operator~! before each formula in~$\Gamma$.}
      $$ \seq {W,\,\Delta,\, \bang{\Gamma}}{Z} $$
 where $W$~and~$Z$ are non-empty {\em tensor products of positive
 literals,}%
 \footnote
 { Henceforth, such products will be called {\em simple products.}}
 $\Gamma$ and $\Delta$ are multisets consisting of {\em Horn implications}
             $$ \limply{X}{Y},$$
 {\em \pH implications}
             $$ \lvariant{X}{Y_1}{Y_2},$$
 {\em \aH implications}
             $$ \lnondet{X_1}{Y_1}{X_2}{Y_2},$$
 and {\em elementary embedded implications}
            $$ \lembed{U}{V}{Y},$$
 here (and henceforth) $X$, $X_1$, $X_2$, $Y$, $Y_1$, $Y_2$,
 $U$, and~$V$ are {\em simple products.}

\begin{definition}  \label{dsimple}
 The tensor product of {\em a positive number} 
 of positive literals is called {\em a simple product.}
 A single literal~$q$ is also called {\em a simple product.}
\end{definition}

\begin{definition}  \label{dmsetprod}
      Taking into account the associativity and commutativity laws, we
 use a natural isomorphism between non-empty finite multisets of positive
 literals and simple products:\\
 A multiset
        $$\{q_1,\ q_2, \ \ldots, \ q_k\}$$
 is represented by the simple product
 $$(q_1 \otimes q_2 \otimes \cdots \otimes q_k),$$
 and vice versa.
\end{definition}

\begin{definition}
 We will write 
        $$ X \cong Y $$
 to indicate that $X$~and~$Y$ represent one and the same multiset~$M$.
\end{definition}

\begin{definition}
 {\em Normalized formulas} are defined as follows:
\bb{a}
\item \mbox{A {\em Horn implication}} is a formula of the form
         $$\limply{X}{Y}.$$
\item \mbox{A {\em $\oplus$-Horn implication}}  is a formula of the form
         $$\lvariant{X}{Y_1}{Y_2}.$$
\item \mbox{An {\em $\&$-Horn implication}} is a formula of the form
             $$ \lnondet{X_1}{Y_1}{X_2}{Y_2}.$$
\item \mbox{An {\em elementary embedded implication}} is a formula of
      the form
            $$ \lembed{U}{V}{Y}.$$
\ee
 Here $X$, $X_1$, $X_2$, $Y$, $Y_1$, $Y_2$, $U$, and~$V$ are
 {\em simple products.}
\end{definition}

 We will consider the {\em Normalized} Fragment of Linear Logic
 \Logic{NLL}{\otimes,\llto,\&,\oplus,\bang{}} that consists of
 such {\em normalized} sequents.

 The most interesting case is as follows.
\begin{theorem} \label{tNLL}
 \mbox{The whole Multiplicative}-Exponential Fragment
 of \mbox{$\bot$-free} Intuitionistic Linear Logic
 \Logic{ILL}{\otimes,\llto,\bang{}} is {\em polynomial-time}
 reducible to its Normalized Fragment
 \Logic{NLL}{\otimes,\llto,\bang{}} containing only
 Horn implications and elementary embedded implications.
 Moreover, under our reduction the depth of implication nesting
 does not increase.
\end{theorem}

 As a first attempt to determine the complexity level
 of the whole~\Logic{LL}{\otimes,\Par,!,?,\bot,\odin}, we have:
\begin{corollary} The derivability problem is decidable for
 the Multiplicative-Exponential Fragment of \mbox{$\bot$-free}
 Intuitionistic Linear Logic \Logic{ILL}{\otimes,\llto,\bang{}}
 consisting of sequents of the form~\seq{\Sigma}{Z}\ 
 where $\Sigma$ contains no embedded implications 
 (unbounded nesting of the storage operator~\bang{} is allowed!).
\end{corollary}

\section {\mbox{Acyclic Programs} \mbox{with Stack}}

 {\em Acyclic programs with stack} will be considered
 as computational counterparts of Linear Logic sequents.

 From the computational point of view, when we intend to use an
 {\em elementary embedded implication} \lembed{U}{V}{Y}, 
 before involving~$Y$ in the computational process, we
 should solve the {\em subtask} of producing~$V$ for the
 given~$U$. It is complicated additionally because we have to
 keep in mind the resource problems related to the current
 value: one part of it should be {\em suspended} together
 with~$Y$, the rest should be incorporated in a solution
 of the foregoing {\em subtask.}

 For these purposes we will use the standard stack operations
 {\em push} and {\em pop}~\cite{Aho} in a {\em resource-fair} manner:
\bb{a}
\item While {\em pushing}, we should indicate explicitly the part
      of the current value that will be involved in a further
      {\em active} computation, the remaining part is {\em suspended}
      in our stack.
      More precisely, the command~\mbox{$PUSH(Y_1; X_2, Y_2)$}
      will mean:
      split the current value into two parts~$X_2$ and, say~$X_1$,
      add the value~\mbox{$(X_1 \otimes Y_1)$} to the top of our
      pushdown store,
      and place the value~\mbox{$(X_2 \otimes Y_2)$} as a new active
      input for a further computation.
\item While {\em popping}, we should indicate explicitly
      that the desired result has been obtained in our
      active computation and, hence, the active memory
      can be cleaned up.
      Formally, the command~\mbox{$POP(V)$} will mean:
      remove the topmost value~$Y$ from our pushdown store
      and place this~$Y$ as a new active input for a further
      computation,
      provided that the desired target~$V$ has been obtained at the
      current point.
\ee
 Without loss of generality, we can confine ourselves to
 studying programs with the following peculiarities:

\begin{definition}
 {\em An acyclic program with stack} is a rooted binary tree such that
\bb{a}
\item Every edge of it is labelled
      either by a Horn implication of the form~\limply{X}{Y},
      or by a {\em push} command of the \mbox{form $PUSH(Y_1; X_2,Y_2)$,}
      or by a {\em pop} command of the form~$POP(V)$.
\item The root of the tree is specified as the {\em input} vertex.
      A~vertex with no outgoing edges will
      be specified as an {\em output}  one.
\item A~vertex~$v$ with exactly two outgoing edges~$(v, w_1)$
      and~$(v, w_2)$ will be called {\em divergent.} These
      two outgoing edges should be labelled by Horn implications
      with one and the same antecedent, say \limply{X}{Y_1} and
      \limply{X}{Y_2}, respectively.
\item On each path~$b$ leading from the input vertex to an output
      vertex, the sequence of~{\em push}'s and~{\em pop}'s should be
      well-blocked: each of~{\em push}'s has the unique partner~{\em pop}.%
      \footnote{ A \mbox{{\em push}-edge} may have different
                 \mbox{{\em pop}-partners} which must belong to
                 different paths.}
\ee
\end{definition}

 Now, we should explain how such a program~$P$ runs for a given
 input~$W$.

\begin{definition} \label{dstrong}
  For a given program~$P$ and any simple product~$W$,
  {\em a strong computation}  is defined by induction
  as follows: we assign a simple product~\OUT{v} and a
  stack~\STACK{v} to each vertex~$v$ of~$P$ in such a way that
\bb{a}
\item For the input vertex~$v$, \STACK{v} is empty and
          $$\OUT{v} = W.$$
\item For any vertex~$v$ and its son~$w$
      with the edge~$(v, w)$ labelled by a Horn implication
       \limply{X}{Y},
      if \OUT{v} is defined and, for some simple product~$V$:
          $$\OUT{v} \cong (X \otimes V),$$
      then
          $$ \OUT{w} = (Y \otimes V) $$
      and $$ \STACK{w} = \STACK{v}.$$
\item For any edge~$(v, w)$ labelled by a {\em push} command of the
        \mbox{form $PUSH(Y_1; X_2,Y_2)$,}
      if \OUT{v} is defined and, for some simple product~$X_1$:
         $$\OUT{v} \cong (X_1 \otimes X_2),$$
      then \STACK{w} is the result of pushing
      \mbox{$(X_1 \otimes Y_1)$} onto the~\STACK{v} and
          $$ \OUT{w} = (X_2 \otimes Y_2).$$
\item For any edge~$(v, w)$ labelled by a {\em pop} command of the
        \mbox{form $POP(V)$,}
      if $$\OUT{v} \cong V$$
      then \STACK{w} is the result of popping a product~$Y$
      from the top of the stack~\STACK{v} and
          $$ \OUT{w} = Y.$$
      Otherwise, \OUT{w} is declared to be undefined.
\ee
\end{definition}

\begin{definition}
  For a program~$P$ and a simple product~$W$,
  we say that $$ P(W) = Z $$
  if and only if, for each output vertex~$v$ of~$P$,
  the stack~\STACK{v} is empty and
      $$\OUT{v} = Z.$$
\end{definition}
    These definitions fall within the paradigm of Linear Logic,
 ensuring that
\bb{a}
 \item the execution of a program does not allow for its
       operators to share their inputs,
 \item after the program has been executed, the pushdown memory
       that was occupied by temporary and auxiliary objects is free.
\ee

 We will describe each of our program constructs by Linear Logic
 formulas. Namely, we will associate a certain formula~$A$
 to each edge~$e$ of a given program~$P$, and say that\\
\centerline{{\em ``This formula~$A$ is used on the edge~$e$.''}}

\begin{definition} Let $P$ be a one-stack acyclic program.
\bb{a}
\item Let $v$ be a non-divergent vertex of~$P$ with an outgoing
      edge~$e$ labelled by a Horn implication~$A$.
      Then we will say that either\\
\centerline{{\em ``Formula~$A$ itself is used on the edge~$e$.''}}\\
      or\\
\centerline{{\em ``Formula~$(A\&B)$ is used on the edge~$e$.''}}\\
      or\\
\centerline{{\em ``Formula~$(B\&A)$ is used on the edge~$e$.''}}\\
      where $B$~is an arbitrary Horn implication.
\item Let $v$ be a divergent vertex of~$P$ with two outgoing
      edges~$e_1$ and~$e_2$ labelled by Horn implications
      \limply{X}{Y_1} and \limply{X}{Y_2}, respectively,
      and let $A$~be the \pH implication
       $$ \lvariant{X}{Y_1}{Y_2}.$$
      Then we will say that\\
\centerline{{\em ``Formula~$A$ is used on~$e_1$.''}}\\
 and\\
\centerline{{\em ``Formula~$A$ is used on~$e_2$.''}}
\item Let $v$ be a non-divergent vertex of~$P$ with an outgoing
      edge~$e$ labelled by a {\em push} command of the
      \mbox{form $PUSH(Y_1; X_2,Y_2)$,}
      and let $A$ be a formula of the form
      $$ \lembed{Y_2}{V}{Y_1}.$$
      We will say that\\
\centerline{{\em ``Formula~$A$ is used on the edge~$e$.''}}\\
      if each of \mbox{{\em pop}-partners} of our \mbox{{\em push}-edge $e$}
      is labelled by a {\em pop} command of the form~$POP(V)$.
\ee
\end{definition}

\begin{definition} A one-stack acyclic program~$P$ is said
 to be {\em a strong solution to}  a sequent of the form
$$ \seq{W_0, W_1, \ldots, W_n,\,\Delta,\,\bang{\Gamma}}{Z} $$
 if
\bb{a}
 \item $ P((W_0\otimes W_1\otimes\cdots\otimes W_n)) = Z.$
 \item For every (non-{\em pop}) edge~$e$ in~$P$, the formula~$A$ used
       on~$e$ is drawn either from~$\Gamma$ or from~$\Delta$.
 \item Whatever path~$b$ leading from the input vertex to an output
       vertex we take, each formula~$A$ from~$\Delta$
       is used once and exactly once on this path~$b$.
\ee
\end{definition}

\begin{theorem}[Completeness]   \label{tlinprog}
 Let $\Gamma$ and $\Delta$ be multisets consisting of normalized
 formulas.\\
 Any sequent of the form
   $$ \seq{W,\,\Delta,\,\bang{\Gamma}}{Z} $$
 is derivable in Linear Logic {\em if and only if} one can construct a
 one-stack acyclic program~$P$ which is a strong solution to
 the given sequent.
\end{theorem}

\proof For a given strong solution~$P$, running from its leaves to its
       root, we can assemble a derivation of our sequent.

 In the other direction we can apply Theorem~\ref{ttobot},
 Lemma~\ref{lEF}, and Theorem~\ref{tprog}.
\QED


\begin{table*}[htp]
$$\begin{array}{|lcl|}\hline&&\\%
 \widetilde{H}_0(p)& = &\cimply{p^\ang{N+2}}{p^\ang{2}},\\[1ex]%
 \widetilde{C}_0(p)& = &\cimply
    {\cimply{\widetilde{H}_0(p)^\ang{2}}{p^\ang{3}}}
                                        {p^\ang{3}},\\[1ex]%
 \widetilde{H}_1(p)& = &\cimply{\widetilde{C}_0(p)^\ang{4}}
                                        {p^\ang{N}},%
\\[1.5ex]\hline\hline&&\\%
 H_{00} & = &\limply{\bot^\good{N+2}}{\bot^\good{2}},\\[1ex]%
 C_{00} & = &\limply{\limply{H_{00}^\good{2}}{\bot^\good{3}}}
                                      {\bot^\good{3}},\\[1ex]%
 H_1 & = &\limply{C_{00}^\good{4}}{\bot^\good{N}},%
\\[1.5ex]\hline&&\\%
 \multicolumn{3}{|c|}
{\begin{array}{@{}lcl}%
 \#_\bot(H_{00}) & = & -N,  \\[1ex]%
 \#_\bot(C_{00}) & = & -2N, \\[1ex]%
 \#_\bot(H_{1})  & = & 9N  =  0 \pmod{9N},\\&&%
\end{array}}%
\\[1.5ex]\hline\hline&&\\%
 H_{01} & = &\limply{\odin^\good{[2]}}{\odin^\good{[N+2]}},\\[1ex]%
 C_{01} & = &\limply{\odin^\good{[3]}}
         {(\odin^\good{[3]}\otimes H_{01}\otimes H_{01})}.\\[1.5ex]\hline
  \end{array}$$
\caption { The basic one-literal and constant-only formulas.}
\label{tbasic}
\end{table*}
\begin{table*}[htp]
\begin{center}
\begin{tabular}{|l|}\hline%
 For literal~$p_m$, we set\\[1ex]%
 \(\begin{array}{@{}lcl}%
 \widetilde{D}_\good{p_m}(p)& = &\cimply
 {\cimply{\widetilde{H}_1(p)}{p^\ang{m+4}}}{p^\ang{m+4}},\\[1ex]%
 D_\good{p_m} & = &\limply{\limply{H_1}{\bot^\good{m+4}}}
                          {\bot^\good{m+4}},\\[1ex]%
 G_\good{p_m} & = &(\odin^\good{[m+4]}\otimes
 \limply{\odin^\good{[m+4]}}{(\odin^\good{[N]}\otimes C_{01}^\good{4})}).%
  \end{array}\)\\ \\ \hline%
 Let a simple product $X$ be of the form\\[1ex]%
\hspace*{5ex}
 \(X = (q_1 \otimes q_2 \otimes\cdots\otimes q_{n-1} \otimes q_n).\)
\\[1ex]%
 Then we set\\[1ex]%
\(\begin{array}{@{}lcl}%
 \widetilde{G}_\good{X}(p)& = &\limply{\widetilde{D}_\good{q_1}(p)}
 {\limply{\widetilde{D}_\good{q_2}(p)}
 {(\ldots \limply{\widetilde{D}_\good{q_{n-1}}(p)}
 {\limply{\widetilde{D}_\good{q_n}(p)}{p}}\ldots)}},\\[1ex]%
 D_\good{X}& = &(D_\good{q_1}\, \otimes \, D_\good{q_2} \,
                 \otimes \cdots \otimes \, D_\good{q_n}),\\[1ex]%
 G_\good{X}& = &(G_\good{q_1}\ \Par \ G_\good{q_2} \
                 \Par \cdots \Par \ G_\good{q_n}).%
  \end{array}\)\\ \\ \hline%
\end{tabular}
\end{center}
\caption { The encoding of literals and tensor products.}
\label{tGD}
\end{table*}
\begin{table*}[htp]
$$\begin{array}{|l|}\hline\\%
\widetilde{E}_\good{X}(p) =
 \limply{\widetilde{C}_0(p)^\ang{6}}{\widetilde{G}_\good{X}(p)},%
\\[2ex]%
\widetilde{F}_\limply{X}{Y}(p) = \limply
 {\widetilde{E}_\good{(p\otimes Y)}(p)}
 {\widetilde{E}_\good{(p\otimes X)}(p)},\\[2ex]%
\widetilde{F}_\good{Y}(p) = \limply
 {\widetilde{E}_\good{(p\otimes Y)}(p)}
 {\widetilde{E}_\good{p}(p)},\\[2ex]%
\widetilde{F}_\lembed{U}{V}{Y}(p)  = \limply
{\limply{\widetilde{F}_\good{Y}(p)}     {p}}
{\limply{\widetilde{F}_\limply{U}{V}(p)}{p}},\\[2ex]%
\widetilde{F}_\lvariant{X}{Y_1}{Y_2}(p) = \limply
 {(\widetilde{E}_\good{(p\otimes Y_1)}(p)\,\&\,
   \widetilde{E}_\good{(p\otimes Y_2)}(p))}
  {\widetilde{E}_\good{(p\otimes X)}(p)},\\[2ex]%
\widetilde{F}_\lnondet{X_1}{Y_1}{X_2}{Y_2}(p) =
 (\widetilde{F}_\limply{X_1}{Y_1}(p) \ \&\
  \widetilde{F}_\limply{X_2}{Y_2}(p)) ,\\[2.5ex]\hline\hline\\%
E_\good{X} = (C_{00}^\good{6} \otimes D_\good{X}),\\[2ex]%
F_\limply{X}{Y} = \limply{E_\good{(p\otimes X)}}
                         {E_\good{(p\otimes Y)}},\\[2ex]%
F_\good{Y} = \limply{E_\good{p}}
                    {E_\good{(p\otimes Y)}},\\[2ex]%
F_\lembed{U}{V}{Y} = \limply{F_\limply{U}{V}}
                            {F_\good{Y}},\\[2ex]%
F_\lvariant{X}{Y_1}{Y_2} = \lvariant{E_\good{(p\otimes X)}}
                      {E_\good{(p\otimes Y_1)}}
                      {E_\good{(p\otimes Y_2)}},\\[2ex]%
F_\lnondet{X_1}{Y_1}{X_2}{Y_2} =
(F_\limply{X_1}{Y_1}\ \&\ F_\limply{X_2}{Y_2}),\\[2.5ex]\hline\hline\\%
E^1_\good{X} = \limply{C_{01}^\good{6}}{G_\good{X}},\\[2ex]%
F^1_\limply{X}{Y} = \limply{E^1_\good{(p\otimes Y)}}
                           {E^1_\good{(p\otimes X)}},\\[2ex]%
F^1_\good{Y} = \limply{E^1_\good{(p\otimes Y)}}
                      {E^1_\good{p}},\\[2ex]%
F^1_\lembed{U}{V}{Y} = \limply{F^1_\limply{U}{V}}
                              {F^1_\good{Y}},\\[2ex]%
F^1_\lvariant{X}{Y_1}{Y_2} = \limply{(E^1_\good{(p\otimes Y_1)}\,\&\,
                                      E^1_\good{(p\otimes Y_2)})}
                                  {E^1_\good{(p\otimes X)}},\\[2ex]%
F^1_\lnondet{X_1}{Y_1}{X_2}{Y_2} =
(F^1_\limply{X_1}{Y_1}\ \&\ F^1_\limply{X_2}{Y_2}).\\[2.5ex]\hline
  \end{array}$$
\caption { Encoding \Logic{ILL}{\otimes,\llto,\oplus,\&}
           into the one-literal, \mbox{$\bot$-only,} and unit-only
           fragments of \mbox{Linear Logic.}}
\label{tEF}
\end{table*}

\section {The Main Encoding}
 Now we demonstrate how to encode normalized sequents into one-literal,
 \mbox{$\bot$-only,} and unit-only fragments of Linear Logic.

\begin{definition} We will use the abbreviation:
$$ A^\good{n} = (\underbrace{A \otimes A \otimes\cdots\otimes A}%
                         _\good{n \ \mbox{times}}). $$
 For\ $n=0$,\ \ $ A^\good{0} = \odin.$\\
 Dually, we will define:
$$ A^\good{[n]} = (\underbrace{A\ \Par\ A \ \Par \cdots \Par\ A}%
                         _\good{n \ \mbox{times}}). $$
 For\ $n=0$,\ \ $ A^\good{[0]} = \bot.$
\end{definition}

\begin{definition} We define
           $$ \limply{A^\ang{n}}{B} $$
 by induction:
$$\begin{array}{rcl}
   \limply{A^\ang{0}}{B} & = & B,\\
 \limply{A^\ang{n+1}}{B} & = & \limply{A}{\limply{A^\ang{n}}{B}}.
\end{array}$$
\end{definition}
 For a given integer~$N$, let
  $$ p_1,\ p_2,\ \ldots,\ p_m,\ \ldots,\ p_{N-7} $$
 be the list of all literals that will be used here and henceforth
 in Linear Logic formulas.

 In particular, we assume that this list includes a certain
 literal~$p$. This {\em leading} literal~$p$ will be involved in
 our encodings only for a more reasonable representation of
 {\em embedded implications} 
            $$ \lembed{U}{V}{Y} $$
 by embedded implications with {\em non-empty} antecedents:
$$ \lembed{(p\otimes U)}{(p\otimes V)}{\limply{p}{(p\otimes Y)}}.$$
 First of all, we specify certain one-literal, \mbox{$\bot$-only},
 and unit-only formulas by Table~\ref{tbasic}.

\begin{definition}
 We will encode each simple tensor product~$X$ by one-literal,
 \mbox{$\bot$-only}, and unit-only
 formulas~$\widetilde{G}_\good{X}(p)$, $D_\good{X}$,
 and~$G_\good{X}$, respectively. \mbox{(See Table~\ref{tGD})}

 According to Table~\ref{tEF}, we encode each normalized formula~$A$
 by one-literal, \mbox{$\bot$-only}, and unit-only
 formulas~$\widetilde{F}_\good{A}(p)$, $F_\good{A}$,
 and~$F^1_\good{A}$, respectively.

 Let $\Gamma$ be a multiset consisting of normalized
 formulas. By $\widetilde{F}_\good{\Gamma}(p)$, $F_\good{\Gamma}$,
 and~$F^1_\good{\Gamma}$ we will denote multisets that are obtained
 from~$\Gamma$ by replacing each formula~$A$ with
 formulas~$\widetilde{F}_\good{A}(p)$, $F_\good{A}$, and~$F^1_\good{A}$,
 respectively.
\end{definition}
\begin{lemma} \label{lEF}
 For any normalized formula~$A$, formulas~$F_\good{A}$,
 $\widetilde{F}_\good{A}(\bot)$, and~$F^1_\good{A}$ are equivalent
 pairwise in Linear Logic.

 As a corollary, the following three sentences are equivalent:
\bb{i}
\item An {\em auxiliary} \mbox{$\bot$-only} sequent of the form
 $$ \seq{E_\good{(p\otimes W)}, F_\good{\Delta},
        \bang{F_\good{\Gamma}}}{E_\good{(p\otimes Z)}}$$
      is derivable in Linear Logic.
\item A \mbox{$\bot$-only} sequent of the form
 $$ \seq{\widetilde{E}_\good{(p\otimes Z)}(\bot),\,
     \widetilde{F}_\good{\Delta}(\bot),
         \, \bang{\widetilde{F}_\good{\Gamma}(\bot)}}
       {\widetilde{E}_\good{(p\otimes W)}(\bot)} $$
      is derivable in Linear Logic, as well.
\item The \mbox{unit-only} sequent
 $$ \seq{E^1_\good{(p \otimes Z)},\, F^1_\good{\Delta},\,
         \bang{F^1_\good{\Gamma}}} {E^1_\good{(p\otimes W)}} $$
      is also derivable in Linear Logic.
\ee
\end{lemma}

\begin{theorem} \label{ttobot}
 Let $\Gamma$ and $\Delta$ be multisets consisting of normalized
 formulas.\\
 If a sequent of the form
 $$ \seq{W,\,\Delta,\,\bang{\Gamma}}{Z} $$
 is derivable in Linear Logic then the following three sequents,
 the one-literal sequent
 $$ \seq{\widetilde{E}_\good{(p\otimes Z)}(p),\,
                  \widetilde{F}_\good{\Delta}(p),
         \, \bang{\widetilde{F}_\good{\Gamma}(p)}}
       {\widetilde{E}_\good{(p\otimes W)}(p)},$$
 the \mbox{$\bot$-only} sequent
 $$ \seq{\widetilde{E}_\good{(p\otimes Z)}(\bot),\,
           \widetilde{F}_\good{\Delta}(\bot),
         \, \bang{\widetilde{F}_\good{\Gamma}(\bot)}}
       {\widetilde{E}_\good{(p\otimes W)}(\bot)},$$
 and the \mbox{unit-only} sequent
 $$ \seq{E^1_\good{(p \otimes Z)},\, F^1_\good{\Delta},\,
         \bang{F^1_\good{\Gamma}}} {E^1_\good{(p \otimes W)}} $$
 are also derivable in Linear Logic.
\end{theorem}

\proof By induction on derivations.
\QED

 Now we should prove the {\em fairness} of our encodings.

 We will kill three birds (one-literal, \mbox{$\bot$-only}, and
 unit-only ones) with one stone.

 Namely, we will prove that {\em all} derivations of an
 {\em auxiliary} \mbox{$\bot$-only} sequent of the form
 $$ \seq{E_\good{(p\otimes W)}, F_\good{\Delta},
        \bang{F_\good{\Gamma}}}{E_\good{(p\otimes Z)}}$$
 {\em must be} {\em regular:} Due to the following key
 technical lemmas, any derivation {\em cannot} develop in
 {\em undesired} directions. Let us demonstrate the crucial
 point of our construction:

\begin{lemma} \label{laxiom}
 Let $\Gamma$ and $\Delta$ be multisets consisting of normalized
 formulas.\\
 Let a sequent of the form
 $$ \seq{(C_{00}^\good{6} \otimes D_\good{W}), F_\good{\Gamma},
    \bang{F_\good{\Delta}}} {(C_{00}^\good{6} \otimes D_\good{Z})}$$
 be derivable in Linear Logic, and let the last step in some
 cut-free derivation of it be performed according
 to rule~{\rm {\em R$\otimes$}}.\\
 Then, as a matter of fact, we meet a trivial {\em axiom} situation:\\
 This multiset~$\Gamma$ must be empty, $\bang{F_\good{\Delta}}$~can
 be produced by rules~{\rm {\em W\bang{}}} and~{\rm {\em C\bang{}}}
 only (there is no applications of rule~{\rm {\em L\bang{}}} in the
 given derivation), and, moreover,
              $$ W \cong Z.$$
\end{lemma}

\proof See \mbox{{\bf Case~of~an Axiom}} in the proof of
       Theorem~\ref{tprog} below.
\QED

 The detailed proof of Lemma~\ref{laxiom} involves a huge number
 of technical lemmas related to derivations of specific sequents.
 All this technical stuff is contained in section~\ref{saxiom}.

 In our proof we exploit the well-known idea that all
 derivable sequents should be {\em well-balanced.}

\begin{definition} The total number~$\#_\bot(A)$
 of positive and negative occurrences of~$\bot$ in~$A$ is defined
 as follows:
$$\begin{array}{lcl}
 \#_\bot(q)     & = & 0, \mbox{for every literal~$q$,}\\
 \#_\bot(\bot)  & = & 1, \\
 \#_\bot(\odin) & = & 0.
 \end{array}$$
 For any formulas~$A$ and~$B$,
$$\begin{array}{lcl}
 \#_\bot((A \otimes B)) & = & \#_\bot(A) + \#_\bot(B),\\
 \#_\bot((A\, \Par\, B) & = & \#_\bot(A) + \#_\bot(B) - 1,\\
 \#_\bot(\limply{B}{A}) & = & \#_\bot(A) - \#_\bot(B),\\
 \#_\bot((A\, \&\, B))  & = & \min \{ \#_\bot(A), \#_\bot(B) \},\\
 \#_\bot((A \oplus B))  & = & \max \{ \#_\bot(A), \#_\bot(B) \}.
 \end{array}$$
\end{definition}
\begin{lemma} \label{lcost}
 For basic \mbox{$\bot$-only} formulas we have:
$$\begin{array}{@{}lcl@{}cl}%
 \#_\bot(H_{00}) & = & -N,&&  \\[1ex]%
 \#_\bot(C_{00}) & = & -2N,&& \\[1ex]%
 \#_\bot(H_{1})  & = & 9N & = & 0 \pmod{9N}.%
\end{array}$$
 For any simple products~$X$, $Y_1$, and~$Y_2$:
$$\begin{array}{@{}lcl}%
 \#_\bot(D_\good{X})  & = & 0   \pmod{9N},\\[1ex]%
 \#_\bot(E_\good{X})  & = & 6N  \pmod{9N},\\[1ex]%
 \#_\bot(E_\good{Y_1} \oplus E_\good{Y_2}) & = & 6N \pmod{9N}.%
\end{array}$$
 For any simple product~$Y$:
$$ \#_\bot(F_\good{Y})  =  0   \pmod{9N}.$$
 For any normalized formula~$A$:
$$ \#_\bot(F_\good{A})  =  0   \pmod{9N}.$$
\end{lemma}
\begin{lemma} \label{L1}
 Let $\Gamma$ and $\Delta$ be multisets consisting of normalized
 formulas, and let \mbox{$A_1, A_2, \ldots, A_k$} and
 \mbox{$B_1, B_2, \ldots, B_m$} be formulas built up of
 constant~$\bot$ by connectives from
 the \mbox{set~$\{ \otimes, \llto \}$.}
 In addition, some of~$A_i$ is allowed to be of the
 form~\mbox{$(E_\good{Y_1} \oplus E_\good{Y_2})$.}\\
 If~a~sequent of the form
$$\seq{A_1, A_2,\ldots, A_k, F_\good{\Gamma},\bang{F_\good{\Delta}}}
      {B_1, B_2, \ldots, B_m} $$
 is derivable in Linear Logic then the following holds:
 $$ \sum_{i=1}^k \#_\bot(A_i) =
    1 - m + \sum_{j=1}^m \#_\bot(B_j)  \pmod{9N}.$$
 In particular, for the empty right-hand side~\mbox{$(m=0):$}
 $$ \sum_{i=1}^k \#_\bot(A_i) = 1  \pmod{9N}.$$
\end{lemma}

\proof  By induction on cut-free derivations.
\QED

 The key {\em fairness} theorem is as follows:

\begin{theorem} \label{tprog}
 Let $\Gamma$ and $\Delta$ be multisets consisting of normalized
 formulas that do not contain a certain literal~$p$.\\
 Let all simple products
    $$ T_1,\ T_2,\ \ldots,\ T_n, \ Z $$
 do not contain this flat literal~$p$, either.\\
 Let~$K$ be a multiset of the form
 $$ K = T_1,\ T_2,\ \ldots,\ T_n.$$
\bb{a}
\item 
 If a \mbox{$\bot$-only} sequent of the form
$$\seq{E_\good{W'},F_\good{K},F_\good{\Delta},\bang{F_\good{\Gamma}}}
       {E_\good{(p\otimes Z)}}$$
 is derivable in Linear Logic then
\bbb{a}{1}
\item the simple product~$W'$ contains exactly one occurrence of
      literal~$p$, and it is of the form
            $$ W' \cong (p\otimes W),$$
\item one can construct a one-stack program~$P$ that is a strong
      solution to the {\em original} sequent
          $$ \seq{W,\,K,\,\Delta,\,\bang{\Gamma}}{Z}.$$
\eee
\item 
      For the case of the {\em 'empty'}~$Z$:

 If a \mbox{$\bot$-only} sequent of the form
$$\seq{E_\good{W'},F_\good{K},F_\good{\Delta},\bang{F_\good{\Gamma}}}
       {E_\good{p}}$$
 is derivable in Linear Logic then
\bbb{a}{1}
\item the simple product~$W'$ consists of one literal~$p$:
            $$ W' = p,$$
\item both multisets~$K$ and~$\Delta$ must be empty,
      and $\bang{F_\good{\Gamma}}$~must be {\em degenerate:}
      $\bang{F_\good{\Gamma}}$~can be produced by
      rules~{\rm {\em W\bang{}}} and~{\rm {\em C\bang{}}} only
      (there is no applications of rule~{\rm {\em L\bang{}}} in the
      derivation above this sequent).
\eee
\ee
\end{theorem}

\proof
 We assemble the desired program~$P$ by induction on a given
 derivation in Linear Logic.

 First of all, regarding to the form of the {\em principal\/} formula
 at a current point of the derivation, we demonstrate inconsistency
 of the following {\em undesirable\/} cases.

 Assume that the {\em principal\/} formula belongs to the left-hand
 side, and it is of the form
 $$ D_\good{q} =
    \limply{\limply{H_1}{\bot^\good{b}}}{\bot^\good{b}} $$
 where
          $$ 4 \ \leq\  b \ \leq\  N-3,$$
 and, according to rule~{\rm {\bf L$\llto$}}, our sequent is
 derived from two sequents of the form
 $$ \seq{C_{00}^\good{k_1}, D_\good{T'}, F_\good{K_1},
    F_\good{\Delta_1}, \bang{F_\good{\Gamma_1}}}
       {\limply{H_1}{\bot^\good{b}}} $$
 and
 $$ \seq{C_{00}^\good{k_2}, D_\good{W_2}, F_\good{K_2},
           \bot^\good{b},
           F_\good{\Delta_2}, \bang{F_\good{\Gamma_2}}}
     {E_\good{(p\otimes Z)}} $$
 where
$$\left\{\begin{array}{lcl}%
    6    & = & k_1 + k_2,\\%
    W'   & = & (q \otimes T' \otimes W_2),\\%
    K    & = & K_1,\,K_2,\\%
 \Delta  & = & \Delta_1,\,\Delta_2,\\%
  \Gamma & = & \Gamma_1,\,\Gamma_2.%
\end{array}\right.$$
 Then Lemma~\ref{L1} and Lemma~\ref{lcost} yield a contradiction:
$$\left\{\begin{array}{lcl}%
       -2Nk_1     & = & b  \pmod{9N},\\%
       -2Nk_2 + b & = & 6N \pmod{9N}.\\%
\end{array}\right.$$
 If our sequent were derived from two sequents of the form
 $$ \seq{C_{00}^\good{k_1}, D_\good{T'}, F_\good{K_1},
    F_\good{\Delta_1}, \bang{F_\good{\Gamma_1}}}
       {\limply{H_1}{\bot^\good{b}}, E_\good{(p\otimes Z)}} $$
 and
 $$ \seq{C_{00}^\good{k_2}, D_\good{W_2}, F_\good{K_2},
           \bot^\good{b},
           F_\good{\Delta_2}, \bang{F_\good{\Gamma_2}}}{ } $$
 then we had a contradiction as well:
$$\left\{\begin{array}{lcl}%
       -2Nk_1     & = & b+6N-1  \pmod{9N},\\%
       -2Nk_2 + b & = & 1 \pmod{9N}.\\%
\end{array}\right.$$

 Assume that the left-hand {\em principal\/} formula is of the form
 $$ C_{00} = \limply{\limply{H_{00}^\good{2}}{\bot^\good{3}}}
                    {\bot^\good{3}},$$
 and, according to rule~{\rm {\bf L$\llto$}}, our sequent is
 derived from two sequents of the form
 $$ \seq{C_{00}^\good{k_1}, D_\good{T'}, F_\good{K_1},
    F_\good{\Delta_1}, \bang{F_\good{\Gamma_1}}}
       {\limply{H_{00}^\good{2}}{\bot^\good{3}}} $$
 and
 $$ \seq{C_{00}^\good{k_2}, D_\good{W_2}, F_\good{K_2},
           \bot^\good{3},
           F_\good{\Delta_2}, \bang{F_\good{\Gamma_2}}}
     {E_\good{(p\otimes Z)}} $$
 where
$$\left\{\begin{array}{lcl}%
    5    & = & k_1 + k_2,\\%
    W'   & = & (T' \otimes W_2),\\%
    K    & = & K_1,\,K_2,\\%
 \Delta  & = & \Delta_1,\,\Delta_2,\\%
  \Gamma & = & \Gamma_1,\,\Gamma_2.%
\end{array}\right.$$
 Then, by Lemma~\ref{L1} and Lemma~\ref{lcost}, the following
 contradiction is immediate:
$$\left\{\begin{array}{lcl}%
       -2Nk_1     & = &  2N + 3 \pmod{9N},\\%
       -2Nk_2 + 3 & = &  6N \pmod{9N}.\\%
\end{array}\right.$$
 If our sequent were derived from two sequents of the form
 $$ \seq{C_{00}^\good{k_1}, D_\good{T'}, F_\good{K_1},
    F_\good{\Delta_1}, \bang{F_\good{\Gamma_1}}}
       {\limply{H_{00}^\good{2}}{\bot^\good{3}},
         E_\good{(p\otimes Z)} } $$
 and
 $$ \seq{C_{00}^\good{k_2}, D_\good{W_2}, F_\good{K_2},
           \bot^\good{3},
           F_\good{\Delta_2}, \bang{F_\good{\Gamma_2}}}{ } $$
 then we had also a contradiction:
$$\left\{\begin{array}{lcl}%
       -2Nk_1     & = &  8N + 2 \pmod{9N},\\%
       -2Nk_2 + 3 & = &  1 \pmod{9N}.\\%
\end{array}\right.$$

 Thus Lemma~\ref{L1} and Lemma~\ref{lcost} contract eventually the
 set of all possible cases to the following set.
 
 \noindent\mbox{{\bf Case~of~a formula from~$F_\good{K}$.}}
 Suppose that the {\em principal\/} formula is from~$F_\good{K}$,
 and it is of the form
       $$ \limply{E_\good{p}}{E_\good{(p\otimes Y)}},$$
 and, according to rule~{\rm {\bf L$\llto$}}, our sequent is
 derived from two sequents of the form
 $$ \seq{C_{00}^\good{k_1}, D_\good{T'}, F_\good{K_1},
    F_\good{\Delta_1}, \bang{F_\good{\Gamma_1}}}
        {E_\good{p}} $$
 and
 $$ \seq{C_{00}^\good{k_2}, D_\good{W_2}, F_\good{K_2},
            E_\good{(p\otimes Y)},
           F_\good{\Delta_2}, \bang{F_\good{\Gamma_2}}}
     {E_\good{(p\otimes Z)}} $$
 where
$$\left\{\begin{array}{lcl}%
    6    & = & k_1 + k_2,\\%
    W'   & = & (T' \otimes W_2),\\%
    K    & = & K_1,\, K_2,\,
               \limply{E_\good{p}}{E_\good{(p\otimes Y)}},\\%
 \Delta  & = & \Delta_1,\,\Delta_2,\\%
  \Gamma & = & \Gamma_1,\,\Gamma_2.%
\end{array}\right.$$
 According to Lemma~\ref{L1} and Lemma~\ref{lcost}:
$$\left\{\begin{array}{lcl}%
      -2Nk_1     & = & 6N \pmod{9N},\\%
      -2Nk_2+6N  & = & 6N \pmod{9N},%
\end{array}\right.$$
 and, hence,
$$\left\{\begin{array}{lcl}%
      k_1  & = & 6,\\%
      k_2  & = & 0.%
\end{array}\right.$$
 In the case of item~(b) we have a contradiction that, by
 the inductive hypothesis:
        $$ (W_2 \otimes p\otimes Y) = p $$
 for {\em non-empty}~$Y$.

 The case of item~(a) is handled in the following way.

 By applying the inductive hypothesis from item~(b), we have:
\bb{1}
\item the simple product~$T'$ is trivial:
            $$ T' = p,$$
\item both multisets~$K_1$ and~$\Delta_1$ must be empty,
\ee
 which results in:
$$\left\{\begin{array}{lcl}%
    W'   & = & (p\otimes W_2),\\%
    K    & = & K_2,\, F_\good{(p\otimes Y)},\\%
 \Delta  & = & \Delta_2,\\%
  \Gamma & = & \Gamma_1,\,\Gamma_2.%
\end{array}\right.$$
 The inductive hypothesis from item~(a) yields:
\bb{1}
\item the simple product~$W_2$ does not contain any occurrence of
      literal~$p$,
\item there exists a one-stack program~$P$ that is a strong
      solution to the sequent
$$ \seq{(W_2\otimes Y),\,K_2,\,\Delta_2,\,\bang{\Gamma_2}}{Z}.$$
\ee
 Just the same program~$P$ will be also a strong solution to the
 sequent
$$ \seq{W_2,\,K,\,\Delta,\,\bang{\Gamma}}{Z}.$$
 Let us note that if, according to rule~{\rm {\bf L$\llto$}}, our
 sequent were derived from two sequents of the form
 $$ \seq{C_{00}^\good{k_1}, D_\good{T'}, F_\good{K_1},
    F_\good{\Delta_1}, \bang{F_\good{\Gamma_1}}}
        {E_\good{p}, E_\good{(p\otimes Z)}} $$
 and
 $$ \seq{C_{00}^\good{k_2}, D_\good{W_2}, F_\good{K_2},
            E_\good{(p\otimes Y)},
           F_\good{\Delta_2}, \bang{F_\good{\Gamma_2}}}{ } $$
 then we got the following contradiction:
$$\left\{\begin{array}{lcl}%
      -2Nk_1     & = & 12N-1 \pmod{9N},\\%
      -2Nk_2+6N  & = & 1 \pmod{9N}.%
\end{array}\right.$$

\newcommand{ \IMPLY }[1]%
{\begin{picture}(0,0) \thicklines
 \Ycur=14
 \put(0,-\Ycur){\Node{\ \ $(W_1\otimes K_1\otimes W_2\otimes K_2)$}
               {\toRight}}
 \put(0,-\Ycur){\South{#1}{\ \fbox{$P_1$}}{\toRight}}
 \advance \Ycur by #1
 \put(0,-\Ycur){\NodeXIV{$v_1$}{\ \ $(X \otimes W_2\otimes K_2)$}
               {\toRight}}
 \advance \Ycur by 14
 \put(0,-\Ycur){\South{#1}{\ \limply{X}{Y}}{\toRight}}
 \advance \Ycur by #1
 \put(0,-\Ycur){\NodeXIV{$v_2$}{\ $(Y \otimes W_2\otimes K_2)$}
               {\toRight}}
 \advance \Ycur by 14
 \put(0,-\Ycur){\Tree{#1}{$P_2$}{$\,Z$}}
\end{picture}}

\begin{figure}[htp]
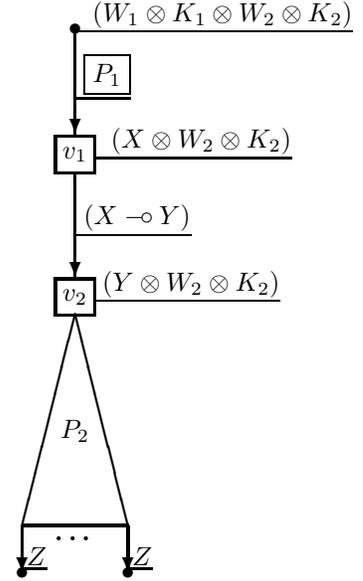

\begin{center}      \vPICTURE {\IMPLY} {4}{4}{1} {40}
\end{center}
\caption { The Horn Implication.}
\label{fIMPLY}
\end{figure}

 \noindent\mbox{{\bf Case~of~a Horn implication.}}
 Our sequent is of the form
$$ \seq{E_\good{W'}, F_\good{K}, F_\limply{X}{Y},
  F_\good{\Delta},\bang{F_\good{\Gamma}}}{E_\good{(p\otimes Z)}} $$
 and, according to rule~{\rm {\bf L$\llto$}}, it is derived from
 the following two sequents:
 $$ \seq{C_{00}^\good{k_1}, D_\good{T'}, F_\good{K_1},
    F_\good{\Delta_1}, \bang{F_\good{\Gamma_1}}}
        {E_\good{(p\otimes X)}} $$
 and
 $$ \seq{C_{00}^\good{k_2}, D_\good{W_2}, F_\good{K_2},
            E_\good{(p\otimes Y)},
           F_\good{\Delta_2}, \bang{F_\good{\Gamma_2}}}
     {E_\good{(p\otimes Z)}} $$
 where
$$\left\{\begin{array}{lcl}%
    6    & = & k_1 + k_2,\\%
    W'   & = & (T' \otimes W_2),\\%
    K    & = & K_1,\,K_2.\\%
 \Delta  & = & \Delta_1,\,\Delta_2,\\%
  \Gamma & = & \Gamma_1,\,\Gamma_2.%
\end{array}\right.$$
 According to Lemma~\ref{L1} and Lemma~\ref{lcost}:
$$\left\{\begin{array}{lcl}%
      -2Nk_1     & = & 6N \pmod{9N},\\%
      -2Nk_2+6N  & = & 6N \pmod{9N},%
\end{array}\right.$$
 and, hence,
$$\left\{\begin{array}{lcl}%
      k_1  & = & 6,\\%
      k_2  & = & 0.%
\end{array}\right.$$
 In the case of item~(b) we have a contradiction that, by
 the inductive hypothesis:
        $$ (W_2 \otimes p\otimes Y) = p $$
 for {\em non-empty}~$Y$.

 For the case of item~(a), both these sequents can be rewritten as
 $$ \seq{E_\good{T'}, F_\good{K_1}, F_\good{\Delta_1},
            \bang{F_\good{\Gamma_1}}}{E_\good{(p\otimes X)}} $$
 and
 $$ \seq{E_\good{(W_2 \otimes p\otimes Y)}, F_\good{K_2},
          F_\good{\Delta_2}, \bang{F_\good{\Gamma_2}}}
        {E_\good{(p\otimes Z)}},$$
 respectively.

 \noindent
 By the inductive hypothesis, for some~$W_1$:
          $$ T' \cong (p\otimes W_1),$$
 and \mbox{$(W_1 \otimes W_2)$} does not contain literal~$p$.

 According to the inductive hypothesis, suppose that $P_1$~is a
 strong solution to a sequent of the form
 $$ \seq{W_1,\,K_1,\,\Delta_1,\,\bang{\Gamma_1}}{X},$$
 and $P_2$ is a strong solution to a sequent of the form
$$\seq{(W_2 \otimes Y),\,K_2,\,\Delta_2,\,\bang{\Gamma_2}}{Z}.$$

 Now a program~$P$ is assembled by the following
 \mbox{(see~Figure~\ref{fIMPLY}):}
\bb{a}
\item For each output vertex, say~$v_1$, of program~$P_1$,
      we connect this vertex~$v_1$ with the root~$v_2$ of~$P_2$
      by a new edge~$(v_1,v_2)$ and label this edge by the Horn
      implication \limply{X}{Y}.
\ee

 It is easily verified that our program~$P$ is a strong solution to
 the sequent
 $$ \seq{(W_1 \otimes W_2), K, \limply{X}{Y},
                            \Delta, \bang{\Gamma}}{Z}.$$

 If our sequent were derived from two sequents of the form
 $$ \seq{C_{00}^\good{k_1}, D_\good{T'}, F_\good{K_1},
    F_\good{\Delta_1}, \bang{F_\good{\Gamma_1}}}
        {E_\good{(p\otimes X)}, E_\good{(p\otimes Z)}} $$
 and
 $$ \seq{C_{00}^\good{k_2}, D_\good{W_2}, F_\good{K_2},
       E_\good{(p\otimes Y)},
     F_\good{\Delta_2}, \bang{F_\good{\Gamma_2}}}{ } $$
 then, by Lemma~\ref{L1} and Lemma~\ref{lcost}, we had
 a contradiction:
$$\left\{\begin{array}{lcl}%
      -2Nk_1      & = & 12N - 1 \pmod{9N},\\%
      -2Nk_2 + 6N & = & 1 \pmod{9N}.%
\end{array}\right.$$

 \noindent\mbox{{\bf Case~of~an \aH implication.}}
 Our sequent is of the form
 $$ \seq{E_\good{W'}, K, F_\lnondet{X_1}{Y_1}{X_2}{Y_2},
   F_\good{\Delta},\bang{F_\good{\Gamma}}}{E_\good{(p\otimes Z)}} $$
 and it is derived either from the sequent
 $$ \seq{E_\good{W'}, K, F_\limply{X_1}{Y_1},
   F_\good{\Delta},\bang{F_\good{\Gamma}}}{E_\good{(p\otimes Z)}} $$
 or from the sequent
 $$ \seq{E_\good{W'}, K, F_\limply{X_2}{Y_2},
   F_\good{\Delta},\bang{F_\good{\Gamma}}}{E_\good{(p\otimes Z)}} $$
 In item~(a) by the inductive hypothesis, for some~$W$:
          $$ W' \cong (p\otimes W),$$
 and we have a program~$P$ that is a strong solution to one of the
 following sequents:
 $$ \seq{W, K, \Delta, \limply{X_1}{Y_1}, \bang{\Gamma}}{Z} $$
 or
 $$ \seq{W, K, \Delta, \limply{X_2}{Y_2}, \bang{\Gamma}}{Z}.$$
 This $P$~will be also a strong solution to the sequent
 $$ \seq{W,K,\lnondet{X_1}{Y_1}{X_2}{Y_2},\Delta,\bang{\Gamma}}{Z}.$$
 For the case of item~(b) we have a contradiction that, by
 the inductive hypothesis, one of these {\em non-empty} multisets
    $$ \Delta, \limply{X_1}{Y_1} $$
 and
    $$ \Delta, \limply{X_2}{Y_2},$$
 must be {\em empty.}

\newcommand{ \PUSHPOP }[1]%
{\begin{picture}(0,0) \thicklines
 \Ycur=14
 \put(0,-\Ycur){\NodeXIV{$v_0$}{\ \ $(X_1 \otimes X_2)$}{\toRight}}
 \advance \Ycur by 14
 \put(0,-\Ycur){\South{#1}{\ $PUSH(Y; X_2,U)$}{\toRight}}
 \advance \Ycur by #1
 \put(0,-\Ycur){\NodeXIV{$v_1$}{\ \ $(X_2 \otimes U)$}{\toRight}}
 \advance \Ycur by 14
 \Xcur=#1 \multiply \Xcur by 2
 \put(0,-\Ycur){\Tree{\Xcur}{$P_2$}{}}
 \advance \Ycur by \Xcur \advance \Ycur by \Xcur  \advance \Ycur by 14
 \put(-#1,-\Ycur){\NodeXIV{$w_1$}{$V$\ \ }{\toLeft}}
 \put( #1,-\Ycur){\NodeXIV{$w_n$}{\ \ $V$}{\toRight}}
 \advance \Ycur by 14
 \put(-#1,-\Ycur){\South{#1}{$POP(V)$\ }{\toLeft}}
 \put(  0,-\Ycur){\makebox(0,0)[t]{{\bf \ldots}}}
 \put( #1,-\Ycur){\South{#1}{\ $POP(V)$}{\toRight}}
 \advance \Ycur by #1
 \put(-#1,-\Ycur){\NodeXIV{$t_1$}{$(X_1 \otimes Y)$\ \ }{\toLeft}}
 \put(  0,-\Ycur){\makebox(0,0)[t]{{\bf \ldots}}}
 \put( #1,-\Ycur){\NodeXIV{$t_n$}{\ \ $(X_1 \otimes Y)$}{\toRight}}
 \advance \Ycur by 14
 \put(-#1,-\Ycur){\Tree{#1}{$P_1$}{$\,Z$}}
 \put( #1,-\Ycur){\Tree{#1}{$P_1$}{$\,Z$}}
\end{picture}}

\begin{figure}[htp]
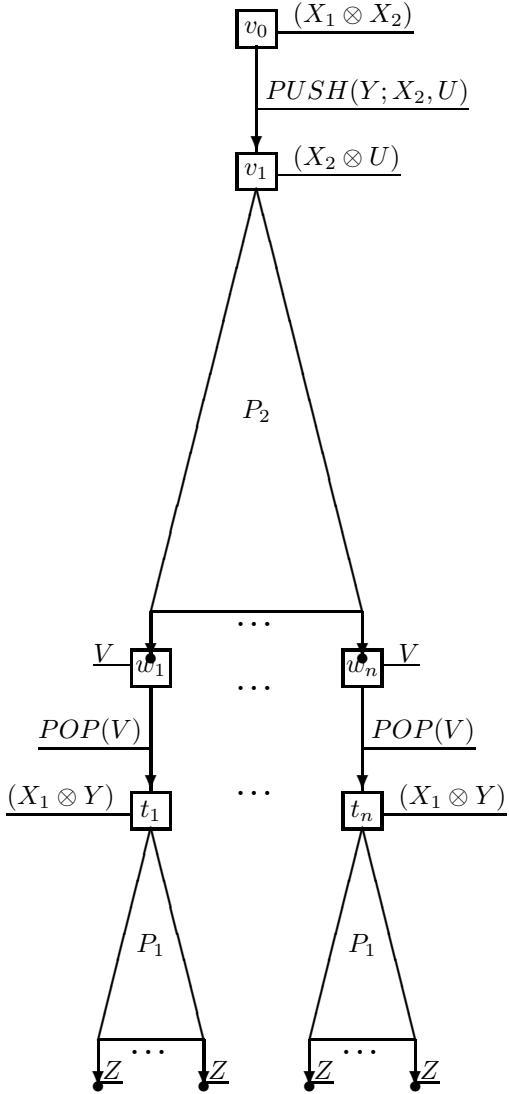

\begin{center}      \vPICTURE {\PUSHPOP} {8}{7}{3} {40}
\end{center}
\caption { The Stack Operations.}
\label{fStack}
\end{figure}

 \noindent\mbox{{\bf Case~of~an embedded implication.}}
 Our sequent is of the form
$$ \seq{E_\good{W'}, F_\good{K}, F_\lembed{U}{V}{Y},
  F_\good{\Delta},\bang{F_\good{\Gamma}}}{E_\good{(p\otimes Z)}} $$
 and it is derived from the following two sequents:
 $$ \seq{C_{00}^\good{k_1}, D_\good{T'}, F_\good{K_1}, F_\good{Y},
    F_\good{\Delta_1}, \bang{F_\good{\Gamma_1}}}
        {E_\good{(p\otimes Z)}} $$
 and
 $$ \seq{C_{00}^\good{k_2}, D_\good{W_2}, F_\good{K_2},
            F_\good{\Delta_2}, \bang{F_\good{\Gamma_2}}}
     {F_\limply{U}{V}} $$
 where
$$\left\{\begin{array}{lcl}%
    6    & = & k_1 + k_2,\\%
    W'   & = & (T' \otimes W_2),\\%
    K    & = & K_1,\,K_2,\\%
 \Delta  & = & \Delta_1,\,\Delta_2,\\%
  \Gamma & = & \Gamma_1,\,\Gamma_2.%
\end{array}\right.$$
 By Lemma~\ref{L1} and Lemma~\ref{lcost}, we have
$$\left\{\begin{array}{lcl}%
      -2Nk_1     & = & 6N \pmod{9N},\\%
      -2Nk_2     & = & 0 \pmod{9N},%
\end{array}\right.$$
 and 
$$\left\{\begin{array}{lcl}%
      k_1  & = & 6,\\%
      k_2  & = & 0.%
\end{array}\right.$$
 In the case of item~(b) we have a contradiction that, by
 the inductive hypothesis, the {\em non-empty} multiset
    $$  F_\good{Y},\ F_\good{\Delta_1} $$
 must be {\em empty.}

 For item~(a), our sequent can be derived from the following two
 sequents:
 $$ \seq{E_\good{T'}, F_\good{K_1}, F_\good{Y},
            F_\good{\Delta_1},\bang{F_\good{\Gamma_1}}}
        {E_\good{(p\otimes Z)}} $$
 and
 $$ \seq{E_\good{(W_2 \otimes p\otimes U)}, F_\good{K_2},
          F_\good{\Delta_2}, \bang{F_\good{\Gamma_2}}}
        {E_\good{(p\otimes V)}}.$$
 By the inductive hypothesis, for some~$W_1$:
          $$ T' \cong (p\otimes W_1),$$
 and \mbox{$(W_1 \otimes W_2)$} does not contain literal~$p$.

 According to the inductive hypothesis, suppose that
 $P_1$ and~$P_2$ are strong solutions to sequents of the form
     $$ \seq{W_1, Y, K_1,\,\Delta_1,\,\bang{\Gamma_1}}{Z},$$
 and
  $$ \seq{(W_2 \otimes U), K_2,\,\Delta_2,\,\bang{\Gamma_2}}{V},$$
 respectively.\\
 Let us set
$$\left\{\begin{array}{lcl}%
    X_1   & = & (W_1 \otimes \bigotimes K_1),\\%
    X_2   & = & (W_2 \otimes \bigotimes K_2).%
\end{array}\right.$$
 Now a one-stack program~$P$ can be assembled as follows
 \mbox{(see~Figure~\ref{fStack}):}
\bb{a}
\item First, we create a new input vertex~$v_0$.
\item After that, we connect this input vertex~$v_0$ with the
      root~$v_1$ of~$P_2$ by a new edge~$(v_0,v_1)$ and label
      this edge by the {\bf push} operation~\mbox{$PUSH(Y; X_2,U)$.}
\item Finally, we connect each output vertex~$w_k$ of program~$P_2$
      with the root~$t_k$ of \mbox{$k$-th copy} of program~$P_1$ by
      a new edge~$(w_k,t_k)$ and label this edge by the {\bf pop}
      operation~\mbox{$POP(V)$.}
\ee
 We can verify that our program~$P$ is a strong solution to the
 sequent
 $$ \seq{(W_1 \otimes W_2), K, \lembed{U}{V}{Y},
             \Delta, \bang{\Gamma}}{Z}.$$

 If our sequent were derived from two sequents of the form
 $$ \seq{C_{00}^\good{k_1}, D_\good{T'}, F_\good{K_1}, F_\good{Y},
    F_\good{\Delta_1}, \bang{F_\good{\Gamma}}}{ } $$
 and
 $$ \seq{C_{00}^\good{k_2}, D_\good{W_2}, F_\good{K_2},
            F_\good{\Delta_2}, \bang{F_\good{\Gamma}}}
     {F_\limply{U}{V}, E_\good{(p\otimes Z)}} $$
 then we got an immediate contradiction:
$$\left\{\begin{array}{lcl}%
      -2Nk_1     & = & 1 \pmod{9N},\\%
      -2Nk_2     & = & 6N-1 \pmod{9N}.%
\end{array}\right.$$

\newcommand{ \DIVERGE }[1]%
{\begin{picture}(0,0) \thicklines
 \Ycur=14
 \put(0,-\Ycur){\Node{\ \ $(W_1\otimes K_1\otimes W_2\otimes K_2)$}
               {\toRight}}
 \put(0,-\Ycur){\South{#1}{\ \fbox{$P_0$}}{\toRight}}
 \advance \Ycur by #1
 \put(0,-\Ycur){\NodeXIV{$v_0$}{\ \ $(X \otimes W_2\otimes K_2)$}
               {\toRight}}
 \advance \Ycur by 14
 \put(0,-\Ycur){\SouthWest{#1}{\limply{X}{Y_1}\ }{\toLeft}}
 \put(0,-\Ycur){\SouthEast{#1}{\ \limply{X}{Y_2}}{\toRight}}
 \advance \Ycur by #1
 \put(-#1,-\Ycur){\NodeXIV{$v_1$}{$(Y_1 \otimes W_2\otimes K_2)$\ }
                 {\toLeft}}
 \put( #1,-\Ycur){\NodeXIV{$v_2$}{\ $(Y_2 \otimes W_2\otimes K_2)$}
                 {\toRight}}
 \advance \Ycur by 14
 \put(-#1,-\Ycur){\Tree{#1}{$P_1$}{$\,Z$}}
 \put( #1,-\Ycur){\Tree{#1}{$P_2$}{$\,Z$}}
\end{picture}}

\begin{figure*}[htp]
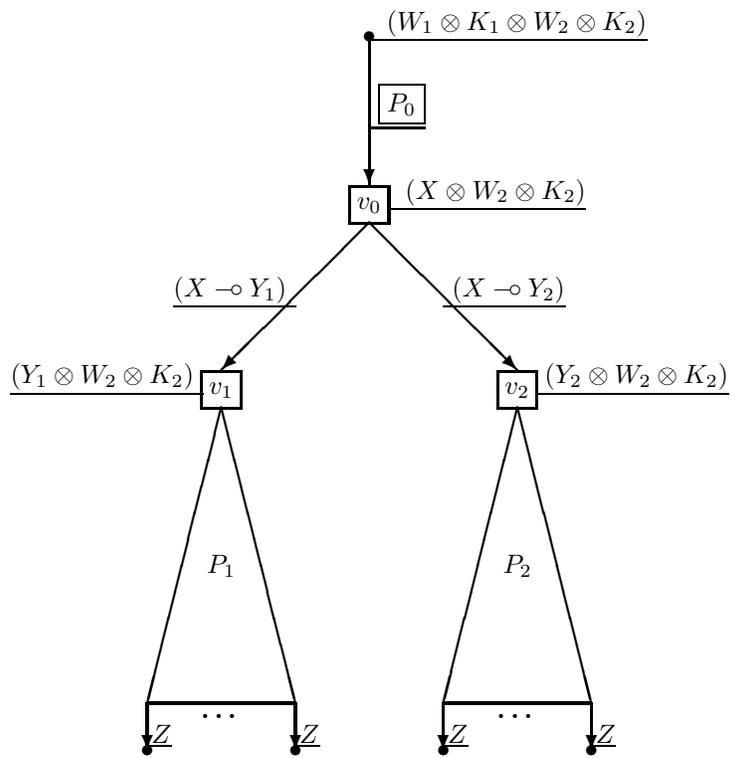

\begin{center}      \vPICTURE {\DIVERGE} {4}{4}{3} {56}
\end{center}
\caption { Strong Forking.}
\label{fDIV}
\end{figure*}

 \noindent\mbox{{\bf Case~of~a \pH implication.}}
 Our sequent is of the form
$$ \seq{E_\good{W'}, F_\good{K}, F_\lvariant{X}{Y_1}{Y_2},
  F_\good{\Delta},\bang{F_\good{\Gamma}}}{E_\good{(p\otimes Z)}} $$
 and, taking into account Lemma~\ref{L1} and Lemma~\ref{lcost},
 it is derived from the following three sequents:
 $$ \seq{E_\good{T'}, F_\good{K_1}, F_\good{\Delta_1},
            \bang{F_\good{\Gamma_1}}}
        {E_\good{(p\otimes X)}},$$
 $$ \seq{E_\good{(W_2 \otimes p\otimes Y_1)}, F_\good{K_2},
          F_\good{\Delta_2}, \bang{F_\good{\Gamma_2}}}
        {E_\good{(p\otimes Z)}},$$
 and
 $$ \seq{E_\good{(W_2 \otimes p\otimes Y_2)}, F_\good{K_2},
          F_\good{\Delta_2}, \bang{F_\good{\Gamma_2}}}
        {E_\good{(p\otimes Z)}},$$
 where
$$\left\{\begin{array}{lcl}%
    W'   & = & (T' \otimes W_2),\\%
    K    & = & K_1,\,K_2,\\%
 \Delta  & = & \Delta_1,\,\Delta_2,\\%
  \Gamma & = & \Gamma_1,\,\Gamma_2.%
\end{array}\right.$$
 In the case of item~(b) we have a contradiction that, by
 the inductive hypothesis:
        $$ (W_2 \otimes p\otimes Y_1) = p $$
 for {\em non-empty}~$Y_1$.

 For item~(a), by the inductive hypothesis, for some~$W_1$:
          $$ T' \cong (p\otimes W_1),$$
 and \mbox{$(W_1 \otimes W_2)$} does not contain literal~$p$.

 Suppose that $P_0$ is a strong solution to a sequent
 of the form
  $$ \seq{W_1,\,K_1,\,\Delta_1,\,\bang{\Gamma_1}}{X},$$
 and $P_1$ and~$P_2$ are strong solutions to sequents
 of the form
$$ \seq{(Y_1 \otimes W_2),\,K_2,\,\Delta_2,\,\bang{\Gamma_2}}{Z} $$
 and
$$ \seq{(Y_2 \otimes W_2),\,K_2,\,\Delta_2,\,\bang{\Gamma_2}}{Z},$$
 respectively.

 Now a program~$P$ can be assembled by the following
 {\bf Strong Forking} \mbox{(see~Figure~\ref{fDIV}):}
\bb{a}
\item For each output vertex, say~$v_0$, of program~$P_0$,
      we connect this vertex~$v_0$ with the
      root~$v_1$ of~$P_1$ by a new edge~$(v_0,v_1)$ and label
      this edge by the Horn implication \limply{X}{Y_1}.
     
\item In its turn, we connect this vertex~$v_0$ with the
      root~$v_2$ of~$P_2$ by a new edge~$(v_0,v_2)$ and label
      this edge by the Horn implication \limply{X}{Y_2}.
\ee

 It is easily verified that our program~$P$ is a strong solution to
 the sequent
 $$ \seq{(W_1 \otimes W_2), K, \lvariant{X}{Y_1}{Y_2},
                            \Delta, \bang{\Gamma}}{Z}.$$

\noindent\mbox{{\bf Case~of~a formula from~\bang{F_\good{\Gamma}}.}}
 Suppose that the {\em principal\/} formula belongs
 to~\bang{F_\good{\Gamma}}, and it is of the form
              $$ \bang{F_\good{A}}.$$ 
 Assume that it is produced by rule~{\rm {\bf L\bang{}}}, and our
 sequent is derived from a sequent of the form
$$\seq{E_\good{W'},F_\good{K},F_\good{\Delta},
       F_\good{A}, \bang{F_\good{\Gamma'}}}
       {E_\good{(p\otimes Z)}}.$$
 Then item~(a) can be completed by the inductive hypothesis.
 As for item~(b), in this case we have a contradiction that
 the {\em non-empty} multiset
      $$ \Delta,\ A $$
 must be {\em empty.}

 The remaining cases of rules~{\rm {\bf W\bang{}}}
 and~{\rm {\bf C\bang{}}} are readily completed by the inductive
 hypothesis.

\newcommand{ \Id }[1]%
{\begin{picture}(0,0)  \thicklines
 \put(0,-14){\Node{\ $Z$}{\toRight}}
\end{picture}}

\begin{figure}[htp]
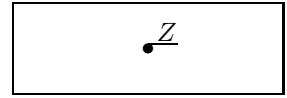

\begin{center} \fbox{\vPICTURE {\Id} {0}{2}{2} {48}}
\end{center}
\caption { The Axiom Case.}
\label{fId}
\end{figure}

 \noindent\mbox{{\bf Case~of~an Axiom.}}
 Suppose that the right-hand formula is {\em principal,\/}
 our sequent is of the form
$$\seq{C_{00}^\good{6},D_\good{W'},F_\good{K},
          F_\good{\Delta},\bang{F_\good{\Gamma}}}
      {(C_{00}^\good{6} \otimes D_\good{(p\otimes Z)})} $$
 and, according to rule~{\rm {\bf R$\otimes$}}, it is derived from
 the following two sequents:
 $$ \seq{C_{00}, D_\good{W_1}, F_\good{K_1},
       F_\good{\Delta_1},\bang{F_\good{\Gamma_1}}}{C_{00}} $$
 and
 $$ \seq{C_{00}^\good{5}, D_\good{T'},
   F_\good{K_2}, F_\good{\Delta_2}, \bang{F_\good{\Gamma_2}}}
        {(C_{00}^\good{5} \otimes D_\good{(p\otimes Z)})} $$
 where
$$\left\{\begin{array}{lcl}%
    W'   & = & (W_1 \otimes T'),\\%
    K    & = & K_1,\,K_2,\\%
 \Delta  & = & \Delta_1,\,\Delta_2,\\%
  \Gamma & = & \Gamma_1,\,\Gamma_2.%
\end{array}\right.$$
 According to Lemma~\ref{LC00}, we have:
$$\left\{\begin{array}{ll}%
    W_1                    & \mbox{must be empty,}\\%
    K_1                    & \mbox{must be empty,}\\%
    \Delta_1               & \mbox{must be empty,}\\%
  \bang{F_\good{\Gamma_1}} & \mbox{must be {\em degenerate.}}
\end{array}\right.$$
 Lemma~\ref{Laxiom} demonstrates that:
$$\left\{\begin{array}{l}%
           T' \cong (p\otimes Z),\\%
         K_2\ \mbox{must be empty,}\\%
    \Delta_2\ \mbox{must be empty,}\\%
  \bang{F_\good{\Gamma_2}}\ \mbox{must be {\em degenerate.}}
\end{array}\right.$$
 Hence, for item~(a) we can conclude that
\bb{a}
\item $ W' \cong (p\otimes Z),$
\item and the most trivial program~$P$ consisting of single vertex
 \mbox{(see~Figure~\ref{fId})} will be a strong solution to the
 corresponding  sequent
 $$ \seq{Z,\,K,\,\Delta,\,\bang{\Gamma}}{Z}.$$
\ee
 In the case of item~(b) we have the desired:
\bb{a}
\item $ W' = p,$
\item and the whole multisets~$K$ and~$\Delta$ are empty,
      and the whole~$\bang{F_\good{\Gamma}}$ is {\em degenerate.}
\ee
 Finally, bringing together all the cases considered, we can
 complete the proof of Theorem~\ref{tprog}.
\QED

\begin{corollary} \label{cprog}
 Let $\Gamma$ and $\Delta$ be multisets consisting of normalized
 formulas.\\
 Let literal~$p$ do not occur in a sequent of the form
         $$ \seq{W,\Delta,\bang{\Gamma}}{Z}.$$
 Then the sequent
         $$ \seq{W,\Delta,\bang{\Gamma}}{Z} $$
 is derivable in Linear Logic {\em if and only if} the
 {\em auxiliary} \mbox{$\bot$-only} sequent
 $$ \seq{E_\good{(p\otimes W)}, F_\good{\Delta},
        \bang{F_\good{\Gamma}}}{E_\good{(p\otimes Z)}} $$
 is also derivable in Linear Logic.
\end{corollary}

\proof The first implication (from the left to the right) can
 be proved by induction on derivations.

 In the other direction, by applying Theorem~\ref{tprog}, we
 construct a strong solution~$P$ to the sequent
         $$ \seq{W,\Delta,\bang{\Gamma}}{Z}.$$
 After that, for such a program~$P$, running from its leaves to its
 root, we assemble a derivation of this sequent.
\QED

\begin{corollary}[Fairness] \label{cfair}
 Let $\Gamma$ and $\Delta$ be multisets consisting of normalized
 formulas.\\
 Let literal~$p$ do not occur in a sequent of the form
         $$ \seq{W,\Delta,\bang{\Gamma}}{Z}.$$
 The following sentences are equivalent pairwise:
\bb{a}
\item A sequent of the form
       $$ \seq{W,\,\Delta,\,\bang{\Gamma}}{Z} $$
      is derivable in Linear Logic.
\item The one-literal sequent
 $$ \seq{\widetilde{E}_\good{(p\otimes Z)}(p),\,
           \widetilde{F}_\good{\Delta}(p),
         \, \bang{\widetilde{F}_\good{\Gamma}(p)}}
       {\widetilde{E}_\good{(p\otimes W)}(p)} $$
      is also derivable in Linear Logic.
\item A \mbox{$\bot$-only} sequent of the form
 $$ \seq{\widetilde{E}_\good{(p\otimes Z)}(\bot),\,
     \widetilde{F}_\good{\Delta}(\bot),
         \, \bang{\widetilde{F}_\good{\Gamma}(\bot)}}
       {\widetilde{E}_\good{(p\otimes W)}(\bot)} $$
      is derivable in Linear Logic, as well.
\item The following \mbox{unit-only} sequent
 $$ \seq{E^1_\good{(p \otimes Z)},\, F^1_\good{\Delta},\,
         \bang{F^1_\good{\Gamma}}} {E^1_\good{(p\otimes W)}} $$
      is derivable in Linear Logic.
\ee
\end{corollary}

\proof One direction is provided by Theorem~\ref{ttobot}.
 The most complicated implications are provided by
 Corollary~\ref{cprog} and Lemma~\ref{lEF}.
\QED

\noindent 
 {\bf Remark.}\ \ In our proof we use also the fact that the derivable
                  sequents in question must be {\bf well-balanced}
                  with respect to the {\em leading literal~$p$} as well.
 In fact, we need this {\em leading literal~$p$} only for simulating
 embedded implications
            $$ \lembed{U}{V}{Y} $$
 by embedded implications with {\em non-empty} antecedents:
 $$ \lembed{(p\otimes U)}{(p\otimes V)}{\limply{p}{(p\otimes Y)}}.$$

\begin{corollary} \label{cmain}
 Theorems~\ref{tONE}, \ref{tBOT}, and~\ref{tUNIT} are valid.
\end{corollary}

\proof According to Corollary~\ref{cfair}, we can apply all complexity
 constructions from~\cite{lics} and Theorem~\ref{tNLL}.
\QED

\section {The Proof of Key Technical Lemmas} \label{saxiom}

\subsection {Lemma~5.1}

\begin{lemma} \label{L2}
 Let $a$~be an integer such that
         $$ 1\ \leq\  a\ \leq\ N + 2.$$
 Let~$\Delta$ consist of formulas of the
 form~$F_\good{A}$, and~$\Gamma$ consist of formulas of the
 form~$H_1$, $D_\good{X}$, $F_\good{A}$, and~$F_\good{Y}$.
\bb{a}
\item Let~$B$ be a formula either of the form~$\bot^\good{a}$,
      or of the form~$\limply{H_1}{\bot^\good{a}}$, or of the
      form~$C_{00}^\good{4}$.

 If~a~sequent of the form
  $$ \seq{\bot^\good{a},\, \Gamma,\, \bang{\Delta}}{B} $$
 occurs in a cut-free derivation in Linear Logic
 then this formula~$B$ must be exactly equal to~$\bot^\good{a}$,
 multiset~$\Gamma$ must be empty, and $\bang{\Delta}$~can be
 produced by rules~{\rm {\bf W\bang{}}} and~{\rm {\bf C\bang{}}}
 only (there is no applications of rule~{\rm {\bf L\bang{}}} in the
 derivation above this sequent).\footnote%
{ We will say that this~$\bang{\Delta}$ is {\em degenerate.}}
\item For the case of the 'empty' formula~$B$:
 
 If~a~sequent of the form
  $$ \seq{\bot, \, \Gamma,\, \bang{\Delta}}{} $$
 occurs in a cut-free derivation in Linear Logic
 then $\Gamma$~must be empty, and $\bang{\Delta}$~can be produced
 by rules~{\rm {\bf W\bang{}}} and~{\rm {\bf C\bang{}}} only.
\item Let~$B$ be a formula of the form~$C_{00}^\good{5}$.

 Any sequent of the form
$$ \seq{C_{00},\,\bot^\good{a},\,\Gamma,\,\bang{\Delta}}{B} $$
 does not occur in any derivations in Linear Logic.
\ee
\end{lemma}

\proof {
 We use induction on a given derivation. Regarding to the form of
 the {\em principal\/} formula at a current point of the derivation,
 we will demonstrate that each of the {\em undesirable\/} cases is
 inconsistent.

\CASE{0} The {\em principal\/} formula belongs to~$\bang{\Delta}$.
 
 Assume that it is produced by rule~{\rm {\bf L\bang{}}},
 and our sequent is derived from a sequent of the form
$$\seq{\bot^\good{a},\,\Gamma,\, F_\good{A},\,\bang{\Delta'}}{B}.$$
 Then, by the inductive hypothesis, the multiset
      $$ \Gamma,\, F_\good{A} $$
 must be empty, which is a contradiction.

 Hence, the only possibility is to apply either~{\rm {\bf W\bang{}}}
 or~{\rm {\bf C\bang{}}}. It remains to use the inductive hypothesis
 for completing this case.

 Item~(c) is handled similarly.

\CASE{1} Formula~$B$ is {\em principal.}

 There are the following subcases to be considered.

\CASE{1.1} Formula~$B$ is of the form~$\bot^\good{a}$, and,
 according to rule~{\rm {\bf R$\otimes$}}, our sequent is derived
 from two sequents of the form
$$\seq{\bot^\good{a_1},\,\Gamma_1,\,\bang{\Delta_1}}{\bot}$$
and
$$\seq{\bot^\good{a_2},\,\Gamma_2,\,\bang{\Delta_2}}
      {\bot^\good{a-1}} $$
where
$$\left\{\begin{array}{lcl}%
      a  & = & a_1 + a_2,\\%
 \Gamma  & = & \Gamma_1,\,\Gamma_2,\\%
 \Delta  & = & \Delta_1,\,\Delta_2.%
\end{array}\right.$$
By Lemma~\ref{L1} we have
$$\left\{\begin{array}{lcl}%
 a_1  & = & 1      \pmod{9N},\\%
 a_2  & = & a - 1  \pmod{9N},%
\end{array}\right.$$
and, therefore,
$$\left\{\begin{array}{lcl}%
 a_1  & = & 1,\\%
 a_2  & = & a - 1.%
\end{array}\right.$$
 By applying the inductive hypothesis to both sequents, we
 get the {\em emptiness} of both~$\Gamma_1$ and~$\Gamma_2$,
 and the {\em degeneracy} of both~$\bang{\Delta_1}$
 and~$\bang{\Delta_2}$, which results in the desired {\em emptiness}
 of the whole~$\Gamma$ and the {\em degeneracy} of the
 whole~$\bang{\Delta}$.

\CASE{1.2} Assume that formula~$B$ is of the
 form~$\limply{H_1}{\bot^\good{a}}$, and,
 by rule~{\rm {\bf R$\llto$}}, our sequent is derived
 from the sequent
$$\seq{\bot^\good{a},\,\Gamma,\,\bang{\Delta},\,H_1}{\bot^\good{a}}.$$
 Then, according to the inductive hypothesis, the multiset
      $$ \Gamma, \  H_1 $$
 must be empty, which is a contradiction.

\CASE{1.3} Assume that formula~$B$ is of the form~$C_{00}^\good{m}$,
           \ \ \mbox{$ m = 4,5 $,}\ \ and, according to
           rule~{\rm {\bf R$\otimes$}}, our sequent is derived from
           two sequents of the form
$$\seq{C_{00}^\good{k_1},\bot^\good{a_1},\Gamma_1,\bang{\Delta_1}}
      {C_{00}}$$
 and
$$\seq{C_{00}^\good{k_2},\bot^\good{a_2},\Gamma_2,\bang{\Delta_2}}
      {C_{00}^\good{m-1}} $$
 where
$$\left\{\begin{array}{lcl}%
      1  & \geq & k_1 + k_2,\\%
      a  & = & a_1 + a_2,\\%
 \Gamma  & = & \Gamma_1,\,\Gamma_2,\\%
 \Delta  & = & \Delta_1,\,\Delta_2.%
\end{array}\right.$$
 Then Lemma~\ref{L1} and Lemma~\ref{lcost} show:
$$\left\{\begin{array}{lcl}%
 -2Nk_1 + a_1  & = & -2N       \pmod{9N},\\%
 -2Nk_2 + a_2  & = & -2N(m-1)  \pmod{9N}.%
\end{array}\right.$$
 The only solution of this system is the following:
         $$\left\{\begin{array}{lcl}%
                       k_2  & = & 0,\\%
                       a_2  & = & N,\\%
                       m    & = & 5,%
                    \end{array}\right.$$
 which yield a contradiction because, according to the inductive
 hypothesis, the latter sequent cannot occur in our derivation.

\CASE{2} Assume that the {\em principal\/} formula belongs
          to~$\Gamma$, and it is of the form~$F_\good{A}$ 
          \mbox{(or~$F_\good{Y}$).}

 The following subcases are to be considered.

\CASE{2.0} The {\em principal\/} formula is of the
           form~\mbox{$(F_\good{A_1} \& F_\good{A_2})$,}
 and, by rule~{\rm {\bf L$\&$}}, our sequent is derived either
 from the sequent
$$\seq{\bot^\good{a},\,\Gamma',\, F_\good{A_1},\,\bang{\Delta}}
      {B}$$
 or from the sequent
$$\seq{\bot^\good{a},\,\Gamma',\, F_\good{A_2},\,\bang{\Delta}}
      {B}.$$
 Then, according to the inductive hypothesis, either 
 the multiset
      $$ \Gamma', \ F_\good{A_1} $$
 or the multiset
      $$ \Gamma', \ F_\good{A_2} $$
 must be empty, which is a contradiction.

 Item~(c) is handled similarly.

\CASE{2.1} Assume that the {\em principal\/} formula is of the
           form~\limply{E_\good{X}}{E_\good{Y}}, and,
 according to rule~{\rm {\bf L$\llto$}}, our sequent is derived
 from two sequents of the form
$$\seq{C_{00}^\good{k_1},\bot^\good{a_1},\Gamma_1,\bang{\Delta_1}}
      {E_\good{X}}$$
and
$$\seq{C_{00}^\good{k_2},\bot^\good{a_2},E_\good{Y},
            \Gamma_2,\bang{\Delta_2}}{B} $$
 where
$$\left\{\begin{array}{lcl}%
      1  &  \geq  & k_1 + k_2,\\%
      a  & =      & a_1 + a_2,\\%
 \Gamma  &\supset & \Gamma_1,\,\Gamma_2,\\%
 \Delta  & = & \Delta_1,\,\Delta_2.%
\end{array}\right.$$
 Then Lemma~\ref{L1} and Lemma~\ref{lcost} yield:
$$\left\{\begin{array}{lcl}%
 -2Nk_1 + a_1       & = &  6N         \pmod{9N},\\%
 -2Nk_2 + a_2 + 6N  & = &  \#_\bot(B) \pmod{9N},%
\end{array}\right.$$
 which is a contradiction.

\CASE{2.2} Assume that the {\em principal\/} formula is of the
           form~\limply{E_\good{X}}{E_\good{Y}}, and,
 by rule~{\rm {\bf L$\llto$}}, our sequent is derived
 from two sequents of the form
$$\seq{C_{00}^\good{k_1},\bot^\good{a_1},\Gamma_1,\bang{\Delta_1}}
      {E_\good{X}, B}$$
and
$$\seq{C_{00}^\good{k_2},\bot^\good{a_2},E_\good{Y},
            \Gamma_2,\bang{\Delta_2}}{ } $$
 Then Lemma~\ref{L1} and Lemma~\ref{lcost} yield:
$$\left\{\begin{array}{lcl}%
 -2Nk_1 + a_1       & = &  6N + \#_\bot(B) - 1 \pmod{9N},\\%
 -2Nk_2 + a_2 + 6N  & = &  1 \pmod{9N},%
\end{array}\right.$$
 which is also a contradiction.

\CASE{2.3} Assume that the {\em principal\/} formula is of the
           form~\limply{F_\good{A}}{F_\good{Y}}, and,
 according to rule~{\rm {\bf L$\llto$}}, our sequent is derived
 from two sequents of the form
$$ \seq{C_{00}^\good{k_1},\bot^\good{a_1},
   \Gamma_1, \bang{\Delta_1}}{F_\good{A}} $$
and
$$\seq{C_{00}^\good{k_2},\bot^\good{a_2},F_\good{Y},
                          \Gamma_2,\bang{\Delta_2}}{B} $$
 where
$$\left\{\begin{array}{lcl}%
      1  &  \geq  & k_1 + k_2,\\%
      a  & = & a_1 + a_2,\\%
 \Gamma &\supset & \Gamma_1,\,\Gamma_2,\\%
 \Delta  & = & \Delta_1,\,\Delta_2.%
\end{array}\right.$$
 Then Lemma~\ref{L1} and Lemma~\ref{lcost} yield:
$$\left\{\begin{array}{lcl}%
  -2Nk_1 + a_1  & = &  0  \pmod{9N},\\%
  -2Nk_2 + a_2  & = &  \#_\bot(B) \pmod{9N}.%
\end{array}\right.$$
 Hence,
         $$ k_1 = 0,$$
 and either
         $$ a_2 = a,$$
 or (for the case of the 'empty'~$B$)
         $$ a_2 = 1.$$
 By the inductive hypothesis, we can get a contradiction that the
 {\em non-empty} multiset
      $$ F_\good{Y},\ \Gamma_2 $$
 must be empty.

\CASE{2.4} Assume that the {\em principal\/} formula is of the
           form~\limply{F_\good{A}}{F_\good{Y}}, and,
 according to rule~{\rm {\bf L$\llto$}}, our sequent is derived
 from two sequents of the form
$$ \seq{C_{00}^\good{k_1},\bot^\good{a_1},
   \Gamma_1, \bang{\Delta_1}}{F_\good{A}, B} $$
and
$$\seq{C_{00}^\good{k_2},\bot^\good{a_2},F_\good{Y},
                          \Gamma_2,\bang{\Delta_2}}{ } $$
 where
$$\left\{\begin{array}{lcl}%
      1  &  \geq  & k_1 + k_2,\\%
      a  & = & a_1 + a_2,\\%
 \Gamma &\supset & \Gamma_1,\,\Gamma_2,\\%
 \Delta  & = & \Delta_1,\,\Delta_2.%
\end{array}\right.$$
 Then Lemma~\ref{L1} and Lemma~\ref{lcost} yield:
$$\left\{\begin{array}{lcl}%
  -2Nk_1 + a_1  & = & \#_\bot(B) - 1 \pmod{9N},\\%
  -2Nk_2 + a_2  & = & 1              \pmod{9N}.%
\end{array}\right.$$
 Therefore, 
         $$\left\{\begin{array}{lcl}%
                       k_2  & = & 0,\\%
                       a_2  & = & 1,%
                    \end{array}\right.$$
 and, by the inductive hypothesis, the multiset
      $$ F_\good{Y},\ \Gamma_2 $$
 must be empty, which is a contradiction as well.

\CASE{2.5} Case of the {\em principal\/} formula of the
 form~\lvariant{E_\good{X}}{E_\good{Y_1}}{E_\good{Y_2}} is handled
 similarly to~\mbox{{\bf Case 2.1}} and~\mbox{{\bf Case 2.2.}}

\CASE{3} Assume that the {\em principal\/} formula belongs
         to~$\Gamma$, and it is of the form
$$D_\good{q} = \limply{\limply{H_1}{\bot^\good{b}}}{\bot^\good{b}}$$
 where
          $$ 4 \ \leq\  b \ \leq\   N-3.$$

\CASE{3.1} According to rule~{\rm {\bf L$\llto$}}, let our sequent
           be derived from two sequents of the form
$$\seq{C_{00}^\good{k_1},\bot^\good{a_1},\Gamma_1,\bang{\Delta_1}}
      {\limply{H_1}{\bot^\good{b}}} $$
and
$$\seq{C_{00}^\good{k_2},\bot^\good{a_2+b},
                   \Gamma_2,\bang{\Delta_2}}{B} $$
 where
$$\left\{\begin{array}{lcl}%
      1  &  \geq  & k_1 + k_2,\\%
      a  & = & a_1 + a_2,\\%
 \Gamma &\supset & \Gamma_1,\,\Gamma_2,\\%
 \Delta  & = & \Delta_1,\,\Delta_2.%
\end{array}\right.$$
 Then Lemma~\ref{L1} and Lemma~\ref{lcost} yield:
$$\left\{\begin{array}{lcl}%
  -2Nk_1 + a_1      & = &  b         \pmod{9N},\\%
  -2Nk_2 + a_2 + b  & = &  \#_\bot(B) \pmod{9N},%
\end{array}\right.$$
 and, hence,
         $$\left\{\begin{array}{lcl}%
                       k_1  & = & 0,\\%
                       a_1  & = & b.%
                    \end{array}\right.$$
 According to the inductive hypothesis, the right-hand side of
 the first sequent must be exactly~$\bot^\good{b}$, and,
 therefore, such a sequent cannot occur in our derivation.

\CASE{3.2} Let our sequent be derived from two sequents of the form
$$\seq{C_{00}^\good{k_1},\bot^\good{a_1},\Gamma_1,\bang{\Delta_1}}
      {\limply{H_1}{\bot^\good{b}}, B} $$
and
$$\seq{C_{00}^\good{k_2},\bot^\good{a_2+b},
                   \Gamma_2,\bang{\Delta_2}}{ } $$
 where
$$\left\{\begin{array}{lcl}%
      1  &  \geq  & k_1 + k_2,\\%
      a  & = & a_1 + a_2,\\%
 \Gamma &\supset & \Gamma_1,\,\Gamma_2,\\%
 \Delta  & = & \Delta_1,\,\Delta_2.%
\end{array}\right.$$
 Then, by Lemma~\ref{L1} and Lemma~\ref{lcost} we have:
$$\left\{\begin{array}{lcl}%
  -2Nk_1 + a_1      & = & b + \#_\bot(B) - 1 \pmod{9N},\\%
  -2Nk_2 + a_2 + b  & = & 1                  \pmod{9N},%
\end{array}\right.$$
 which is also a contradiction because of
       $$ 4 \ \leq\ a_2 + b \ \leq\ 2N-1.$$

\CASE{4} Assume that the {\em principal\/} formula belongs
         to~$\Gamma$, and it is of the form
  $$ H_1 = \limply{C_{00}^\good{4}}{\bot^\good{N}}.$$

\CASE{4.1} According to rule~{\rm {\bf L$\llto$}}, let our
           sequent be derived from two sequents of the form
$$\seq{C_{00}^\good{k_1},\bot^\good{a_1},\Gamma_1,\bang{\Delta_1}}
      {C_{00}^\good{4}} $$
 and
$$\seq{C_{00}^\good{k_2},\bot^\good{a_2+N},
                  \Gamma_2,\bang{\Delta_2}}{B} $$
 where
$$\left\{\begin{array}{lcl}%
      1  &  \geq  & k_1 + k_2,\\%
      a  & = & a_1 + a_2,\\%
 \Gamma &\supset & \Gamma_1,\,\Gamma_2,\\%
 \Delta  & = & \Delta_1,\,\Delta_2.%
\end{array}\right.$$
 Then Lemma~\ref{L1} and Lemma~\ref{lcost} show:
$$\left\{\begin{array}{lcl}%
  -2Nk_1 + a_1      & = &  -8N       \pmod{9N},\\%
  -2Nk_2 + a_2 + N  & = &  \#_\bot(B) \pmod{9N}.%
\end{array}\right.$$
 Therefore,
         $$\left\{\begin{array}{lcl}%
                       k_1  & = & 0,\\%
                       a_1  & = & N,%
                    \end{array}\right.$$
 and, according to the inductive hypothesis, the first sequent
 with its {\em wrong} right-hand side cannot occur in our
 derivation.

\CASE{4.2}
          Let our sequent be derived from two sequents of the form
$$\seq{C_{00}^\good{k_1},\bot^\good{a_1},\Gamma_1,\bang{\Delta_1}}
      {C_{00}^\good{4}, B} $$
 and
$$\seq{C_{00}^\good{k_2},\bot^\good{a_2+N},
                  \Gamma_2,\bang{\Delta_2}}{ } $$
 where
$$\left\{\begin{array}{lcl}%
      1  &  \geq  & k_1 + k_2,\\%
      a  & = & a_1 + a_2,\\%
 \Gamma &\supset & \Gamma_1,\,\Gamma_2,\\%
 \Delta  & = & \Delta_1,\,\Delta_2.%
\end{array}\right.$$
 By Lemma~\ref{L1} and Lemma~\ref{lcost} we have:
$$\left\{\begin{array}{lcl}%
         -2N(k_1+k_2) + a  & = & \#_\bot(B) \pmod{9N},\\%
         -2Nk_2 + a_2 + N  & = & 1          \pmod{9N}.%
\end{array}\right.$$
 Assuming that\ \ \mbox{$k_2 = 0$,}\ \ we get a contradiction:
             $$ a_2 + (N-1) = 0.$$
 \mbox{ For\ \ $k_2 = 1$,}\ \ we get also a contradiction as follows:
         $$\left\{\begin{array}{lcl}%
                          B & = & C_{00}^\good{5},\\%
                 \#_\bot(B) & = & -10N,           \\%
                       a    & = & N,              \\%
                       a_2  & = & N+1.%
                    \end{array}\right.$$

\CASE{5} Finally, for item~(c), assume that the left-hand
         {\em principal\/} formula is of the form
 $$ C_{00} = \limply{\limply{H_{00}^\good{2}}{\bot^\good{3}}}
                    {\bot^\good{3}},$$
 and, according to rule~{\rm {\bf L$\llto$}}, our sequent
 of item~(c) is derived from two sequents of the form
$$ \seq{\bot^\good{a_1},\,\Gamma_1,\,\bang{\Delta_1}}
       {\limply{H_{00}^\good{2}}{\bot^\good{3}}} $$
 and
$$ \seq{\bot^\good{a_2+3},\,\Gamma_2,\,\bang{\Delta_2}}{B} $$
 where
$$\left\{\begin{array}{lcl}%
       a  & = & a_1 + a_2,\\%
  \Gamma  & = & \Gamma_1,\,\Gamma_2,\\%
  \Delta  & = & \Delta_1,\,\Delta_2.%
\end{array}\right.$$
 Then, by Lemma~\ref{L1} and Lemma~\ref{lcost}, the following
 contradiction is immediate:
$$\left\{\begin{array}{lcl}%
              a_1     & = & 2N + 3  \pmod{9N},\\%
              a_2 + 3 & = & -10N    \pmod{9N}.%
\end{array}\right.$$
 If our sequent of item~(c) were derived from two sequents of the form
$$ \seq{\bot^\good{a_1},\,\Gamma_1,\,\bang{\Delta_1}}
       {\limply{H_{00}^\good{2}}{\bot^\good{3}}, B} $$
 and
$$ \seq{\bot^\good{a_2+3},\,\Gamma_2,\,\bang{\Delta_2}}{ } $$
 then we got a contradiction as well:
$$\left\{\begin{array}{lcl}%
              a_1     & = & -8N + 2 \pmod{9N},\\%
              a_2 + 3 & = & 1       \pmod{9N}.%
\end{array}\right.$$
 Now, bringing together all the cases considered, we can
 complete the proof of Lemma~\ref{L2}.
\QED }

\subsection {Lemma~5.2}

\begin{lemma} \label{LC00}
 Let~$\Delta$ consist of formulas of the
 form~$F_\good{A}$, and~$\Gamma$ consist of formulas of the
 form~$H_1$, $D_\good{X}$, $F_\good{A}$, and~$F_\good{Y}$.
\bb{a}
\item
 Let~$K$ be a multiset either of the form
        $$ H_{00},$$
 or of the form
        $$ H_{00},\ H_{00},$$
 or of the form
 $$ \underbrace{C_{00},\ C_{00},\ \ldots,\ C_{00}}%
               _\good{k \ \mbox{times}} $$
 \mbox{where\ \ $1 \leq k \leq 5 $.}\\
 Let~$B$ be a formula either of the form~\mbox{$H_{00}$,}
 or of the form~\mbox{$H_{00}^\good{2}$,} or of the
 form~\mbox{$C_{00}^\good{m}$,} \mbox{where\ \ $1 \leq  m \leq 5 $.}

 If~a~sequent of the form
  $$ \seq{K,\, \Gamma,\, \bang{\Delta}}{B} $$
 occurs in a cut-free derivation in Linear Logic
 then $\Gamma$~must be empty, and $\bang{\Delta}$~can be produced by
 rules~{\rm {\bf W\bang{}}} and~{\rm {\bf C\bang{}}} only (there is
 no applications of rule~{\rm {\bf L\bang{}}} in the derivation above
 this sequent).\footnote%
 { We say that this~$\bang{\Delta}$ is {\em degenerate.}}
\item
 If~a~sequent of the form
$$ \seq{\limply{H_{00}^\good{2}}{\bot^\good{3}},\,
         \Gamma,\, \bang{\Delta}}
    {\limply{H_{00}^\good{2}}{\bot^\good{3}}} $$
 occurs in a cut-free derivation in Linear Logic
 then $\Gamma$~must be empty, and $\bang{\Delta}$~can be produced
 by rules~{\rm {\bf W\bang{}}} and~{\rm {\bf C\bang{}}} only.
\item
 Let~$K$ be a multiset either of the form
        $$ C_{00} $$
 or of the form
        $$ H_{00},\ H_{00}.$$
 If~a~sequent of the form
  $$ \seq{K,\,\limply{H_{00}^\good{2}}{\bot^\good{3}},\,
          \Gamma,\, \bang{\Delta}}{\bot^\good{3}} $$
 occurs in a cut-free derivation in Linear Logic
 then $\Gamma$~must be empty, and $\bang{\Delta}$~can be produced
 by rules~{\rm {\bf W\bang{}}} and~{\rm {\bf C\bang{}}} only.
\item
 Let $a$~be an integer such that
               $$ 1 \leq a \leq 2.$$

 If~a~sequent of the form
  $$ \seq{H_{00}, \,\bot^\good{N+a},\,\Gamma,\,\bang{\Delta}}
         {\bot^\good{a}} $$
 occurs in a cut-free derivation in Linear Logic
 then this integer~$a$ must be equal exactly to~$2$,
 multiset~$\Gamma$ must be empty, and $\bang{\Delta}$~can be
 produced by rules~{\rm {\bf W\bang{}}} and~{\rm {\bf C\bang{}}} only.
\item Any sequent of the form 
 $$ \seq{H_{00}, \,\bot^\good{N+1},\,\Gamma,\,\bang{\Delta}}{ } $$
 does not occur in derivations in Linear Logic.
\ee
\end{lemma}

\proof {
 We use induction on a given derivation. Regarding to the form of
 the {\em principal\/} formula at a current point of the derivation,
 we will demonstrate that each of the {\em undesirable\/} cases is
 inconsistent.

\CASE{0} The {\em principal\/} formula belongs to~$\bang{\Delta}$.
 
 Assume that it is produced by rule~{\rm {\bf L\bang{}}},
 and our sequent of the form
$$\seq{\ldots,\,\Gamma,\,\bang{\Delta}}{\ldots} $$
 is derived from a sequent of the form
$$\seq{\ldots,\,\Gamma,\, F_\good{A},\,\bang{\Delta'}}{\ldots} $$
 Then, by the inductive hypothesis, the multiset
      $$ \Gamma,\, F_\good{A} $$
 must be empty, which is a contradiction.

 Hence, the only possibility is to apply either~{\rm {\bf W\bang{}}}
 or~{\rm {\bf C\bang{}}}. It remains to use the inductive hypothesis
 for completing this case.

\CASE{1} The right-side formula is {\em principal.}

 There are the following cases to be considered.

\CASE{1.a} For item~(a), let us note that for any subset~$K'$
           of multiset~$K$:
$$\left[\begin{array}{@{}l}%
  \mbox{either}\ \#_\bot(K') = -N,\\%
  \mbox{or}\ \#_\bot(K') = -2Nk',\ \mbox{for some $k'\leq 5$.}%
\end{array}\right.$$
 Let us consider four possible versions of the {\em principal\/}
 formula~$B$.

\CASE{1.a.1} The {\em principal\/} formula~$B$ is of the form
 $$ H_{00}  = \limply{\bot^\good{N+2}}{\bot^\good{2}},$$
 and, according to rule~{\rm {\bf R$\llto$}}, our sequent
 of item~(a) is derived from the sequent
 $$ \seq{H_{00},\,\bot^\good{N+2},\,\Gamma,\,\bang{\Delta}}
        {\bot^\good{2}}.$$
 Here we have accounted that, by Lemma~\ref{L1},
      $$ \#_\bot(K) = \#_\bot(B) = -N.$$
 Now we can apply the inductive hypothesis from item~(d).

\CASE{1.a.2} The {\em principal\/} formula~$B$ is of the
              form~$H_{00}^\good{2}$, and, according to
              rule~{\rm {\bf R$\otimes$}} and Lemmas~\ref{lcost}
              and~\ref{L1}, our sequent of item~(a) is derived from
              two sequents of the form
$$ \seq{H_{00},\,\Gamma_1,\,\bang{\Delta_1}}{H_{00}} $$
 and
$$ \seq{H_{00},\,\Gamma_2,\,\bang{\Delta_2}}{H_{00}} $$
 where
$$\left\{\begin{array}{lcl}%
    \Gamma  & = & \Gamma_1,\,\Gamma_2,\\%
    \Delta  & = & \Delta_1,\,\Delta_2.%
         \end{array}\right.$$
 By applying the inductive hypothesis from item~(c), we get the
 {\em emptiness} of both~$\Gamma_1$ and~$\Gamma_2$, and the
 {\em degeneracy} of both~$\bang{\Delta_1}$ and~$\bang{\Delta_2}$,
 which results in the desired {\em emptiness} of the whole~$\Gamma$
 and the {\em degeneracy} of the whole~$\bang{\Delta}$.

\CASE{1.a.3} The {\em principal\/} formula~$B$ is of the form
 $$ C_{00}  = \limply{\limply{H_{00}^\good{2}}{\bot^\good{3}}}
                                      {\bot^\good{3}},$$
 and, according to rule~{\rm {\bf R$\llto$}}, our sequent
 of item~(a) is derived from the sequent
 $$ \seq{K,\,\limply{H_{00}^\good{2}}{\bot^\good{3}},\,
          \Gamma,\, \bang{\Delta}}{\bot^\good{3}}.$$
 By Lemma~\ref{L1}
     $$ \#_\bot(K) = \#_\bot(B) = -2N.$$
 Therefore, we can complete this case by applying the inductive
 hypothesis from item~(c).

\CASE{1.a.4} The {\em principal\/} formula is of the
              form~$C_{00}^\good{m}$, and, according to
              rule~{\rm {\bf R$\otimes$}}, our sequent of item~(a)
              is derived from two sequents of the form
$$ \seq{K_1,\,\Gamma_1,\,\bang{\Delta_1}}{C_{00}} $$
 and
$$ \seq{K_2,\,\Gamma_2,\,\bang{\Delta_2}}{C_{00}^\good{m-1}} $$
 where
$$\left\{\begin{array}{lcl}%
         K  & = & K_1,\,K_2,\\%
    \Gamma  & = & \Gamma_1,\,\Gamma_2,\\%
    \Delta  & = & \Delta_1,\,\Delta_2.%
         \end{array}\right.$$
 By applying the inductive hypothesis, we can get the
 {\em emptiness} of both~$\Gamma_1$ and~$\Gamma_2$, and the
 {\em degeneracy} of both~$\bang{\Delta_1}$ and~$\bang{\Delta_2}$,
 which results in the desired {\em emptiness} of the whole~$\Gamma$
 and the {\em degeneracy} of the whole~$\bang{\Delta}$.

\CASE{1.b} The {\em principal\/} formula is of the
           form~\limply{H_{00}^\good{2}}{\bot^\good{3}},
 and, according to rule~{\rm {\bf R$\llto$}}, the corresponding
 sequent of item~(b) is derived from the sequent
$$ \seq{H_{00}^\good{2},\,\limply{H_{00}^\good{2}}{\bot^\good{3}},\,
       \Gamma,\,\bang{\Delta}}{\bot^\good{3}}.$$
 It remains to apply the inductive hypothesis from item~(c).

\CASE{1.c} Assume that the corresponding sequent of item~(c) is
 derived from two sequents of the form
$$\seq{K_1,\,\limply{H_{00}^\good{2}}{\bot^\good{3}}^\good{h_1},\,
  \Gamma_1,\,\bang{\Delta_1}}{\bot}$$
 and
$$\seq{K_2,\,\limply{H_{00}^\good{2}}{\bot^\good{3}}^\good{h_2},\,
  \Gamma_2,\,\bang{\Delta_2}}{\bot^\good{2}} $$
 where
$$\left\{\begin{array}{lcl}%
       K  & = & K_1,\,K_2,\\%
       1  & = & h_1 + h_2,\\%
  \Gamma  & = & \Gamma_1,\,\Gamma_2,\\%
  \Delta  & = & \Delta_1,\,\Delta_2.%
\end{array}\right.$$
 By Lemma~\ref{L1} and Lemma~\ref{lcost} we have:
$$\left\{\begin{array}{lcl}%
  \#_\bot(K_1) + (2N + 3)h_1  & = &  1 \pmod{9N},\\%
  \#_\bot(K_2) + (2N + 3)h_2  & = &  2 \pmod{9N},\\%
\multicolumn{3}{l}%
 {\#_\bot(K_1) = -Nk',\ \mbox{for some $k'\leq 2$,}}%
\end{array}\right.$$
 which is a contradiction.

\CASE{1.d.1} Assume that $a=1$, and, by rule~{\rm {\bf R$\bot$}},
 the corresponding sequent of item~(d) is derived from the sequent
$$ \seq{H_{00},\,\bot^\good{N+1},\,\Gamma,\,\bang{\Delta}}{ } $$
 But, according to item~(e), the latter sequent is not derivable.

\CASE{1.d.2} Assume that $a=2$, and, by rule~{\rm {\bf R$\otimes$}},
 our sequent of item~(d) is derived from two sequents of the form
$$\seq{H_{00},\,\bot^\good{a_1},\,\Gamma_1,\,\bang{\Delta_1}}{\bot}$$
 and
$$ \seq{\bot^\good{a_2},\,\Gamma_2,\,\bang{\Delta_2}}{\bot} $$
 where
$$\left\{\begin{array}{lcl}%
   N + 2  & = & a_1 + a_2,\\%
  \Gamma  & = & \Gamma_1,\,\Gamma_2,\\%
  \Delta  & = & \Delta_1,\,\Delta_2.%
\end{array}\right.$$
 By Lemma~\ref{L1} and Lemma~\ref{lcost} we have:
$$\left\{\begin{array}{rcl}%
  -N + a_1 & = & 1 \pmod{9N},\\%
       a_2 & = & 1 \pmod{9N},%
\end{array}\right.$$
 and, therefore, 
 $$ a_1 = N + 1,$$
 which leads to a contradiction because, according to the inductive
 hypothesis, the first sequent cannot occur in our derivation.

\CASE{2} Assume that the {\em principal\/} formula belongs
         to~$\Gamma$, and it is of the form~$F_\good{A}$
         \mbox{(or~$F_\good{Y}$).}
 The following subcases are to be considered.

\CASE{2.0} The {\em principal\/} formula is of the
           form~\mbox{$(F_\good{A_1} \& F_\good{A_2})$,}
 and, by rule~{\rm {\bf L$\&$}}, the corresponding sequent
 of the form
$$\seq{\ldots,\,\Gamma,\,\bang{\Delta}}{\ldots} $$
 is derived either from the sequent
$$\seq{\ldots,\,\Gamma',\, F_\good{A_1},\,\bang{\Delta}}
      {\ldots} $$
 or from the sequent
$$\seq{\ldots,\,\Gamma',\, F_\good{A_2},\,\bang{\Delta}}
      {\ldots} $$
 Then, according to the inductive hypothesis, either 
 the multiset
      $$ \Gamma', \ F_\good{A_1} $$
 or the multiset
      $$ \Gamma', \ F_\good{A_2} $$
 must be empty, which is a contradiction.

\CASE{2.1.a} Assume that the {\em principal\/} formula is of the
              form~\limply{E_\good{X}}{E_\good{Y}}, and,
 according to rule~{\rm {\bf L$\llto$}}, the sequent
 of item~(a) is derived from two sequents of the form
$$ \seq{K_1,\,\Gamma_1,\,\bang{\Delta_1}}{E_\good{X}} $$
 and
$$ \seq{K_2,\,E_\good{Y},\,\Gamma_2,\,\bang{\Delta_2}}{B} $$
 where
$$\left\{\begin{array}{lcl}%
       K  & = & K_1,\,K_2,\\%
  \Gamma &\supset & \Gamma_1,\,\Gamma_2,\\%
  \Delta  & = & \Delta_1,\,\Delta_2.%
\end{array}\right.$$
 Then Lemma~\ref{L1} and Lemma~\ref{lcost} yield:
$$\left\{\begin{array}{lcl}%
 \#_\bot(K_1)      & = & 6N  \pmod{9N},\\%
 \#_\bot(K_2) + 6N & = & \#_\bot(B) \pmod{9N},\\%
 \multicolumn{3}{l}%
 {\begin{array}{@{}l}%
  \mbox{either}\ \#_\bot(K_1) = -N,\\%
  \mbox{or}\ \#_\bot(K_1) = -2Nk',\ \mbox{for some $k'\leq 5$,}%
  \end{array}}%
\end{array}\right.$$
 which is a contradiction.

\CASE{2.2.a} Assume that the {\em principal\/} formula is of the
           form~\limply{E_\good{X}}{E_\good{Y}}, and,
 by rule~{\rm {\bf L$\llto$}}, our sequent of item~(a) is derived
 from two sequents of the form
$$\seq{K_1,\,\Gamma_1,\,\bang{\Delta_1}}{E_\good{X}, B} $$
 and
$$\seq{K_2,\,E_\good{Y},\,\Gamma_2,\,\bang{\Delta_2}}{ } $$
 Then Lemma~\ref{L1} and Lemma~\ref{lcost} yield:
$$\left\{\begin{array}{lcl}%
 \#_\bot(K_1)       & = &  6N + \#_\bot(B) - 1 \pmod{9N},\\%
 \#_\bot(K_2) + 6N  & = &  1 \pmod{9N},\\%
\multicolumn{3}{l}%
 {\begin{array}{@{}l}%
  \mbox{either}\ \#_\bot(K_2) = -N,\\%
  \mbox{or}\ \#_\bot(K_2) = -2Nk',\ \mbox{for some $k'\leq 5$,}%
  \end{array}}%
\end{array}\right.$$
 which is also a contradiction.

\CASE{2.3.a} Assume that the {\em principal\/} formula is of the
           form~\limply{F_\good{A}}{F_\good{Y}}, and,
 according to rule~{\rm {\bf L$\llto$}}, our sequent of item~(a)
 is derived from two sequents of the form
$$ \seq{K_1,\,\Gamma_1,\,\bang{\Delta_1}}{F_\good{A}} $$
 and
$$ \seq{K_2,\,F_\good{Y},\,\Gamma_2,\,\bang{\Delta_2}}{B} $$
 where
$$\left\{\begin{array}{lcl}%
       K  & = & K_1,\,K_2,\\%
  \Gamma &\supset & \Gamma_1,\,\Gamma_2,\\%
  \Delta  & = & \Delta_1,\,\Delta_2.%
\end{array}\right.$$
 Then, by Lemma~\ref{L1} and Lemma~\ref{lcost}, we have:
$$\left\{\begin{array}{lcl}%
 \#_\bot(K)   & = & \#_\bot(B) \pmod{9N}.\\%
 \#_\bot(K_1) & = & 0 \pmod{9N},\\%
 \#_\bot(K_2) & = & \#_\bot(B) \pmod{9N},%
\end{array}\right.$$
 Hence,
         $$ K_2 = K,$$
 and, by the inductive hypothesis, the multiset
      $$ F_\good{Y},\ \Gamma_2 $$
 must be empty, which is a contradiction.

\CASE{2.4.a} Assume that the {\em principal\/} formula is of the
           form~\limply{F_\good{A}}{F_\good{Y}}, and,
 according to rule~{\rm {\bf L$\llto$}}, our sequent
 of item~(a) is derived from two sequents of the form
$$ \seq{K_1,\,\Gamma_1,\,\bang{\Delta_1}}{F_\good{A}, B} $$
 and
$$ \seq{K_2,\,F_\good{Y},\,\Gamma_2,\,\bang{\Delta_2}}{ } $$
 Then Lemma~\ref{L1} and Lemma~\ref{lcost} show that
$$\left\{\begin{array}{lcl}%
 \#_\bot(K_1) & =  & \#_\bot(B) - 1 \pmod{9N},\\%
 \#_\bot(K_2) & =  & 1 \pmod{9N},\\%
 \multicolumn{3}{l}%
 {\begin{array}{@{}l}%
  \mbox{either}\ \#_\bot(K_2) = -N,\\%
  \mbox{or}\ \#_\bot(K_2) = -2Nk',\ \mbox{for some $k'\leq 5$,}%
  \end{array}}%
\end{array}\right.$$
 which is a contradiction as well.

\CASE{2.1.b}  Assume that the {\em principal\/} formula is of the
              form~\limply{E_\good{X}}{E_\good{Y}}, and
 our sequent of item~(b) is derived from two sequents of the form
$$\seq{\limply{H_{00}^\good{2}}{\bot^\good{3}}^\good{h_1},
       \Gamma_1,\bang{\Delta_1}}{E_\good{X}} $$
 and
$$\seq{\limply{H_{00}^\good{2}}{\bot^\good{3}}^\good{h_2},
        E_\good{Y},\Gamma_2,\bang{\Delta_2}}
      {\limply{H_{00}^\good{2}}{\bot^\good{3}}} $$
 where
$$\left\{\begin{array}{lcl}%
       1  & = & h_1 + h_2,\\%
  \Gamma &\supset & \Gamma_1,\,\Gamma_2,\\%
  \Delta  & = & \Delta_1,\,\Delta_2.%
\end{array}\right.$$
 By Lemma~\ref{L1} and Lemma~\ref{lcost} we have:
$$\left\{\begin{array}{lcl}%
  (2N + 3)h_1  & = &  6N \pmod{9N},\\%
  (2N + 3)h_2 + 6N & = &  2N + 3 \pmod{9N},%
\end{array}\right.$$
 which is a contradiction.

\CASE{2.2.b} Assume that the {\em principal\/} formula is of the
           form~\limply{E_\good{X}}{E_\good{Y}}, and now our
 sequent of item~(b) is derived from two sequents of the form
$$\seq{\limply{H_{00}^\good{2}}{\bot^\good{3}}^\good{h_1},
       \Gamma_1,\bang{\Delta_1}}
  {E_\good{X}, \limply{H_{00}^\good{2}}{\bot^\good{3}}} $$
 and
$$\seq{\limply{H_{00}^\good{2}}{\bot^\good{3}}^\good{h_2},
        E_\good{Y},\Gamma_2,\bang{\Delta_2}}{ } $$
 Then Lemma~\ref{L1} and Lemma~\ref{lcost} yield:
$$\left\{\begin{array}{lcl}%
  (2N + 3)h_1  & = &  8N + 2 \pmod{9N},\\%
  (2N + 3)h_2 + 6N & = &  1 \pmod{9N},%
\end{array}\right.$$
 which is also a contradiction.

\CASE{2.3.b} Assume that the {\em principal\/} formula is of the
           form~\limply{F_\good{A}}{F_\good{Y}}, and our
 sequent of item~(b) is derived from two sequents of the form
$$\seq{\limply{H_{00}^\good{2}}{\bot^\good{3}}^\good{h_1},
       \Gamma_1, \bang{\Delta_1}}{F_\good{A}} $$
 and
$$\seq{\limply{H_{00}^\good{2}}{\bot^\good{3}}^\good{h_2},
        F_\good{Y},\Gamma_2,\bang{\Delta_2}}
      {\limply{H_{00}^\good{2}}{\bot^\good{3}}} $$
 where
$$\left\{\begin{array}{lcl}%
       1  & = & h_1 + h_2,\\%
  \Gamma &\supset & \Gamma_1,\,\Gamma_2,\\%
  \Delta  & = & \Delta_1,\,\Delta_2.%
\end{array}\right.$$
 According to Lemma~\ref{L1} and Lemma~\ref{lcost}, we have:
$$\left\{\begin{array}{lcl}%
  (2N + 3)h_1  & = &  0 \pmod{9N},\\%
  (2N + 3)h_2  & = &  2N + 3 \pmod{9N}.%
\end{array}\right.$$
 Hence,
        $$ h_2 =  1,$$
 and, by the inductive hypothesis, the multiset
      $$ F_\good{Y},\ \Gamma_2 $$
 must be empty, which is a contradiction.

\CASE{2.4.b} Assume that the {\em principal\/} formula is of the
           form~\limply{F_\good{A}}{F_\good{Y}}, and now
 our sequent of item~(b) is derived from two sequents of the form
$$ \seq{\limply{H_{00}^\good{2}}{\bot^\good{3}}^\good{h_1},
       \Gamma_1, \bang{\Delta_1}}
   {F_\good{A}, \limply{H_{00}^\good{2}}{\bot^\good{3}}} $$
 and
$$\seq{\limply{H_{00}^\good{2}}{\bot^\good{3}}^\good{h_2},
        F_\good{Y}, \Gamma_2,\,\bang{\Delta_2}}{ } $$
 Then Lemma~\ref{L1} and Lemma~\ref{lcost} yield:
$$\left\{\begin{array}{lcl}%
  (2N + 3)h_1  & = &  2N + 2 \pmod{9N},\\%
  (2N + 3)h_2  & = &  1 \pmod{9N},%
\end{array}\right.$$
 which is a contradiction.

\CASE{2.1.c} Assume that the {\em principal\/} formula is of the
              form~\limply{E_\good{X}}{E_\good{Y}}, and
 our sequent of item~(c) is derived from two sequents of the form
$$\seq{K_1,\limply{H_{00}^\good{2}}{\bot^\good{3}}^\good{h_1},
       \Gamma_1,\bang{\Delta_1}}{E_\good{X}} $$
 and
$$\seq{K_2,\limply{H_{00}^\good{2}}{\bot^\good{3}}^\good{h_2},
        E_\good{Y},\Gamma_2,\bang{\Delta_2}}{\bot^\good{3}} $$
 where
$$\left\{\begin{array}{lcl}%
       K  & = & K_1,\,K_2,\\%
       1  & = & h_1 + h_2,\\%
  \Gamma &\supset & \Gamma_1,\,\Gamma_2,\\%
  \Delta  & = & \Delta_1,\,\Delta_2.%
\end{array}\right.$$
 By Lemma~\ref{L1} and Lemma~\ref{lcost} we have:
$$\left\{\begin{array}{lcl}%
  \#_\bot(K_1) + (2N + 3)h_1       & = &  6N \pmod{9N},\\%
  \#_\bot(K_2) + (2N + 3)h_2 + 6N  & = &  3 \pmod{9N},\\%
\multicolumn{3}{l}%
 {\#_\bot(K_1) =  -Nk',\ \mbox{for some $k'\leq 2$,}}%
\end{array}\right.$$
 which is a contradiction.

\CASE{2.2.c} Assume that the {\em principal\/} formula is of the
           form~\limply{E_\good{X}}{E_\good{Y}}, and now our
 sequent of item~(c) is derived from two sequents of the form
$$\seq{K_1,\limply{H_{00}^\good{2}}{\bot^\good{3}}^\good{h_1},
       \Gamma_1,\bang{\Delta_1}}{E_\good{X}, \bot^\good{3}} $$
 and
$$\seq{K_2,\limply{H_{00}^\good{2}}{\bot^\good{3}}^\good{h_2},
        E_\good{Y},\Gamma_2,\bang{\Delta_2}}{ } $$
 Then Lemma~\ref{L1} and Lemma~\ref{lcost} yield:
$$\left\{\begin{array}{lcl}%
  \#_\bot(K_1) + (2N + 3)h_1  & = &  6N + 2 \pmod{9N},\\%
  \#_\bot(K_2) + (2N + 3)h_2 + 6N & = &  1 \pmod{9N},\\%
\multicolumn{3}{l}%
 {\#_\bot(K_2) = -Nk',\ \mbox{for some $k'\leq 2$,}}%
\end{array}\right.$$
 which is also a contradiction.

\CASE{2.3.c} Assume that the {\em principal\/} formula is of the
           form~\limply{F_\good{A}}{F_\good{Y}}, and our
 sequent of item~(c) is derived from two sequents of the form
$$\seq{K_1,\limply{H_{00}^\good{2}}{\bot^\good{3}}^\good{h_1},
       \Gamma_1, \bang{\Delta_1}}{F_\good{A}} $$
 and
$$\seq{K_2,\limply{H_{00}^\good{2}}{\bot^\good{3}}^\good{h_2},
        F_\good{Y}, \Gamma_2, \bang{\Delta_2}}{\bot^\good{3}} $$
 where
$$\left\{\begin{array}{lcl}%
       K  & = & K_1,\,K_2,\\%
       1  & = & h_1 + h_2,\\%
  \Gamma &\supset & \Gamma_1,\,\Gamma_2,\\%
  \Delta  & = & \Delta_1,\,\Delta_2.%
\end{array}\right.$$
 Then Lemma~\ref{L1} and Lemma~\ref{lcost} show:
$$\left\{\begin{array}{lcl}%
  \#_\bot(K)   + (2N + 3)     & = &  3 \pmod{9N},\\%
  \#_\bot(K_1) + (2N + 3)h_1  & = &  0 \pmod{9N},\\%
  \#_\bot(K_2) + (2N + 3)h_2  & = &  3 \pmod{9N},\\%
\multicolumn{3}{l}%
 {\#_\bot(K_2) = -Nk',\ \mbox{for some $k'\leq 2$.}}%
\end{array}\right.$$
 Hence,
$$\left\{\begin{array}{lcl}%
         h_2 & = & 1,\\%
         K_2 & = & K,%
\end{array}\right.$$
 and, by the inductive hypothesis, the multiset
      $$ F_\good{Y},\ \Gamma_2 $$
 must be empty, which is a contradiction.

\CASE{2.4.c} Assume that the {\em principal\/} formula is of the
           form~\limply{F_\good{A}}{F_\good{Y}}, and now
 our sequent of item~(c) is derived from two sequents of the form
$$ \seq{K_1,\limply{H_{00}^\good{2}}{\bot^\good{3}}^\good{h_1},
       \Gamma_1, \bang{\Delta_1}}{F_\good{A}, \bot^\good{3}}$$
 and
$$\seq{K_2,\limply{H_{00}^\good{2}}{\bot^\good{3}}^\good{h_2},
        F_\good{Y}, \Gamma_2,\,\bang{\Delta_2}}{ } $$
 Then Lemma~\ref{L1} and Lemma~\ref{lcost} yield:
$$\left\{\begin{array}{lcl}%
  \#_\bot(K_1) + (2N + 3)h_1  & = &  2 \pmod{9N},\\%
  \#_\bot(K_2) + (2N + 3)h_2  & = &  1 \pmod{9N},\\%
\multicolumn{3}{l}%
 {\#_\bot(K_2) = -Nk',\ \mbox{for some $k'\leq 2$,}}%
\end{array}\right.$$
 which is a contradiction.

\CASE{2.1.d+e} Assume that the {\em principal\/} formula is of the
               form~\limply{E_\good{X}}{E_\good{Y}}, and
 the corresponding sequent of item~(d) or~(e) is derived from two
 sequents of the form
$$\seq{H_{00}^\good{h_1},\bot^\good{a_1},\Gamma_1,\bang{\Delta_1}}
   {E_\good{X}} $$
 and
$$\seq{H_{00}^\good{h_2},\bot^\good{a_2},
   E_\good{Y},\Gamma_2,\bang{\Delta_2}}{\bot^\good{a}} $$
 where
$$\left\{\begin{array}{lcl}%
       1  & = & h_1 + h_2,\\%
   N + a  & = & a_1 + a_2,\\%
  \Gamma &\supset & \Gamma_1,\,\Gamma_2,\\%
  \Delta  & = & \Delta_1,\,\Delta_2.%
\end{array}\right.$$
 By Lemma~\ref{L1} and Lemma~\ref{lcost} we have:
$$\left\{\begin{array}{lcl}%
   -Nh_1 + a_1 & = &  6N \pmod{9N},\\%
   -Nh_2 + a_2 + 6N  & = &  a \pmod{9N},%
\end{array}\right.$$
 which is a contradiction.

\CASE{2.2.d} Assume that the {\em principal\/} formula is of the
           form~\limply{E_\good{X}}{E_\good{Y}}, and now our
 sequent of item~(d) is derived from two sequents of the form
$$\seq{H_{00}^\good{h_1},\bot^\good{a_1},\Gamma_1,\bang{\Delta_1}}
   {E_\good{X}, \bot^\good{a}} $$
 and
$$\seq{H_{00}^\good{h_2},\bot^\good{a_2},
   E_\good{Y},\Gamma_2,\bang{\Delta_2}}{ } $$
 Then Lemma~\ref{L1} and Lemma~\ref{lcost} yield:
$$\left\{\begin{array}{lcl}%
   -Nh_1 + a_1 & = &  6N + a - 1 \pmod{9N},\\%
   -Nh_2 + a_2 + 6N  & = &  1 \pmod{9N},%
\end{array}\right.$$
 which is also a contradiction.

\CASE{2.3.d+e} Let the {\em principal\/} formula be of the
           form~\limply{F_\good{A}}{F_\good{Y}}, and let our
 sequent of item~(d) or~(e) be derived from two sequents of the form
$$\seq{H_{00}^\good{h_1},\bot^\good{a_1},\Gamma_1,\bang{\Delta_1}}
      {F_\good{A}} $$
 and
$$\seq{H_{00}^\good{h_2},\bot^\good{a_2},
        F_\good{Y},\Gamma_2,\bang{\Delta_2}}{\bot^\good{a}} $$
 where
$$\left\{\begin{array}{lcl}%
       1  & = & h_1 + h_2,\\%
   N + a  & = & a_1 + a_2,\\%
  \Gamma &\supset & \Gamma_1,\,\Gamma_2,\\%
  \Delta  & = & \Delta_1,\,\Delta_2.%
\end{array}\right.$$
 Then Lemma~\ref{L1} and Lemma~\ref{lcost} show:
$$\left\{\begin{array}{lcl}%
   -Nh_1 + a_1 = 0 \pmod{9N},\\%
   -Nh_2 + a_2 = a \pmod{9N}.%
\end{array}\right.$$

 Assume that\ \ $h_2 = 0$.

 Then \ \ $a_2 = a$, \ \ and, according to Lemma~\ref{L2},
 the multiset
      $$ F_\good{Y},\ \Gamma_2 $$
 must be empty, which is a contradiction.

 Assume that\ \ $h_2 = 1$.

 Then \ \ $a_2 = N + a$. \ \ Now the {\em non-empty} multiset
      $$ F_\good{Y},\ \Gamma_2 $$
 must be empty by the inductive hypothesis.

\CASE{2.4.d} Let the {\em principal\/} formula be of the
           form~\limply{F_\good{A}}{F_\good{Y}}, and let our
 sequent of item~(d) be derived from two sequents of the form
$$\seq{H_{00}^\good{h_1},\bot^\good{a_1},\Gamma_1,\bang{\Delta_1}}
      {F_\good{A}, \bot^\good{a}} $$
 and
$$\seq{H_{00}^\good{h_2},\bot^\good{a_2},
        F_\good{Y},\Gamma_2,\bang{\Delta_2}}{ } $$
 where
$$\left\{\begin{array}{lcl}%
       1  & = & h_1 + h_2,\\%
   N + a  & = & a_1 + a_2,\\%
  \Gamma &\supset & \Gamma_1,\,\Gamma_2,\\%
  \Delta  & = & \Delta_1,\,\Delta_2.%
\end{array}\right.$$
 Then Lemma~\ref{L1} and Lemma~\ref{lcost} yield:
$$\left\{\begin{array}{lcl}%
   -Nh_1 + a_1 = a - 1 \pmod{9N},\\%
   -Nh_2 + a_2 = 1 \pmod{9N}.%
\end{array}\right.$$

 Assume that\ \ $h_2 = 0$.

 Then \ \ $a_2 = 1$, \ \ and, according to Lemma~\ref{L2},
 the multiset
      $$ F_\good{Y},\ \Gamma_2 $$
 must be empty, which is a contradiction.

 Assume that\ \ $h_2 = 1$.

 Then \ \ $a_2 = N + 1$,\ \ and we get a contradiction to item~(e).

\CASE{2.5} Case of the {\em principal\/} formula of the
 form~\lvariant{E_\good{X}}{E_\good{Y_1}}{E_\good{Y_2}}
 is handled similarly to~\mbox{{\bf Cases 2.1.abcde}}
 and~\mbox{{\bf Cases 2.2.abcde.}}

\CASE{3} Assume that the {\em principal\/} formula belongs
         to~$\Gamma$, and it is of the form
 $$ D_\good{q} =
    \limply{\limply{H_1}{\bot^\good{b}}}{\bot^\good{b}} $$
 where
          $$ 4 \ \leq\  b \ \leq\  N-3.$$

\CASE{3.1.a} According to rule~{\rm {\bf L$\llto$}}, let our sequent
             of item~(a) be derived from two sequents of the form
$$ \seq{K_1,\,\Gamma_1,\,\bang{\Delta_1}}
       {\limply{H_1}{\bot^\good{b}}} $$
 and
$$ \seq{K_2,\,\bot^\good{b},\,\Gamma_2,\,\bang{\Delta_2}}{B} $$
 where
$$\left\{\begin{array}{lcl}%
       K  & = & K_1,\,K_2,\\%
  \Gamma &\supset & \Gamma_1,\,\Gamma_2,\\%
  \Delta  & = & \Delta_1,\,\Delta_2.%
\end{array}\right.$$
 Then Lemma~\ref{L1} and Lemma~\ref{lcost} yield:
$$\left\{\begin{array}{lcl}%
 \#_\bot(K_1)      & = & b  \pmod{9N},\\%
 \#_\bot(K_2) + b  & = & \#_\bot(B) \pmod{9N},\\%
 \multicolumn{3}{l}%
 {\begin{array}{@{}l}%
  \mbox{either}\ \#_\bot(K_1) = -N,\\%
  \mbox{or}\ \#_\bot(K_1) = -2Nk',\ \mbox{for some $k'\leq 5$,}%
  \end{array}}%
\end{array}\right.$$
 which is a contradiction.

\CASE{3.2.a} Now let our sequent of item~(a) be derived from two
 sequents of the form
$$ \seq{K_1,\,\Gamma_1,\,\bang{\Delta_1}}
       {\limply{H_1}{\bot^\good{b}}, B} $$
 and
$$ \seq{K_2,\,\bot^\good{b},\,\Gamma_2,\,\bang{\Delta_2}}{ } $$
 Then, by Lemma~\ref{L1} and Lemma~\ref{lcost}, we have:
$$\left\{\begin{array}{lcl}%
 \#_\bot(K_1)       & = &  b + \#_\bot(B) - 1 \pmod{9N},\\%
 \#_\bot(K_2) + b   & = &  1 \pmod{9N},\\%
\multicolumn{3}{l}%
 {\begin{array}{@{}l}%
  \mbox{either}\ \#_\bot(K_2) = -N,\\%
  \mbox{or}\ \#_\bot(K_2) = -2Nk',\ \mbox{for some $k'\leq 5$,}%
  \end{array}}%
\end{array}\right.$$
 which is a contradiction as well.

\CASE{3.1.b} According to rule~{\rm {\bf L$\llto$}}, assume that
             the corresponding sequent of item~(b) is derived from
             two sequents of the form
$$\seq{\limply{H_{00}^\good{2}}{\bot^\good{3}}^\good{h_1},
       \Gamma_1,\bang{\Delta_1}}
      {\limply{H_1}{\bot^\good{b}}} $$
 and
$$\seq{\limply{H_{00}^\good{2}}{\bot^\good{3}}^\good{h_2},
       \bot^\good{b},\Gamma_2,\bang{\Delta_2}}
       {\limply{H_{00}^\good{2}}{\bot^\good{3}}} $$
 where
$$\left\{\begin{array}{lcl}%
       1  & = & h_1 + h_2,\\%
  \Gamma &\supset & \Gamma_1,\,\Gamma_2,\\%
  \Delta  & = & \Delta_1,\,\Delta_2.%
\end{array}\right.$$
 Then Lemma~\ref{L1} and Lemma~\ref{lcost} yield the following
 contradiction:
$$\left\{\begin{array}{lcl}%
  (2N + 3)h_1 & = &  b \pmod{9N},\\%
  (2N + 3)h_2 + b & = & 2N + 3 \pmod{9N}.%
\end{array}\right.$$

\CASE{3.2.b} Now assume that our sequent of item~(b) is derived from
             two sequents of the form
$$\seq{\limply{H_{00}^\good{2}}{\bot^\good{3}}^\good{h_1},
       \Gamma_1,\bang{\Delta_1}}
      {\limply{H_1}{\bot^\good{b}},
        \limply{H_{00}^\good{2}}{\bot^\good{3}}} $$
 and
$$\seq{\limply{H_{00}^\good{2}}{\bot^\good{3}}^\good{h_2},
       \bot^\good{b},\Gamma_2,\bang{\Delta_2}}{ } $$
 Then, by Lemma~\ref{L1} and Lemma~\ref{lcost}, we have:
$$\left\{\begin{array}{lcl}%
  (2N + 3)h_1 & = &  b + 2N + 2 \pmod{9N},\\%
  (2N + 3)h_2 + b & = & 1 \pmod{9N},\\%
\end{array}\right.$$
 which is a contradiction as well.

\CASE{3.1.c} According to rule~{\rm {\bf L$\llto$}}, let our sequent
             of item~(c) be derived from two sequents of the form
$$\seq{K_1,\limply{H_{00}^\good{2}}{\bot^\good{3}}^\good{h_1},
       \Gamma_1,\bang{\Delta_1}}
       {\limply{H_1}{\bot^\good{b}}} $$
 and
$$\seq{K_2,\limply{H_{00}^\good{2}}{\bot^\good{3}}^\good{h_2},
      \bot^\good{b},\Gamma_2,\bang{\Delta_2}}{\bot^\good{3}} $$
 where
$$\left\{\begin{array}{lcl}%
       K  & = & K_1,\,K_2,\\%
       1  & = & h_1 + h_2,\\%
  \Gamma &\supset & \Gamma_1,\,\Gamma_2,\\%
  \Delta  & = & \Delta_1,\,\Delta_2.%
\end{array}\right.$$
 By Lemma~\ref{L1} and Lemma~\ref{lcost} we have:
$$\left\{\begin{array}{lcl}%
  \#_\bot(K_1) + (2N + 3)h_1 & = &  b \pmod{9N},\\%
  \#_\bot(K_2) + (2N + 3)h_2 + b & = & 3 \pmod{9N},\\%
\multicolumn{3}{l}%
 {\#_\bot(K_1) = -Nk',\ \mbox{for some $k'\leq 2$,}}%
\end{array}\right.$$
 which is a contradiction.

\CASE{3.2.c} Now assume that our sequent of item~(c) is derived from
             two sequents of the form
$$\seq{K_1,\limply{H_{00}^\good{2}}{\bot^\good{3}}^\good{h_1},
       \Gamma_1,\bang{\Delta_1}}
       {\limply{H_1}{\bot^\good{b}}, \bot^\good{3}} $$
 and
$$\seq{K_2,\limply{H_{00}^\good{2}}{\bot^\good{3}}^\good{h_2},
      \bot^\good{b},\Gamma_2,\bang{\Delta_2}}{ } $$
 Then Lemma~\ref{L1} and Lemma~\ref{lcost} yield:
$$\left\{\begin{array}{lcl}%
  \#_\bot(K_1) + (2N + 3)h_1 & = &  b + 2 \pmod{9N},\\%
  \#_\bot(K_2) + (2N + 3)h_2 + b & = & 1 \pmod{9N},\\%
\multicolumn{3}{l}%
 {\#_\bot(K_2) = -Nk',\ \mbox{for some $k'\leq 2$,}}%
\end{array}\right.$$
 which is also a contradiction.

\CASE{3.1.d+e} Let the corresponding sequent of item~(d)
               or~(e) be derived from two sequents of the form
$$\seq{H_{00}^\good{h_1},\bot^\good{a_1},\Gamma_1,\bang{\Delta_1}}
      {\limply{H_1}{\bot^\good{b}}} $$
 and
$$\seq{H_{00}^\good{h_2},\bot^\good{a_2+b},
        \Gamma_2,\bang{\Delta_2}}{\bot^\good{a}} $$
 where
$$\left\{\begin{array}{lcl}%
       1  & = & h_1 + h_2,\\%
   N + a  & = & a_1 + a_2,\\%
  \Gamma &\supset & \Gamma_1,\,\Gamma_2,\\%
  \Delta  & = & \Delta_1,\,\Delta_2.%
\end{array}\right.$$
 By Lemma~\ref{L1} and Lemma~\ref{lcost} we have:
$$\left\{\begin{array}{lcl}%
   -Nh_1 + a_1     & = &  b \pmod{9N},\\%
   -Nh_2 + a_2 + b & = & a \pmod{9N}.%
\end{array}\right.$$

 Assuming that\ \ \mbox{$h_1 = 0$,}\ \ we can conclude that\ \
 \mbox{$a_1 = b$,}\ \ which gives a contradiction because,
 by Lemma~\ref{L2}, the first sequent with its {\em wrong} right-hand
 side cannot occur in our derivation.

 \mbox{For\ \ $h_2 = 0$,}\ \ we can get also a contradiction
 because of
     $$ 2 \ \leq\ a_2 + (b - a) \ \leq\ 2N-1.$$

\CASE{3.2.d} Let the corresponding sequent of item~(d)
             be derived from two sequents of the form
$$\seq{H_{00}^\good{h_1},\bot^\good{a_1},\Gamma_1,\bang{\Delta_1}}
      {\limply{H_1}{\bot^\good{b}}, \bot^\good{a}} $$
 and
$$\seq{H_{00}^\good{h_2},\bot^\good{a_2+b},\Gamma_2,\bang{\Delta_2}}
      { } $$
 where
$$\left\{\begin{array}{lcl}%
       1  & = & h_1 + h_2,\\%
   N + a  & = & a_1 + a_2,\\%
  \Gamma &\supset & \Gamma_1,\,\Gamma_2,\\%
  \Delta  & = & \Delta_1,\,\Delta_2.%
\end{array}\right.$$
 By Lemma~\ref{L1} and Lemma~\ref{lcost} we have:
$$\left\{\begin{array}{lcl}%
   -Nh_1 + a_1      & = & b + a - 1 \pmod{9N},\\%
   -Nh_2 + a_2 + b  & = & 1 \pmod{9N}.%
\end{array}\right.$$
 Assuming that\ \ \mbox{$h_2 = 0$,}\ \ we get a contradiction
 because of
     $$ 3 \ \leq\ a_2 + (b - 1) \ \leq\ 2N-2.$$
 \mbox{For\ \ $h_2 = 1$,}\ \ we have that \ \
 \mbox{$a_2 + b = N + 1$,} \ \  which yields a contradiction because,
 according to the inductive hypothesis from item~(e), the latter
 sequent cannot occur in our derivation.

\CASE{4} Assume that the {\em principal\/} formula belongs
         to~$\Gamma$, and it is of the form
  $$ H_1 = \limply{C_{00}^\good{4}}{\bot^\good{N}}.$$

\CASE{4.1.a} According to rule~{\rm {\bf L$\llto$}}, assume that
             the corresponding sequent of item~(a) is derived from
             two sequents of the form
$$ \seq{K_1,\,\Gamma_1,\,\bang{\Delta_1}}{C_{00}^\good{4}} $$
 and
$$ \seq{K_2,\,\bot^\good{N},\,\Gamma_2,\,\bang{\Delta_2}}{B} $$
 where
$$\left\{\begin{array}{lcl}%
       K  & = & K_1,\,K_2,\\%
  \Gamma &\supset & \Gamma_1,\,\Gamma_2,\\%
  \Delta  & = & \Delta_1,\,\Delta_2.%
\end{array}\right.$$
 Then Lemma~\ref{L1} and Lemma~\ref{lcost} yield:
$$\left\{\begin{array}{lcl}%
 \#_\bot(K)       & = & \#_\bot(B) \pmod{9N},\\%
 \#_\bot(K_1)     & = & -8N        \pmod{9N},\\%
 \multicolumn{3}{l}%
 {\begin{array}{@{}l}%
  \mbox{either}\ \#_\bot(K_1) = -N,\\%
  \mbox{or}\ \#_\bot(K_1) = -2Nk',\ \mbox{for some $k'\leq 5$.}%
  \end{array}}%
\end{array}\right.$$
 The only solution of this system is as follows:
$$\left\{\begin{array}{lcl}%
       K_1  & = & C_{00}, C_{00}, C_{00}, C_{00}\\%
       B    & = & C_{00}^\good{m}\ \mbox{(where\ $ m = 4, 5$).}          %
\end{array}\right.$$
 Hence, the latter sequent is of the following form:
$$\seq{C_{00}^\good{m-4},\,\bot^\good{N},\,\Gamma,\,\bang{\Delta}}
      {C_{00}^\good{m}}.$$
 According to Lemma~\ref{L2}, such a sequent cannot occur in any
 derivation in Linear Logic.

\CASE{4.2.a} Now assume that our sequent of item~(a) is derived from
             two sequents of the form
$$ \seq{K_1,\,\Gamma_1,\,\bang{\Delta_1}}{C_{00}^\good{4}, B} $$
 and
$$ \seq{K_2,\,\bot^\good{N},\,\Gamma_2,\,\bang{\Delta_2}}{ } $$
 Then Lemma~\ref{L1} and Lemma~\ref{lcost} yield:
$$\left\{\begin{array}{l}%
 \#_\bot(K_2) + N  =  1                    \pmod{9N},\\%
\begin{array}{@{}l}%
  \mbox{either}\ \#_\bot(K_2) = -N,\\%
  \mbox{or}\ \#_\bot(K_2) = -2Nk',\ \mbox{for some $k'\leq 5$,}%
  \end{array}%
\end{array}\right.$$
 which is a contradiction as well.

\CASE{4.1.b} Assume that the corresponding sequent of item~(b) is
             derived from two sequents of the form
$$\seq{\limply{H_{00}^\good{2}}{\bot^\good{3}}^\good{h_1},
       \Gamma_1,\bang{\Delta_1}}{C_{00}^\good{4}} $$
 and
$$\seq{\limply{H_{00}^\good{2}}{\bot^\good{3}}^\good{h_2},
       \bot^\good{N},\Gamma_2,\bang{\Delta_2}}
       {\limply{H_{00}^\good{2}}{\bot^\good{3}}} $$
 where
$$\left\{\begin{array}{lcl}%
       1  & = & h_1 + h_2,\\%
  \Gamma &\supset & \Gamma_1,\,\Gamma_2,\\%
  \Delta  & = & \Delta_1,\,\Delta_2.%
\end{array}\right.$$
 Then Lemma~\ref{L1} and Lemma~\ref{lcost} yield a contradiction
 as follows:
$$\left\{\begin{array}{lcl}%
  (2N + 3)h_1 & = &  -8N \pmod{9N},\\%
  (2N + 3)h_2 + N & = & 2N + 3 \pmod{9N}.%
\end{array}\right.$$

\CASE{4.2.b} Now assume that our sequent of item~(b) is derived from
             two sequents of the form
$$\seq{\limply{H_{00}^\good{2}}{\bot^\good{3}}^\good{h_1},
       \Gamma_1,\bang{\Delta_1}}
  {C_{00}^\good{4}, \limply{H_{00}^\good{2}}{\bot^\good{3}}} $$
 and
$$\seq{\limply{H_{00}^\good{2}}{\bot^\good{3}}^\good{h_2},
       \bot^\good{N},\Gamma_2,\bang{\Delta_2}}{ } $$
 Then, by Lemma~\ref{L1} and Lemma~\ref{lcost}, we have a
 contradiction as well:
$$\left\{\begin{array}{lcl}%
  (2N + 3)h_1     & = & -6N + 2 \pmod{9N},\\%
  (2N + 3)h_2 + N & = & 1 \pmod{9N}.\\%
\end{array}\right.$$

\CASE{4.1.c} According to rule~{\rm {\bf L$\llto$}}, let our
 sequent of item~(c) be derived from two sequents of the form
$$\seq{K_1,\limply{H_{00}^\good{2}}{\bot^\good{3}}^\good{h_1},
       \Gamma_1,\bang{\Delta_1}} {C_{00}^\good{4}} $$
 and
$$\seq{K_2,\limply{H_{00}^\good{2}}{\bot^\good{3}}^\good{h_2},
      \bot^\good{N},\Gamma_2,\bang{\Delta_2}}{\bot^\good{3}} $$
 where
$$\left\{\begin{array}{lcl}%
       K  & = & K_1,\,K_2,\\%
       1  & = & h_1 + h_2,\\%
  \Gamma &\supset & \Gamma_1,\,\Gamma_2,\\%
  \Delta  & = & \Delta_1,\,\Delta_2.%
\end{array}\right.$$
 By Lemma~\ref{L1} and Lemma~\ref{lcost} we have a contradiction:
$$\left\{\begin{array}{lcl}%
  \#_\bot(K_1) + (2N + 3)h_1     & = & -8N \pmod{9N},\\%
  \#_\bot(K_2) + (2N + 3)h_2 + N & = & 3   \pmod{9N},\\%
\multicolumn{3}{l}%
 {\#_\bot(K_1) = -Nk',\ \mbox{for some $k'\leq 2$.}}%
\end{array}\right.$$

\CASE{4.2.c} Now assume that our sequent of item~(c) is derived from
             two sequents of the form
$$\seq{K_1,\limply{H_{00}^\good{2}}{\bot^\good{3}}^\good{h_1},
  \Gamma_1,\bang{\Delta_1}}{C_{00}^\good{4}, \bot^\good{3}} $$
 and
$$\seq{K_2,\limply{H_{00}^\good{2}}{\bot^\good{3}}^\good{h_2},
      \bot^\good{N},\Gamma_2,\bang{\Delta_2}}{ } $$
 Then Lemma~\ref{L1} and Lemma~\ref{lcost} yield:
$$\left\{\begin{array}{lcl}%
  \#_\bot(K_1) + (2N + 3)h_1     & = & -8N + 2 \pmod{9N},\\%
  \#_\bot(K_2) + (2N + 3)h_2 + N & = & 1 \pmod{9N},\\%
\multicolumn{3}{l}%
 {\#_\bot(K_2) = -Nk',\ \mbox{for some $k'\leq 2$,}}%
\end{array}\right.$$
 which is also a contradiction.

\CASE{4.1.d+e} Let the corresponding sequent of item~(d)
               or~(e) be derived from two sequents of the form
$$\seq{H_{00}^\good{h_1},\bot^\good{a_1},\Gamma_1,\bang{\Delta_1}}
      {C_{00}^\good{4}} $$
 and
$$\seq{H_{00}^\good{h_2},\bot^\good{a_2+N},
        \Gamma_2,\bang{\Delta_2}}{\bot^\good{a}} $$
 where
$$\left\{\begin{array}{lcl}%
       1  & = & h_1 + h_2,\\%
   N + a  & = & a_1 + a_2,\\%
  \Gamma &\supset & \Gamma_1,\,\Gamma_2,\\%
  \Delta  & = & \Delta_1,\,\Delta_2.%
\end{array}\right.$$
 By Lemma~\ref{L1} and Lemma~\ref{lcost} we have:
$$\left\{\begin{array}{lcl}%
   -Nh_1 + a_1     & = & -8N \pmod{9N},\\%
   -Nh_2 + a_2 + N & = & a \pmod{9N}.%
\end{array}\right.$$

 Assuming that\ \ \mbox{$h_1 = 0$,}\ \ we can conclude that\ \
 \mbox{$a_1 = N$,}\ \ which gives a contradiction because,
 by Lemma~\ref{L2}, the first sequent with its {\em wrong} right-hand
 side cannot occur in our derivation.

 \mbox{For\ \ $h_2 = 0$,}\ \ we can get also a contradiction
 because of
     $$ N-2 \ \leq\ a_2 + (N - a) \ \leq\ 2N+1.$$

\CASE{4.2.d} Let the corresponding sequent of item~(d)
             be derived from two sequents of the form
$$\seq{H_{00}^\good{h_1},\bot^\good{a_1},\Gamma_1,\bang{\Delta_1}}
      {C_{00}^\good{4}, \bot^\good{a}} $$
 and
$$\seq{H_{00}^\good{h_2},\bot^\good{a_2+N},\Gamma_2,\bang{\Delta_2}}
      { } $$
 where
$$\left\{\begin{array}{lcl}%
       1  & = & h_1 + h_2,\\%
   N + a  & = & a_1 + a_2,\\%
  \Gamma &\supset & \Gamma_1,\,\Gamma_2,\\%
  \Delta  & = & \Delta_1,\,\Delta_2.%
\end{array}\right.$$
 By Lemma~\ref{L1} and Lemma~\ref{lcost} we have:
$$\left\{\begin{array}{lcl}%
   -Nh_1 + a_1      & = & -8N + a - 1 \pmod{9N},\\%
   -Nh_2 + a_2 + N  & = & 1 \pmod{9N}.%
\end{array}\right.$$
 Assuming that\ \ \mbox{$h_2 = 0$,}\ \ we get a contradiction
 because of
     $$ N-1 \ \leq\ a_2 + (N - 1) \ \leq\ 2N+1.$$
 \mbox{For\ \ $h_2 = 1$,}\ \ we have that \ \
 \mbox{$a_2 + N = N + 1$,} \ \ which gives a contradiction because,
 according to the inductive hypothesis from item~(e), the latter
 sequent cannot occur in our derivation.

\CASE{5} Lastly, let the left-hand {\em principal\/} formula belong
         neither to~$\Gamma$ nor to~$\bang{\Delta}$.

\CASE{5.a.1} Assume that the {\em principal\/} formula is of
             the form
 $$ C_{00} = \limply{\limply{H_{00}^\good{2}}{\bot^\good{3}}}
                    {\bot^\good{3}},$$
 and, according to rule~{\rm {\bf L$\llto$}}, the corresponding
 sequent of item~(a) is derived from two sequents of the form
$$ \seq{K_1,\,\Gamma_1,\,\bang{\Delta_1}}
       {\limply{H_{00}^\good{2}}{\bot^\good{3}}} $$
 and
$$ \seq{K_2,\,\bot^\good{3},\,\Gamma_2,\,\bang{\Delta_2}}{B} $$
 where
$$\left\{\begin{array}{lcl}%
       K  &\supset & K_1,\,K_2,\\%
  \Gamma  & = & \Gamma_1,\,\Gamma_2,\\%
  \Delta  & = & \Delta_1,\,\Delta_2.%
\end{array}\right.$$
 Then, by Lemma~\ref{L1} and Lemma~\ref{lcost}, the following
 contradiction is immediate:
$$\left\{\begin{array}{lcl}%
          \#_\bot(K_1)     & = & 2N + 3     \pmod{9N},\\%
          \#_\bot(K_2) + 3 & = & \#_\bot(B) \pmod{9N},\\%
 \multicolumn{3}{l}%
 {\begin{array}{@{}l}%
  \mbox{either}\ \#_\bot(K_1) = -N,\\%
  \mbox{or}\ \#_\bot(K_1) = -2Nk',\ \mbox{for some $k'\leq 4$.}%
  \end{array}}%
\end{array}\right.$$
 If our sequent of item~(a) were derived from two sequents of the form
$$ \seq{K_1,\,\Gamma_1,\,\bang{\Delta_1}}
       {\limply{H_{00}^\good{2}}{\bot^\good{3}}, B} $$
 and
$$ \seq{K_2,\,\bot^\good{3},\,\Gamma_2,\,\bang{\Delta_2}}{ } $$
 then we had a contradiction as well:
$$\left\{\begin{array}{lcl}%
          \#_\bot(K_1)     & = & 2N + 2 + \#_\bot(B) \pmod{9N},\\%
          \#_\bot(K_2) + 3 & = & 1  \pmod{9N},\\%
 \multicolumn{3}{l}%
 {\begin{array}{@{}l}%
  \mbox{either}\ \#_\bot(K_2) = -N,\\%
  \mbox{or}\ \#_\bot(K_2) = -2Nk',\ \mbox{for some $k'\leq 4$.}%
  \end{array}}%
\end{array}\right.$$

\CASE{5.a.2} Assume that the {\em principal\/} formula is of the form
 $$ H_{00} = \limply{\bot^\good{N+2}}{\bot^\good{2}},$$
 and our sequent of item~(a) is derived from two sequents of the form
$$ \seq{H_{00}^\good{h_1},\,\Gamma_1,\,\bang{\Delta_1}}
       {\bot^\good{N+2}} $$
 and
$$ \seq{H_{00}^\good{h_2},\,\bot^\good{2},\,
                 \Gamma_2,\,\bang{\Delta_2}}{B} $$
 where
$$\left\{\begin{array}{lcl}%
           1  & \geq & h_1 + h_2,\\%
      \Gamma  & = & \Gamma_1,\,\Gamma_2,\\%
      \Delta  & = & \Delta_1,\,\Delta_2.%
\end{array}\right.$$
 Then Lemma~\ref{L1} and Lemma~\ref{lcost} yield the following
 contradiction:
$$\left\{\begin{array}{lcl}%
       -Nh_1     & = & N + 2 \pmod{9N},\\%
       -Nh_2 + 2 & = & \#_\bot(B) \pmod{9N}.\\%
\end{array}\right.$$
 If our sequent of item~(a) were derived from two sequents of the form
$$ \seq{H_{00}^\good{h_1},\,\Gamma_1,\,\bang{\Delta_1}}
       {\bot^\good{N+2}, B} $$
 and
$$ \seq{H_{00}^\good{h_2},\,\bot^\good{2},\,
            \Gamma_2,\,\bang{\Delta_2}}{ } $$
 then we got a contradiction as follows:
$$\left\{\begin{array}{lcl}%
       -Nh_1      & = &  N + 1 + \#_\bot(B) \pmod{9N},\\%
       -Nh_2 + 2  & = &  1 \pmod{9N}.\\%
\end{array}\right.$$

\CASE{5.b} Assume that the {\em principal\/} formula is of the form
           $$ \limply{H_{00}^\good{2}}{\bot^\good{3}},$$
           and the corresponding sequent of item~(b) is derived from
           two sequents of the form
$$\seq{\Gamma_1,\,\bang{\Delta_1}}{H_{00}^\good{2}} $$
 and
$$\seq{\bot^\good{3},\,\Gamma_2,\,\bang{\Delta_2}}
       {\limply{H_{00}^\good{2}}{\bot^\good{3}}} $$
 where
$$\left\{\begin{array}{lcl}%
  \Gamma  & = & \Gamma_1,\,\Gamma_2,\\%
  \Delta  & = & \Delta_1,\,\Delta_2.%
\end{array}\right.$$
 Then, by Lemma~\ref{L1} and Lemma~\ref{lcost}, a contradiction is
 immediate:
$$\left\{\begin{array}{lcl}%
       0 & = & -2N \pmod{9N},\\%
       3 & = &  2N + 3 \pmod{9N}.\\%
\end{array}\right.$$
 If our sequent of item~(b) is derived from two sequents of the form
$$\seq{\Gamma_1,\bang{\Delta_1}}
      {H_{00}^\good{2}, \limply{H_{00}^\good{2}}{\bot^\good{3}}} $$
 and
$$\seq{\bot^\good{3},\,\Gamma_2,\,\bang{\Delta_2}}{ } $$
 then we get also a contradiction:
$$\left\{\begin{array}{lcl}%
       0 & = & 2 \pmod{9N},\\%
       3 & = & 1 \pmod{9N}.\\%
\end{array}\right.$$

\CASE{5.c.1} Suppose that the {\em principal\/} formula is of
             the form
 $$ C_{00} = \limply{\limply{H_{00}^\good{2}}{\bot^\good{3}}}
                    {\bot^\good{3}},$$
 and, according to rule~{\rm {\bf L$\llto$}}, the corresponding
 sequent of item~(c) is derived from two sequents of the form
$$ \seq{\limply{H_{00}^\good{2}}{\bot^\good{3}}^\good{h_1},
        \Gamma_1,\bang{\Delta_1}}
       {\limply{H_{00}^\good{2}}{\bot^\good{3}}} $$
 and
$$\seq{\limply{H_{00}^\good{2}}{\bot^\good{3}}^\good{h_2},
      \bot^\good{3},\Gamma_2,\bang{\Delta_2}}{\bot^\good{3}} $$
 where
$$\left\{\begin{array}{lcl}%
       1  & = & h_1 + h_2,\\%
  \Gamma  & = & \Gamma_1,\,\Gamma_2,\\%
  \Delta  & = & \Delta_1,\,\Delta_2.%
\end{array}\right.$$
 Then Lemma~\ref{L1} and Lemma~\ref{lcost} yield:
$$\left\{\begin{array}{lcl}%
  (2N + 3)h_1     & = & 2N + 3 \pmod{9N},\\%
  (2N + 3)h_2 + 3 & = & 3 \pmod{9N}.%
\end{array}\right.$$
 The only solution of this system is as follows:
$$\left\{\begin{array}{l}%
          h_1 = 1,\\%
          h_2 = 0.%
\end{array}\right.$$
 By applying the inductive hypothesis from item~(b) and Lemma~\ref{L2},
 we can get the {\em emptiness} of both~$\Gamma_1$ and~$\Gamma_2$,
 and the {\em degeneracy} of both~$\bang{\Delta_1}$
 and~$\bang{\Delta_2}$, which results in the desired {\em emptiness}
 of the whole~$\Gamma$ and the {\em degeneracy} of the
 whole~$\bang{\Delta}$.

 If our sequent of item~(c) were derived from two sequents of the form
$$ \seq{\limply{H_{00}^\good{2}}{\bot^\good{3}}^\good{h_1},
        \Gamma_1,\bang{\Delta_1}}
       {\limply{H_{00}^\good{2}}{\bot^\good{3}}, \bot^\good{3}} $$
 and
$$\seq{\limply{H_{00}^\good{2}}{\bot^\good{3}}^\good{h_2},
      \bot^\good{3},\Gamma_2,\bang{\Delta_2}}{ } $$
 then we had a contradiction:
$$\left\{\begin{array}{lcl}%
  (2N + 3)h_1     & = & 2N + 5 \pmod{9N},\\%
  (2N + 3)h_2 + 3 & = & 1 \pmod{9N}.%
\end{array}\right.$$

\CASE{5.c.2} Assume that the {\em principal\/} formula is of the form
 $$ H_{00} = \limply{\bot^\good{N+2}}{\bot^\good{2}},$$
 and our sequent of item~(c) is derived from two sequents of the form
$$ \seq{H_{00}^\good{k_1},
        \limply{H_{00}^\good{2}}{\bot^\good{3}}^\good{h_1},
        \Gamma_1,\bang{\Delta_1}} {\bot^\good{N+2}} $$
 and
$$ \seq{H_{00}^\good{k_2},
        \limply{H_{00}^\good{2}}{\bot^\good{3}}^\good{h_2},
        \bot^\good{2},\Gamma_2,\bang{\Delta_2}}{\bot^\good{3}} $$
 where
$$\left\{\begin{array}{lcl}%
           1  & = & k_1 + k_2,\\%
           1  & = & h_1 + h_2,\\%
      \Gamma  & = & \Gamma_1,\,\Gamma_2,\\%
      \Delta  & = & \Delta_1,\,\Delta_2.%
\end{array}\right.$$
 Then Lemma~\ref{L1} and Lemma~\ref{lcost} yield the following
 contradiction:
$$\left\{\begin{array}{lcl}%
       -Nk_1 + (2N + 3)h_1      & = &  N + 2 \pmod{9N},\\%
       -Nk_2 + (2N + 3)h_2 + 2  & = &  3 \pmod{9N}.%
\end{array}\right.$$
 If our sequent of item~(c) were derived from two sequents of the form
$$ \seq{H_{00}^\good{k_1},
        \limply{H_{00}^\good{2}}{\bot^\good{3}}^\good{h_1},
  \Gamma_1,\bang{\Delta_1}} {\bot^\good{N+2}, \bot^\good{3}} $$
 and
$$ \seq{H_{00}^\good{k_2},
        \limply{H_{00}^\good{2}}{\bot^\good{3}}^\good{h_2},
        \bot^\good{2},\Gamma_2,\bang{\Delta_2}}{ } $$
 then we got a contradiction as follows:
$$\left\{\begin{array}{lcl}%
       -Nk_1 + (2N + 3)h_1      & = &  N + 4 \pmod{9N},\\%
       -Nk_2 + (2N + 3)h_2 + 2  & = &  1 \pmod{9N}.%
\end{array}\right.$$

\CASE{5.c.3} Suppose that the {\em principal\/} formula is of the form
           $$ \limply{H_{00}^\good{2}}{\bot^\good{3}},$$
           and our sequent of item~(c) is derived from two sequents of
           the form
$$\seq{K_1,\,\Gamma_1,\,\bang{\Delta_1}}{H_{00}^\good{2}} $$
 and
$$\seq{K_2,\,\bot^\good{3},\,\Gamma_2,\,\bang{\Delta_2}}
      {\bot^\good{3}} $$
 where
$$\left\{\begin{array}{lcl}%
       K  & = & K_1,\,K_2,\\%
  \Gamma  & = & \Gamma_1,\,\Gamma_2,\\%
  \Delta  & = & \Delta_1,\,\Delta_2.%
\end{array}\right.$$
 Then Lemma~\ref{L1} and Lemma~\ref{lcost} yield:
$$\left\{\begin{array}{lcl}%
  \#_\bot(K) + 2N + 3 & = &  3 \pmod{9N},\\%
  \#_\bot(K_1)        & = & -2N \pmod{9N},\\%
 \multicolumn{3}{l}%
 {\#_\bot(K_1) = -Nk',\ \mbox{for some $k'\leq 2$.}}%
\end{array}\right.$$
 The only solution of this system is the following:
               $$ K_1 = K.$$
 By applying the inductive hypothesis from item~(a) and Lemma~\ref{L2}
 to our both sequents, we can get the {\em emptiness} of
 both~$\Gamma_1$ and~$\Gamma_2$, and the {\em degeneracy} of
 both~$\bang{\Delta_1}$ and~$\bang{\Delta_2}$, which results in the
 desired {\em emptiness} of the whole~$\Gamma$ and the
 {\em degeneracy} of the whole~$\bang{\Delta}$.

 If our sequent of item~(c) were derived from two sequents of the form
$$ \seq{K_1,\,\Gamma_1,\,\bang{\Delta_1}}
       {H_{00}^\good{2}, \bot^\good{3}} $$
 and
$$\seq{K_2,\,\bot^\good{3},\,\Gamma_2,\,\bang{\Delta_2}}{ } $$
 then we had an immediate contradiction:
$$\left\{\begin{array}{lcl}%
  \#_\bot(K_1)        & = & -2N + 2 \pmod{9N},\\%
  \#_\bot(K_2) + 3    & = & 1 \pmod{9N},\\%
 \multicolumn{3}{l}%
 {\#_\bot(K_2) = -Nk',\ \mbox{for some $k'\leq 2$.}}%
\end{array}\right.$$

\CASE{5.d+e} Suppose that the {\em principal\/} formula is of the form
      $$ H_{00} = \limply{\bot^\good{N+2}}{\bot^\good{2}},$$
 and our sequent of item~(d) or~(e) is derived from two sequents of
 the form
$$ \seq{\bot^\good{a_1},\,\Gamma_1,\,\bang{\Delta_1}}
       {\bot^\good{N+2}} $$
 and
$$ \seq{\bot^\good{2+a_2},\,\Gamma_2,\,\bang{\Delta_2}}
       {\bot^\good{a}} $$
 where
$$\left\{\begin{array}{lcl}%
       N + a  & = & a_1 + a_2,\\%
      \Gamma  & = & \Gamma_1,\,\Gamma_2,\\%
      \Delta  & = & \Delta_1,\,\Delta_2.%
\end{array}\right.$$
 By Lemma~\ref{L1} and Lemma~\ref{lcost}, we have:
$$\left\{\begin{array}{lcl}%
      a_1     & =  & N + 2 \pmod{9N},\\%
      a_2 + 2 & =  & a \pmod{9N}.%
\end{array}\right.$$
  The only solution is as follows:
$$\left\{\begin{array}{lcl}%
      a       & =  & 2,\\%
      a_1     & =  & N + 2,\\%
      a_2 + 2 & =  & a.%
\end{array}\right.$$
 Then, by applying Lemma~\ref{L2} to our both sequents, we can get
 the desired {\em emptiness} of the whole~$\Gamma$ and the
 {\em degeneracy} of the whole~$\bang{\Delta}$.
 
 If our sequent of item~(d) were derived from two sequents of
 the form
$$ \seq{\bot^\good{a_1},\,\Gamma_1,\,\bang{\Delta_1}}
       {\bot^\good{N+2}, \bot^\good{a}} $$
 and
$$ \seq{\bot^\good{2+a_2},\,\Gamma_2,\,\bang{\Delta_2}}{ } $$
 then we got an immediate contradiction:
$$\left\{\begin{array}{lcl}%
      a_1     & =  & N + 1 + a \pmod{9N},\\%
      a_2 + 2 & =  & 1 \pmod{9N}.%
\end{array}\right.$$

 Now, extracting the possible cases from this huge amount of
 inconsistency, we can complete Lemma~\ref{LC00}.
\QED }

\subsection {Lemma~5.3}

\begin{lemma} \label{LDq}
 Let~$\Delta$ consist of formulas of the
 form~$F_\good{A}$, and~$\Gamma$ consist of formulas of the
 form~$H_1$, $D_\good{q}$, $F_\good{A}$, and~$F_\good{Y}$.

 Let $a$~be an integer such that
             $$ 4 \ \leq\ a \ \leq\ N-3.$$
\bb{a}
\item
 Let~$B$ be a formula either of the form
  $$ H_1 = \limply{C_{00}^\good{4}}{\bot^\good{N}} $$
 or of the form
  $$ \limply{\limply{H_1}{\bot^\good{a}}}{\bot^\good{a}} $$
 If~a~sequent of the form
  $$ \seq{\Gamma,\, \bang{\Delta}}{B} $$
 occurs in a cut-free derivation in Linear Logic then $\Gamma$~must
 be a singleton of the form either
$$ \Gamma = H_1 = \limply{C_{00}^\good{4}}{\bot^\good{N}} $$
 or
$$ \Gamma = \limply{\limply{H_1}{\bot^\good{a}}}{\bot^\good{a}},$$
 and $\bang{\Delta}$~can be produced by rules~{\rm {\bf W\bang{}}}
 and~{\rm {\bf C\bang{}}} only (there is no applications of
 rule~{\rm {\bf L\bang{}}} in the derivation above this sequent).%
\footnote{ We say that such a~$\bang{\Delta}$ is {\em degenerate.}}
\item 
 If~a~sequent of the form
$$ \seq{\limply{H_1}{\bot^\good{a}},\,\Gamma,\,\bang{\Delta}}
       {\limply{H_1}{\bot^\good{a}}} $$
 occurs in a cut-free derivation in Linear Logic then $\Gamma$~must
 be empty, and $\bang{\Delta}$~can be produced by
 rules~{\rm {\bf W\bang{}}} and~{\rm {\bf C\bang{}}} only.
\item 
 If~a~sequent of the form
$$ \seq{\limply{H_1}{\bot^\good{a}},\,\Gamma,\,\bang{\Delta}}
       {\bot^\good{a}} $$
 occurs in a cut-free derivation in Linear Logic then $\Gamma$~must
 be a singleton of the form either
 $$ \Gamma = H_1 = \limply{C_{00}^\good{4}}{\bot^\good{N}} $$
 or
$$\Gamma = \limply{\limply{H_1}{\bot^\good{a}}}{\bot^\good{a}},$$
 and $\bang{\Delta}$~can be produced by rules~{\rm {\bf W\bang{}}}
 and~{\rm {\bf C\bang{}}} only.
\item 
 If~a~sequent of the form
$$ \seq{C_{00}^\good{4},\,\Gamma,\,\bang{\Delta}}{\bot^\good{N}} $$
 occurs in a cut-free derivation in Linear Logic then $\Gamma$~must
 be a singleton of the form
 $$ \Gamma = H_1 =  \limply{C_{00}^\good{4}}{\bot^\good{N}},$$
 and $\bang{\Delta}$~can be produced by rules~{\rm {\bf W\bang{}}}
 and~{\rm {\bf C\bang{}}} only.
\ee
\end{lemma}

\proof {
 We use induction on a given derivation. Regarding to the form of
 the {\em principal\/} formula at a current point of the derivation,
 we will demonstrate that each of the {\em undesirable\/} cases is
 inconsistent.

\CASE{0} The {\em principal\/} formula belongs to~$\bang{\Delta}$.
 
 Assume that it is produced by rule~{\rm {\bf L\bang{}}},
 and our sequent of the form
$$\seq{\ldots,\,\Gamma,\,\bang{\Delta}}{\ldots} $$
 is derived from a sequent of the form
$$\seq{\ldots,\,\Gamma,\, F_\good{A},\,\bang{\Delta'}}{\ldots} $$
 Then we get a contradiction because, according to the inductive
 hypothesis, the form of the multiset
      $$ \Gamma,\, F_\good{A} $$
 is not correct.

 Hence, the only possibility is to apply either~{\rm {\bf W\bang{}}}
 or~{\rm {\bf C\bang{}}}. It remains to use the inductive hypothesis
 for completing this case.

\CASE{1} The right-side formula is {\em principal.}

 There are the following cases to be considered.

\CASE{1.a} For item~(a), let us consider two possible versions of
 the {\em principal\/} formula~$B$.

\CASE{1.a.1} The {\em principal\/} formula~$B$ is of the form
  $$ H_1 = \limply{C_{00}^\good{4}}{\bot^\good{N}} $$
 and, according to rule~{\rm {\bf R$\llto$}}, our sequent
 of item~(a) is derived from the sequent
$$ \seq{C_{00}^\good{4},\,\Gamma,\,\bang{\Delta}}{\bot^\good{N}},$$
 Now we can apply the inductive hypothesis from item~(d).

\CASE{1.a.2} The {\em principal\/} formula~$B$ is of the form
  $$ \limply{\limply{H_1}{\bot^\good{a}}}{\bot^\good{a}} $$
 and, according to rule~{\rm {\bf R$\llto$}}, our sequent
 of item~(a) is derived from the sequent
$$ \seq{\limply{H_1}{\bot^\good{a}},\,\Gamma,\,\bang{\Delta}}
       {\bot^\good{a}}.$$
 It remains to apply the inductive hypothesis from item~(c).

\CASE{1.b} The {\em principal\/} formula is of the
           form~\limply{H_1}{\bot^\good{a}}, and, according to
           rule~{\rm {\bf R$\llto$}}, the corresponding
 sequent of item~(b) is derived from the sequent
$$ \seq{H_1,\,\limply{H_1}{\bot^\good{a}},\,\Gamma,\,\bang{\Delta}}
       {\bot^\good{a}}.$$
 By applying the inductive hypothesis from item~(c), we prove
 that $\bang{\Delta}$~is {\em degenerate} and that the multiset
     $$ H_1,\ \Gamma $$
 should be a singleton that means the {\em emptiness} of~$\Gamma$.

\CASE{1.c} Assume that the corresponding sequent of item~(c) is
           derived from two sequents of the form
$$\seq{\limply{H_1}{\bot^\good{a}}^\good{h_1},\,
         \Gamma_1,\,\bang{\Delta_1}}{\bot}$$
 and
$$\seq{\limply{H_1}{\bot^\good{a}}^\good{h_2},\,
         \Gamma_2,\,\bang{\Delta_2}}{\bot^\good{a-1}} $$
 where
$$\left\{\begin{array}{lcl}%
       1  & = & h_1 + h_2,\\%
  \Gamma  & = & \Gamma_1,\,\Gamma_2,\\%
  \Delta  & = & \Delta_1,\,\Delta_2.%
\end{array}\right.$$
 By Lemma~\ref{L1} and Lemma~\ref{lcost} we have a contradiction:
$$\left\{\begin{array}{lcl}%
       ah_1 & = & 1 \pmod{9N},\\%
       ah_2 & = & a-1 \pmod{9N}.%
\end{array}\right.$$

\CASE{1.d} Assume that the corresponding sequent of item~(d) is
           derived from two sequents of the form
$$\seq{C_{00}^\good{k_1},\,\Gamma_1,\,\bang{\Delta_1}}{\bot}$$
 and
$$\seq{C_{00}^\good{k_2},\,\Gamma_2,\,\bang{\Delta_2}}
      {\bot^\good{N-1}} $$
 where
$$\left\{\begin{array}{lcl}%
       4  & = & k_1 + k_2,\\%
  \Gamma  & = & \Gamma_1,\,\Gamma_2,\\%
  \Delta  & = & \Delta_1,\,\Delta_2.%
\end{array}\right.$$
 Then a contradiction is immediate:
$$\left\{\begin{array}{lcl}%
       -2Nk_1 & = & 1 \pmod{9N},\\%
       -2Nk_2 & = & N-1 \pmod{9N}.%
\end{array}\right.$$

\CASE{2} Assume that the {\em principal\/} formula belongs
         to~$\Gamma$, and it is of the form~$F_\good{A}$
         \mbox{(or~$F_\good{Y}$).}

 The following subcases are to be considered.

\CASE{2.0} The {\em principal\/} formula is of the
           form~\mbox{$(F_\good{A_1} \& F_\good{A_2})$,}
 and, by rule~{\rm {\bf L$\&$}}, the corresponding sequent
 of the form
$$\seq{\ldots,\,\Gamma,\,\bang{\Delta}}{\ldots} $$
 is derived either from the sequent
$$\seq{\ldots,\,\Gamma',\, F_\good{A_1},\,\bang{\Delta}}
      {\ldots} $$
 or from the sequent
$$\seq{\ldots,\,\Gamma',\, F_\good{A_2},\,\bang{\Delta}}
      {\ldots} $$
 Then we have a contradiction because, according to the inductive
 hypothesis, either the form of the multiset
      $$ \Gamma', \ F_\good{A_1} $$
 is not correct, or the form of the multiset
      $$ \Gamma', \ F_\good{A_2} $$
 is not correct.

\CASE{2.1.a} Assume that the {\em principal\/} formula is of the
              form~\limply{E_\good{X}}{E_\good{Y}}, and,
 according to rule~{\rm {\bf L$\llto$}}, the sequent
 of item~(a) is derived from two sequents of the form
$$ \seq{\Gamma_1,\,\bang{\Delta_1}}{E_\good{X}} $$
 and
$$ \seq{E_\good{Y},\,\Gamma_2,\,\bang{\Delta_2}}{B} $$
 where
$$\left\{\begin{array}{lcl}%
  \Gamma &\supset & \Gamma_1,\,\Gamma_2,\\%
  \Delta  & = & \Delta_1,\,\Delta_2.%
\end{array}\right.$$
 Then Lemma~\ref{L1} and Lemma~\ref{lcost} yield:
$$\left\{\begin{array}{lcl}%
         0  & = & 6N  \pmod{9N},\\%
         6N & = & \#_\bot(B) \pmod{9N},%
\end{array}\right.$$
 which is a contradiction.

\CASE{2.2.a} Assume that the {\em principal\/} formula is of the
             form~\limply{E_\good{X}}{E_\good{Y}}, and our sequent
             of item~(a) is derived from two sequents of the form
$$\seq{\Gamma_1,\,\bang{\Delta_1}}{E_\good{X}, B} $$
 and
$$\seq{E_\good{Y},\,\Gamma_2,\,\bang{\Delta_2}}{ } $$
 Then Lemma~\ref{L1} and Lemma~\ref{lcost} yield a contradiction:
$$\left\{\begin{array}{lcl}%
         0  & = & 6N + \#_\bot(B) - 1 \pmod{9N},\\%
         6N & = & 1 \pmod{9N}.%
\end{array}\right.$$

\CASE{2.3.a} Assume that the {\em principal\/} formula is of the
             form~\limply{F_\good{A}}{F_\good{Y}}, and,
             according to rule~{\rm {\bf L$\llto$}}, our sequent
             of item~(a) is derived from two sequents of the form
$$ \seq{\Gamma_1,\,\bang{\Delta_1}}{F_\good{A}} $$
 and
$$ \seq{F_\good{Y},\,\Gamma_2,\,\bang{\Delta_2}}{B} $$
 where
$$\left\{\begin{array}{lcl}%
  \Gamma &\supset & \Gamma_1,\,\Gamma_2,\\%
  \Delta  & = & \Delta_1,\,\Delta_2.%
\end{array}\right.$$
 Then we can get a contradiction because, according to the inductive
 hypothesis, the multiset
      $$ F_\good{Y},\, \Gamma_2 $$
 must be a {\em singleton} of the differing form.

\CASE{2.4.a} Assume that the {\em principal\/} formula is of the
             form~\limply{F_\good{A}}{F_\good{Y}}, and our sequent
             of item~(a) is derived from two sequents of the form
$$ \seq{\Gamma_1,\,\bang{\Delta_1}}{F_\good{A}, B} $$
 and
$$ \seq{F_\good{Y},\,\Gamma_2,\,\bang{\Delta_2}}{ } $$
 where
$$\left\{\begin{array}{lcl}%
  \Gamma &\supset & \Gamma_1,\,\Gamma_2,\\%
  \Delta  & = & \Delta_1,\,\Delta_2.%
\end{array}\right.$$
 Then Lemma~\ref{L1} and Lemma~\ref{lcost} show the following
 contradiction:
$$\left\{\begin{array}{lcl}%
      0 & =  & \#_\bot(B) - 1 \pmod{9N},\\%
      0 & =  & 1 \pmod{9N}.%
\end{array}\right.$$

\CASE{2.1.b}  Assume that the {\em principal\/} formula is of the
              form~\limply{E_\good{X}}{E_\good{Y}}, and our sequent
              of item~(b) is derived from two sequents of the form
$$\seq{\limply{H_1}{\bot^\good{a}}^\good{h_1},
               \Gamma_1,\bang{\Delta_1}}{E_\good{X}} $$
 and
$$\seq{\limply{H_1}{\bot^\good{a}}^\good{h_2},
            E_\good{Y},\Gamma_2,\bang{\Delta_2}}
      {\limply{H_1}{\bot^\good{a}}} $$
 where
$$\left\{\begin{array}{lcl}%
       1  & = & h_1 + h_2,\\%
  \Gamma &\supset & \Gamma_1,\,\Gamma_2,\\%
  \Delta  & = & \Delta_1,\,\Delta_2.%
\end{array}\right.$$
 By Lemma~\ref{L1} and Lemma~\ref{lcost} we have:
$$\left\{\begin{array}{lcl}%
       ah_1      & = & 6N \pmod{9N},\\%
       ah_2 + 6N & = & a \pmod{9N},%
\end{array}\right.$$
 which is a contradiction.

\CASE{2.2.b} Assume that the {\em principal\/} formula is of the
             form~\limply{E_\good{X}}{E_\good{Y}}, and now our
 sequent of item~(b) is derived from two sequents of the form
$$\seq{\limply{H_1}{\bot^\good{a}}^\good{h_1},
                         \Gamma_1,\bang{\Delta_1}}
         {E_\good{X}, \limply{H_1}{\bot^\good{a}}} $$
 and
$$\seq{\limply{H_1}{\bot^\good{a}}^\good{h_2},
            E_\good{Y},\Gamma_2,\bang{\Delta_2}}{ } $$
 Then Lemma~\ref{L1} and Lemma~\ref{lcost} yield also a contradiction:
$$\left\{\begin{array}{lcl}%
       ah_1      & = & 6N + a -1\pmod{9N},\\%
       ah_2 + 6N & = & 1 \pmod{9N}.%
\end{array}\right.$$

\CASE{2.3.b} Assume that the {\em principal\/} formula is of the
           form~\limply{F_\good{A}}{F_\good{Y}}, and our
 sequent of item~(b) is derived from two sequents of the form
$$\seq{\limply{H_1}{\bot^\good{a}}^\good{h_1},
       \Gamma_1, \bang{\Delta_1}}{F_\good{A}} $$
 and
$$\seq{\limply{H_1}{\bot^\good{a}}^\good{h_2},
        F_\good{Y},\Gamma_2,\bang{\Delta_2}}
      {\limply{H_1}{\bot^\good{a}}} $$
 where
$$\left\{\begin{array}{lcl}%
       1  & = & h_1 + h_2,\\%
  \Gamma &\supset & \Gamma_1,\,\Gamma_2,\\%
  \Delta  & = & \Delta_1,\,\Delta_2.%
\end{array}\right.$$
 According to Lemma~\ref{L1} and Lemma~\ref{lcost}, we have:
$$\left\{\begin{array}{lcl}%
       ah_1      & = & 0 \pmod{9N},\\%
       ah_2      & = & a \pmod{9N}.%
\end{array}\right.$$
 Hence,
        $$ h_2 =  1,$$
 and we can get a contradiction because, by the inductive hypothesis,
 the {\em non-empty} multiset
      $$ F_\good{Y},\ \Gamma_2 $$
 must be empty.

\CASE{2.4.b} Assume that the {\em principal\/} formula is of the
             form~\limply{F_\good{A}}{F_\good{Y}}, and now
 our sequent of item~(b) is derived from two sequents of the form
$$\seq{\limply{H_1}{\bot^\good{a}}^\good{h_1},
                      \Gamma_1,\bang{\Delta_1}}
      {F_\good{A}, \limply{H_1}{\bot^\good{a}}} $$
 and
$$\seq{\limply{H_1}{\bot^\good{a}}^\good{h_2},
        F_\good{Y},\Gamma_2,\bang{\Delta_2}}{ } $$
 Then Lemma~\ref{L1} and Lemma~\ref{lcost} yield a contradiction
 as well:
$$\left\{\begin{array}{lcl}%
       ah_1      & = & a-1 \pmod{9N},\\%
       ah_2      & = & 1 \pmod{9N}.%
\end{array}\right.$$

\CASE{2.1.c} Assume that the {\em principal\/} formula is of the
             form~\limply{E_\good{X}}{E_\good{Y}}, and our sequent
             of item~(c) is derived from two sequents of the form
$$\seq{\limply{H_1}{\bot^\good{a}}^\good{h_1},
               \Gamma_1,\bang{\Delta_1}}{E_\good{X}} $$
 and
$$\seq{\limply{H_1}{\bot^\good{a}}^\good{h_2},
            E_\good{Y},\Gamma_2,\bang{\Delta_2}}{\bot^\good{a}} $$
 where
$$\left\{\begin{array}{lcl}%
       1  & = & h_1 + h_2,\\%
  \Gamma &\supset & \Gamma_1,\,\Gamma_2,\\%
  \Delta  & = & \Delta_1,\,\Delta_2.%
\end{array}\right.$$
 Then a contradiction is as follows:
$$\left\{\begin{array}{lcl}%
       ah_1      & = & 6N \pmod{9N},\\%
       ah_2 + 6N & = & a \pmod{9N}.%
\end{array}\right.$$

\CASE{2.2.c} Assume that the {\em principal\/} formula is of the
             form~\limply{E_\good{X}}{E_\good{Y}}, and our sequent
             of item~(c) is derived from two sequents of the form
$$\seq{\limply{H_1}{\bot^\good{a}}^\good{h_1},
    \Gamma_1,\bang{\Delta_1}}{E_\good{X}, \bot^\good{a}} $$
 and
$$\seq{\limply{H_1}{\bot^\good{a}}^\good{h_2},
            E_\good{Y},\Gamma_2,\bang{\Delta_2}}{ } $$
 Then we have an immediate contradiction:
$$\left\{\begin{array}{lcl}%
       ah_1      & = & 6N+a-1 \pmod{9N},\\%
       ah_2 + 6N & = & 1 \pmod{9N}.%
\end{array}\right.$$

\CASE{2.3.c} Assume that the {\em principal\/} formula is of the
            form~\limply{F_\good{A}}{F_\good{Y}}, and our
 sequent of item~(c) is derived from two sequents of the form
$$\seq{\limply{H_1}{\bot^\good{a}}^\good{h_1},
       \Gamma_1, \bang{\Delta_1}}{F_\good{A}} $$
 and
$$\seq{\limply{H_1}{\bot^\good{a}}^\good{h_2},
        F_\good{Y},\Gamma_2,\bang{\Delta_2}}{\bot^\good{a}} $$
 where
$$\left\{\begin{array}{lcl}%
       1  & = & h_1 + h_2,\\%
  \Gamma &\supset & \Gamma_1,\,\Gamma_2,\\%
  \Delta  & = & \Delta_1,\,\Delta_2.%
\end{array}\right.$$
 According to Lemma~\ref{L1} and Lemma~\ref{lcost}, we have:
$$\left\{\begin{array}{lcl}%
       ah_1      & = & 0 \pmod{9N},\\%
       ah_2      & = & a \pmod{9N}.%
\end{array}\right.$$
 Then
        $$ h_2 =  1,$$
 and we get a contradiction because, according to the inductive
 hypothesis, the multiset
      $$ F_\good{Y},\, \Gamma_2 $$
 must be a {\em singleton} of the differing form.

\CASE{2.4.c} Assume that the {\em principal\/} formula is of the
             form~\limply{F_\good{A}}{F_\good{Y}}, and our
 sequent of item~(c) is derived from two sequents of the form
$$\seq{\limply{H_1}{\bot^\good{a}}^\good{h_1},
       \Gamma_1, \bang{\Delta_1}}{F_\good{A}, \bot^\good{a}} $$
 and
$$\seq{\limply{H_1}{\bot^\good{a}}^\good{h_2},
        F_\good{Y},\Gamma_2,\bang{\Delta_2}}{ } $$
 Then a contradiction is immediate:
$$\left\{\begin{array}{lcl}%
       ah_1      & = & a-1 \pmod{9N},\\%
       ah_2      & = & 1 \pmod{9N}.%
\end{array}\right.$$

\CASE{2.1.d} Assume that the {\em principal\/} formula is of the
             form~\limply{E_\good{X}}{E_\good{Y}}, and the
             corresponding sequent of item~(d) is derived from
             two sequents of the form
$$ \seq{C_{00}^\good{k_1},\Gamma_1,\bang{\Delta_1}}{E_\good{X}} $$
 and
$$ \seq{C_{00}^\good{k_2}, E_\good{Y}, \Gamma_2,\bang{\Delta_2}}
       {\bot^\good{N}} $$
 where
$$\left\{\begin{array}{lcl}%
       4  & = & k_1 + k_2,\\%
  \Gamma &\supset & \Gamma_1,\,\Gamma_2,\\%
  \Delta  & = & \Delta_1,\,\Delta_2.%
\end{array}\right.$$
 Then, by Lemma~\ref{L1} and Lemma~\ref{lcost} we have:
$$\left\{\begin{array}{lcl}%
     -2Nk_1      & = & 6N \pmod{9N},\\%
     -2Nk_2 + 6N & = & N \pmod{9N},%
\end{array}\right.$$
 which is a contradiction.

\CASE{2.2.d} Assume that the {\em principal\/} formula is of the
             form~\limply{E_\good{X}}{E_\good{Y}}, and our sequent
             of item~(d) is derived from two sequents of the form
$$ \seq{C_{00}^\good{k_1},\Gamma_1,\bang{\Delta_1}}
       {E_\good{X}, \bot^\good{N}} $$
 and
$$ \seq{C_{00}^\good{k_2},E_\good{Y},\Gamma_2,\bang{\Delta_2}}{ } $$
 Then we get also a contradiction:
$$\left\{\begin{array}{lcl}%
     -2Nk_1      & = & 7N-1 \pmod{9N},\\%
     -2Nk_2 + 6N & = & 1 \pmod{9N}.%
\end{array}\right.$$
  
\CASE{2.3.d} Assume that the {\em principal\/} formula is of the
             form~\limply{F_\good{A}}{F_\good{Y}}, and our sequent
             of item~(d) is derived from two sequents of the form
$$ \seq{C_{00}^\good{k_1},\Gamma_1,\bang{\Delta_1}}{F_\good{A}} $$
 and
$$ \seq{C_{00}^\good{k_2},F_\good{Y},\Gamma_2,\bang{\Delta_2}}
       {\bot^\good{N}} $$
 where
$$\left\{\begin{array}{lcl}%
       4  & = & k_1 + k_2,\\%
  \Gamma &\supset & \Gamma_1,\,\Gamma_2,\\%
  \Delta  & = & \Delta_1,\,\Delta_2.%
\end{array}\right.$$
 According to Lemma~\ref{L1} and Lemma~\ref{lcost}, we have:
$$\left\{\begin{array}{lcl}%
     -2Nk_1      & = & 0 \pmod{9N},\\%
     -2Nk_2      & = & N \pmod{9N}.%
\end{array}\right.$$
 Then
        $$ k_2 =  4,$$
 and we get a contradiction because, according to the inductive
 hypothesis, the multiset
      $$ F_\good{Y},\, \Gamma_2 $$
 must be a {\em singleton} of the differing form.
  
\CASE{2.4.d} Assume that the {\em principal\/} formula is of the
             form~\limply{F_\good{A}}{F_\good{Y}}, and our sequent
             of item~(d) is derived from two sequents of the form
$$ \seq{C_{00}^\good{k_1},\Gamma_1,\bang{\Delta_1}}
       {F_\good{A}, \bot^\good{N}} $$
 and
$$\seq{C_{00}^\good{k_2},F_\good{Y},\Gamma_2,\bang{\Delta_2}}{ }$$
 Then we can get a contradiction as follows:
$$\left\{\begin{array}{lcl}%
     -2Nk_1      & = & N-1 \pmod{9N},\\%
     -2Nk_2      & = & 1 \pmod{9N}.%
\end{array}\right.$$

\CASE{2.5} Case of the {\em principal\/} formula of the
           form~\lvariant{E_\good{X}}{E_\good{Y_1}}{E_\good{Y_2}}
           is handled similarly to~\mbox{{\bf Cases 2.1.abcd}}
            and~\mbox{{\bf Cases 2.2.abcd.}}

\CASE{3} Assume that the {\em principal\/} formula belongs
         to~$\Gamma$, and it is of the form
$$D_\good{q} = \limply{\limply{H_1}{\bot^\good{b}}}{\bot^\good{b}}$$
 where
          $$ 4 \ \leq\  b \ \leq\  N-3.$$

\CASE{3.1.a} According to rule~{\rm {\bf L$\llto$}}, let our sequent
             of item~(a) be derived from two sequents of the form
$$ \seq{\Gamma_1,\,\bang{\Delta_1}}{\limply{H_1}{\bot^\good{b}}} $$
 and
$$ \seq{\bot^\good{b},\,\Gamma_2,\,\bang{\Delta_2}}{B} $$
 where
$$\left\{\begin{array}{lcl}%
  \Gamma &\supset & \Gamma_1,\,\Gamma_2,\\%
  \Delta  & = & \Delta_1,\,\Delta_2.%
\end{array}\right.$$
 Then Lemma~\ref{L1} and Lemma~\ref{lcost} yield a contradiction:
$$\left\{\begin{array}{lcl}%
          0 & = & b  \pmod{9N},\\%
          b & = & \#_\bot(B) \pmod{9N}.%
\end{array}\right.$$

\CASE{3.2.a} Now let our sequent of item~(a) be derived from two
             sequents of the form
$$ \seq{\Gamma_1,\,\bang{\Delta_1}}{\limply{H_1}
       {\bot^\good{b}}, B} $$
 and
$$ \seq{\bot^\good{b},\,\Gamma_2,\,\bang{\Delta_2}}{ } $$
 Then we have the following contradiction:
$$\left\{\begin{array}{lcl}%
          0 & = & b+\#_\bot(B)-1 \pmod{9N},\\%
          b & = & 1 \pmod{9N}.%
\end{array}\right.$$

\CASE{3.1.b} According to rule~{\rm {\bf L$\llto$}}, assume that
             the corresponding sequent of item~(b) is derived from
             two sequents of the form
$$\seq{\limply{H_1}{\bot^\good{a}}^\good{h_1},
        \Gamma_1,\bang{\Delta_1}}{\limply{H_1}{\bot^\good{b}}} $$
 and
$$\seq{\limply{H_1}{\bot^\good{a}}^\good{h_2},
            \bot^\good{b},\Gamma_2,\bang{\Delta_2}}
      {\limply{H_1}{\bot^\good{a}}} $$
 where
$$\left\{\begin{array}{lcl}%
       1  &=& h_1 + h_2,\\%
  \Gamma &=& \Gamma_1,\Gamma_2,%
          \limply{\limply{H_1}{\bot^\good{b}}}{\bot^\good{b}},\\%
  \Delta  &=& \Delta_1,\,\Delta_2.%
\end{array}\right.$$
 By Lemma~\ref{L1} and Lemma~\ref{lcost} we have:
$$\left\{\begin{array}{lcl}%
       ah_1      & = & b \pmod{9N},\\%
       ah_2 + b  & = & a \pmod{9N}.%
\end{array}\right.$$
 Hence,
$$\left\{\begin{array}{lcl}%
        h_2 & = & 0,\\%
        b   & = & a.%
\end{array}\right.$$
 And we have a contradiction because, according to Lemma~\ref{L2},
 the latter sequent with its {\em wrong} right-hand side cannot
 occur in our derivation.

\CASE{3.2.b} Assume that our sequent of item~(b) is derived from
             two sequents of the form
$$\seq{\limply{H_1}{\bot^\good{a}}^\good{h_1},
        \Gamma_1,\bang{\Delta_1}}
   {\limply{H_1}{\bot^\good{b}}, \limply{H_1}{\bot^\good{a}}} $$
 and
$$\seq{\limply{H_1}{\bot^\good{a}}^\good{h_2},
            \bot^\good{b},\Gamma_2,\bang{\Delta_2}}{ } $$
 A contradiction is immediate:
$$\left\{\begin{array}{lcl}%
       ah_1      & = & b+a-1 \pmod{9N},\\%
       ah_2 + b  & = & 1 \pmod{9N}.%
\end{array}\right.$$

\CASE{3.1.c} According to rule~{\rm {\bf L$\llto$}}, suppose that
             the corresponding sequent of item~(c) is derived from
             two sequents of the form
$$\seq{\limply{H_1}{\bot^\good{a}}^\good{h_1},
      \Gamma_1,\bang{\Delta_1}} {\limply{H_1}{\bot^\good{b}}} $$
 and
$$\seq{\limply{H_1}{\bot^\good{a}}^\good{h_2},
     \bot^\good{b},\Gamma_2,\bang{\Delta_2}}{\bot^\good{a}} $$
 where
$$\left\{\begin{array}{lcl}%
       1  &=& h_1 + h_2,\\%
  \Gamma &=& \Gamma_1,\Gamma_2,%
          \limply{\limply{H_1}{\bot^\good{b}}}{\bot^\good{b}},\\%
  \Delta  &=& \Delta_1,\,\Delta_2.%
\end{array}\right.$$
 By Lemma~\ref{L1} and Lemma~\ref{lcost} we have:
$$\left\{\begin{array}{lcl}%
       ah_1      & = & b \pmod{9N},\\%
       ah_2 + b  & = & a \pmod{9N}.%
\end{array}\right.$$
 Hence,
$$\left\{\begin{array}{lcl}%
        h_1 & = & 1,\\%
        b   & = & a.%
\end{array}\right.$$
 By applying the inductive hypothesis from item~(b) and Lemma~\ref{L2}
 to both sequents, we prove the {\em emptiness} of both~$\Gamma_1$
 and~$\Gamma_2$, and the {\em degeneracy} of both~$\bang{\Delta_1}$
 and~$\bang{\Delta_2}$. Therefore, the whole~$\bang{\Delta}$ is
 {\em degenerate,} and the whole~$\Gamma$ is a {\em singleton}
 of the form
$$ \Gamma = \limply{\limply{H_1}{\bot^\good{b}}}{\bot^\good{b}}.$$

\CASE{3.2.c} Assume that our sequent of item~(c) is derived from
             two sequents of the form
$$\seq{\limply{H_1}{\bot^\good{a}}^\good{h_1},
                      \Gamma_1,\bang{\Delta_1}}
      {\limply{H_1}{\bot^\good{b}}, \bot^\good{a}} $$
 and
$$\seq{\limply{H_1}{\bot^\good{a}}^\good{h_2},
            \bot^\good{b},\Gamma_2,\bang{\Delta_2}}{ } $$
 A contradiction is immediate:
$$\left\{\begin{array}{lcl}%
       ah_1      & = & b+a-1 \pmod{9N},\\%
       ah_2 + b  & = & 1 \pmod{9N}.%
\end{array}\right.$$

\CASE{3.1.d} Let the corresponding sequent of item~(d) be derived
             from two sequents of the form
$$ \seq{C_{00}^\good{k_1},\Gamma_1,\bang{\Delta_1}}
       {\limply{H_1}{\bot^\good{b}}} $$
 and
$$ \seq{C_{00}^\good{k_2},\bot^\good{b},\Gamma_2,\bang{\Delta_2}}
       {\bot^\good{N}} $$
 where
$$\left\{\begin{array}{lcl}%
       4  & = & k_1 + k_2,\\%
  \Gamma &\supset & \Gamma_1,\,\Gamma_2,\\%
  \Delta  & = & \Delta_1,\,\Delta_2.%
\end{array}\right.$$
 Then, by Lemma~\ref{L1} and Lemma~\ref{lcost} we have:
$$\left\{\begin{array}{lcl}%
     -2Nk_1      & = & b \pmod{9N},\\%
     -2Nk_2 + b  & = & N \pmod{9N},%
\end{array}\right.$$
 which is a contradiction.

\CASE{3.2.d} Let our sequent of item~(d) be derived from two sequents
              of the form
$$ \seq{C_{00}^\good{k_1},\Gamma_1,\bang{\Delta_1}}
       {\limply{H_1}{\bot^\good{b}}, \bot^\good{N}} $$
 and
$$ \seq{C_{00}^\good{k_2},\bot^\good{b},\Gamma_2,\bang{\Delta_2}}
       { } $$
 Then we get also a contradiction:
$$\left\{\begin{array}{lcl}%
     -2Nk_1      & = & b+N-1 \pmod{9N},\\%
     -2Nk_2 + b  & = & 1 \pmod{9N}.%
\end{array}\right.$$

\CASE{4} Assume that the {\em principal\/} formula belongs
         to~$\Gamma$, and it is of the form
  $$ H_1 = \limply{C_{00}^\good{4}}{\bot^\good{N}}.$$

\CASE{4.1.a} According to rule~{\rm {\bf L$\llto$}}, assume that
             the corresponding sequent of item~(a) is derived from
             two sequents of the form
$$ \seq{\Gamma_1,\,\bang{\Delta_1}}{C_{00}^\good{4}} $$
 and
$$ \seq{\bot^\good{N},\,\Gamma_2,\,\bang{\Delta_2}}{B} $$
 where
$$\left\{\begin{array}{lcl}%
  \Gamma &\supset & \Gamma_1,\,\Gamma_2,\\%
  \Delta  & = & \Delta_1,\,\Delta_2.%
\end{array}\right.$$
 Then Lemma~\ref{L1} and Lemma~\ref{lcost} yield:
$$\left\{\begin{array}{lcl}%
     0  & = & -8N        \pmod{9N},\\%
     N  & = & \#_\bot(B) \pmod{9N},%
\end{array}\right.$$
  which is a contradiction.

\CASE{4.2.a} Now assume that our sequent of item~(a) is derived from
             two sequents of the form
$$ \seq{\Gamma_1,\,\bang{\Delta_1}}{C_{00}^\good{4}, B} $$
 and
$$ \seq{\bot^\good{N},\,\Gamma_2,\,\bang{\Delta_2}}{ } $$
 Then we get contradiction as well:
$$\left\{\begin{array}{lcl}%
     0  & = & -8N+\#_\bot(B)-1 \pmod{9N},\\%
     N  & = & 1 \pmod{9N}.%
\end{array}\right.$$

\CASE{4.1.b} Assume that the corresponding sequent of item~(b) is
             derived from two sequents of the form
$$\seq{\limply{H_1}{\bot^\good{a}}^\good{h_1},
       \Gamma_1,\bang{\Delta_1}}{C_{00}^\good{4}} $$
 and
$$\seq{\limply{H_1}{\bot^\good{a}}^\good{h_2},
       \bot^\good{N},\Gamma_2,\bang{\Delta_2}}
       {\limply{H_1}{\bot^\good{a}}} $$
 where
$$\left\{\begin{array}{lcl}%
       1  & = & h_1 + h_2,\\%
  \Gamma &\supset & \Gamma_1,\,\Gamma_2,\\%
  \Delta  & = & \Delta_1,\,\Delta_2.%
\end{array}\right.$$
 Then Lemma~\ref{L1} and Lemma~\ref{lcost} yield a contradiction
 as follows:
$$\left\{\begin{array}{lcl}%
  ah_1 =  -8N \pmod{9N},\\%
  ah_2 + N = a \pmod{9N}.%
\end{array}\right.$$

\CASE{4.2.b} Now assume that our sequent of item~(b) is derived from
             two sequents of the form
$$\seq{\limply{H_1}{\bot^\good{a}}^\good{h_1},
       \Gamma_1,\bang{\Delta_1}}
      {C_{00}^\good{4}, \limply{H_1}{\bot^\good{a}}} $$
 and
$$\seq{\limply{H_1}{\bot^\good{a}}^\good{h_2},
       \bot^\good{N},\Gamma_2,\bang{\Delta_2}}{ } $$
 Then, by Lemma~\ref{L1} and Lemma~\ref{lcost}, we have a
 contradiction as well:
$$\left\{\begin{array}{lcl}%
  ah_1     & = & -8N+a-1 \pmod{9N},\\%
  ah_2 + N & = & 1 \pmod{9N}.\\%
\end{array}\right.$$

\CASE{4.1.c} According to rule~{\rm {\bf L$\llto$}}, let our sequent
             of item~(c) be derived from two sequents of the form
$$\seq{\limply{H_1}{\bot^\good{a}}^\good{h_1},
       \Gamma_1,\bang{\Delta_1}} {C_{00}^\good{4}} $$
 and
$$\seq{\limply{H_1}{\bot^\good{a}}^\good{h_2},
      \bot^\good{N},\Gamma_2,\bang{\Delta_2}}{\bot^\good{a}} $$
 where
$$\left\{\begin{array}{lcl}%
       1  & = & h_1 + h_2,\\%
  \Gamma &\supset & \Gamma_1,\,\Gamma_2,\\%
  \Delta  & = & \Delta_1,\,\Delta_2.%
\end{array}\right.$$
 By Lemma~\ref{L1} and Lemma~\ref{lcost} we have a contradiction:
$$\left\{\begin{array}{lcl}%
  ah_1 & = &  -8N \pmod{9N},\\%
  ah_2 + N & = & a \pmod{9N}.%
\end{array}\right.$$

\CASE{4.2.c} Now assume that our sequent of item~(c) is derived from
             two sequents of the form
$$\seq{\limply{H_1}{\bot^\good{a}}^\good{h_1},
   \Gamma_1,\bang{\Delta_1}} {C_{00}^\good{4}, \bot^\good{a}} $$
 and
$$\seq{\limply{H_1}{\bot^\good{a}}^\good{h_2},
      \bot^\good{N},\Gamma_2,\bang{\Delta_2}}{ } $$
 Then Lemma~\ref{L1} and Lemma~\ref{lcost} yield:
$$\left\{\begin{array}{lcl}%
  ah_1     & = & -8N+a-1 \pmod{9N},\\%
  ah_2 + N & = & 1 \pmod{9N},\\%
\end{array}\right.$$
 which is also a contradiction.

\CASE{4.1.d} Suppose that the corresponding sequent of item~(d)
             is derived from two sequents of the form
$$\seq{C_{00}^\good{k_1},\Gamma_1,\bang{\Delta_1}}{C_{00}^\good{4}}$$
 and
$$\seq{C_{00}^\good{k_2},\bot^\good{N},\Gamma_2,\bang{\Delta_2}}
      {\bot^\good{N}} $$
 where
$$\left\{\begin{array}{lcl}%
       4  &=& k_1 + k_2,\\%
  \Gamma  &=& \Gamma_1,\Gamma_2,%
              \limply{C_{00}^\good{4}}{\bot^\good{N}},\\%
  \Delta  &=& \Delta_1,\,\Delta_2.%
\end{array}\right.$$
 By Lemma~\ref{L1} and Lemma~\ref{lcost} we have:
$$\left\{\begin{array}{lcl}%
   -2Nk_1     & = & -8N \pmod{9N},\\%
   -2Nk_2 + N & = & N \pmod{9N}.%
\end{array}\right.$$
 The only solution of this system is as follows:
$$\left\{\begin{array}{lcl}%
      k_1 & =  & 4,\\%
      k_2 & =  & 0.%
\end{array}\right.$$
 Then, by applying Lemma~\ref{LC00} and Lemma~\ref{L2} to both
 sequents, we prove the {\em emptiness} of both~$\Gamma_1$
 and~$\Gamma_2$, and the {\em degeneracy} of both~$\bang{\Delta_1}$
 and~$\bang{\Delta_2}$. Therefore, the whole~$\bang{\Delta}$ is
 {\em degenerate,} and the whole~$\Gamma$ is a {\em singleton}
 of the form
$$ \Gamma = \limply{C_{00}^\good{4}}{\bot^\good{N}}.$$ 

\CASE{4.2.d} Assuming that our sequent of item~(d) is derived from
             two sequents of the form
$$ \seq{C_{00}^\good{k_1},\Gamma_1,\bang{\Delta_1}}
       {C_{00}^\good{4}, \bot^\good{N}}$$
 and
$$\seq{C_{00}^\good{k_2},\bot^\good{N},\Gamma_2,\bang{\Delta_2}}{}$$
 we have a contradiction:
$$\left\{\begin{array}{lcl}%
   -2Nk_1     & = & -7N-1 \pmod{9N},\\%
   -2Nk_2 + N & = & 1 \pmod{9N}.%
\end{array}\right.$$

\CASE{5} Finally, let us consider the case where the left-hand
         {\em principal\/} formula belongs neither to~$\Gamma$
         nor to~$\bang{\Delta}$.

\CASE{5.b} Assume that the {\em principal\/} formula is of the form
           $$ \limply{H_1}{\bot^\good{a}},$$
           and the corresponding sequent of item~(b) is derived from
           two sequents of the form
$$ \seq{\Gamma_1,\,\bang{\Delta_1}}{H_1} $$
 and
$$ \seq{\bot^\good{a},\,\Gamma_2,\,\bang{\Delta_2}}
       {\limply{H_1}{\bot^\good{a}}} $$
 where
$$\left\{\begin{array}{lcl}%
  \Gamma  & = & \Gamma_1,\,\Gamma_2,\\%
  \Delta  & = & \Delta_1,\,\Delta_2.%
\end{array}\right.$$
 And we have a contradiction because, according to Lemma~\ref{L2},
 the latter sequent with its {\em wrong} right-hand side cannot
 occur in our derivation.

 If our sequent of item~(b) were derived from two sequents of the form
$$ \seq{\Gamma_1,\,\bang{\Delta_1}}
       {H_1, \limply{H_1}{\bot^\good{a}}} $$
 and
$$ \seq{\bot^\good{a},\,\Gamma_2,\,\bang{\Delta_2}}{ } $$
 then, by Lemma~\ref{L1} and Lemma~\ref{lcost}, we got
 a contradiction as well:
$$\left\{\begin{array}{lcl}%
       0 & = & a-1 \pmod{9N},\\%
       a & = & 1 \pmod{9N}.\\%
\end{array}\right.$$

\CASE{5.c} Suppose that the {\em principal\/} formula is of the form
           $$ \limply{H_1}{\bot^\good{a}},$$
           and the corresponding sequent of item~(c) is derived from
           two sequents of the form
$$ \seq{\Gamma_1,\,\bang{\Delta_1}}{H_1} $$
 and
$$\seq{\bot^\good{a},\,\Gamma_2,\,\bang{\Delta_2}}{\bot^\good{a}} $$
 where
$$\left\{\begin{array}{lcl}%
  \Gamma  & = & \Gamma_1,\,\Gamma_2,\\%
  \Delta  & = & \Delta_1,\,\Delta_2.%
\end{array}\right.$$
 Then, by applying the inductive hypothesis from item~(a) and
 Lemma~\ref{L2} to our both sequents, we can prove that
\bb{1}
\item multiset~$\Gamma_1$ must be a singleton of the form either
$$ \Gamma_1 = H_1 = \limply{C_{00}^\good{4}}{\bot^\good{N}} $$
 or
$$ \Gamma_1 = 
      \limply{\limply{H_1}{\bot^\good{a'}}}{\bot^\good{a'}},$$
\item multiset~$\Gamma_2$ must be empty,
\item both~$\bang{\Delta_1}$ and~$\bang{\Delta_2}$ must be
      {\em degenerate}, which results in the desired
      {\em degeneracy} of the whole~$\bang{\Delta}$.
\ee
 Hence, the whole~$\Gamma$ is of the required form
         $$ \Gamma = \Gamma_1.$$
 If our sequent of item~(c) were derived from two sequents of the form
$$ \seq{\Gamma_1,\,\bang{\Delta_1}} {H_1, \bot^\good{a}} $$
 and
$$ \seq{\bot^\good{a},\,\Gamma_2,\,\bang{\Delta_2}}{ } $$
 then, by Lemma~\ref{L1} and Lemma~\ref{lcost}, we had
 a contradiction:
$$\left\{\begin{array}{lcl}%
       0 & = & a-1 \pmod{9N},\\%
       a & = & 1 \pmod{9N}.\\%
\end{array}\right.$$

\CASE{5.d} Assume that the {\em principal\/} formula is of the form
 $$ C_{00} = \limply{\limply{H_{00}^\good{2}}{\bot^\good{3}}}
                    {\bot^\good{3}},$$
 and, according to rule~{\rm {\bf L$\llto$}}, the corresponding
 sequent of item~(d) is derived from two sequents of the form
$$ \seq{C_{00}^\good{k_1},\,\Gamma_1,\,\bang{\Delta_1}}
       {\limply{H_{00}^\good{2}}{\bot^\good{3}}} $$
 and
$$ \seq{C_{00}^\good{k_2},\,\bot^\good{3},\,
        \Gamma_2,\,\bang{\Delta_2}}{\bot^\good{N}} $$
 where
$$\left\{\begin{array}{lcl}%
       3  & = & k_1 + k_2,\\%
  \Gamma  & = & \Gamma_1,\,\Gamma_2,\\%
  \Delta  & = & \Delta_1,\,\Delta_2.%
\end{array}\right.$$
 Then, by Lemma~\ref{L1} and Lemma~\ref{lcost}, the following
 contradiction is immediate:
$$\left\{\begin{array}{lcl}%
          -2Nk_1     & = & 2N + 3     \pmod{9N},\\%
          -2Nk_2 + 3 & = & N \pmod{9N}.%
\end{array}\right.$$
 If our sequent of item~(d) were derived from two sequents of the form
$$ \seq{C_{00}^\good{k_1},\,\Gamma_1,\,\bang{\Delta_1}}
       {\limply{H_{00}^\good{2}}{\bot^\good{3}}, \bot^\good{N}} $$
 and
$$ \seq{C_{00}^\good{k_2},\,\bot^\good{3},\,
        \Gamma_2,\,\bang{\Delta_2}}{ } $$
 then we got a contradiction as well:
$$\left\{\begin{array}{lcl}%
          -2Nk_1     & = & 3N + 2     \pmod{9N},\\%
          -2Nk_2 + 3 & = & 1 \pmod{9N}.%
\end{array}\right.$$
 Now, bringing together all the cases considered, we can complete
 the proof of Lemma~\ref{LDq}.
\QED }

\subsection {Lemma~5.4}

\begin{lemma} \label{Laxiom}
 Let~$\Delta$ consist of formulas of the form~$F_\good{A}$,
 and~$\Gamma$ consist of formulas of the form~$D_\good{q}$,
 $F_\good{A}$, and~$F_\good{Y}$.\\
 Let $k$ and~$m$ be integers such that
$$\left\{\begin{array}{lclcl}%
      0  & \leq & k & \leq & 6,\\%
      0  & \leq & m & \leq & 5.%
\end{array}\right.$$
 Let~$K$~be a multiset of the form
 $$ \underbrace{C_{00},\ C_{00},\ \ldots,\ C_{00}}%
               _\good{k \ \mbox{times}}.$$
 \mbox{(For\ \ $k=0$,}\ \ $K$~is the empty multiset.)\\
 Let~$B$ be a formula of the form
 $$ B = (C_{00}^\good{m} \otimes D_\good{q_1} \otimes D_\good{q_2}
          \otimes \cdots \otimes D_\good{q_n}).$$
 If~a~sequent of the form
  $$ \seq{K,\, \Gamma,\, \bang{\Delta}}{B} $$
 occurs in a cut-free derivation in Linear Logic then
\bb{1}
\item  $ k = m,$
\item this~$\Gamma$ must be an \mbox{$n$-element} multiset of the
      following form
$$ \Gamma =  D_\good{q_1}, D_\good{q_2}, \ldots, D_\good{q_n},$$
\item and $\bang{\Delta}$~can be produced by
      rules~{\rm {\bf W\bang{}}} and~{\rm {\bf C\bang{}}} only
      (there is no applications of rule~{\rm {\bf L\bang{}}} in the
       derivation above this sequent).\footnote%
      { We will say that such a~$\bang{\Delta}$ is {\em degenerate.}}
\ee
\end{lemma}

\proof {
 First of all, by Lemma~\ref{L1} and Lemma~\ref{lcost}:
          $$ -2Nk = -2Nm \pmod{9N}.$$
 Hence,
          $$ k = m \leq 5.$$
 Now we will develop induction on a given derivation. Regarding to
 the form of the {\em principal\/} formula at a current point of
 the derivation, we will demonstrate that each of the
 {\em undesirable\/} cases is inconsistent.

\CASE{0} The {\em principal\/} formula belongs to~$\bang{\Delta}$.
 
 Assume that it is produced by rule~{\rm {\bf L\bang{}}},
 and our sequent is derived from a sequent of the form
$$\seq{K,\,\Gamma,\, F_\good{A},\,\bang{\Delta'}}{B} $$
 Then we can get a contradiction because, according to the inductive
 hypothesis, the multiset
      $$ F_\good{A},\, \Gamma $$
 must be a multiset of the differing form.

 Hence, the only possibility is to apply either~{\rm {\bf W\bang{}}}
 or~{\rm {\bf C\bang{}}}. It remains to use the inductive hypothesis
 for completing this case.

\CASE{1} The right-side formula~$B$ is {\em principal.}
 Let us consider four possible versions of the {\em principal\/}
 formula~$B$.

\CASE{1.1} The {\em principal\/} formula~$B$ is of the form
            $$ B = (C_{00}^\good{m} \otimes D_\good{Z}) $$
           where
 $$ D_\good{Z} = (D_\good{q_1} \otimes D_\good{q_2}
                \otimes \cdots \otimes D_\good{q_n}),$$
 and, according to rule~{\rm {\bf R$\otimes$}}, our sequent is
 derived from two sequents of the form
$$ \seq{C_{00}^\good{k_1},\,\Gamma_1,\,\bang{\Delta_1}}{C_{00}} $$
 and
$$ \seq{C_{00}^\good{k_2},\,\Gamma_2,\,\bang{\Delta_2}}
       {(C_{00}^\good{m-1} \otimes D_\good{Z})} $$
 where
$$\left\{\begin{array}{lcl}%
     k      & = & k_1 + k_2,\\%
    \Gamma  & = & \Gamma_1,\,\Gamma_2,\\%
    \Delta  & = & \Delta_1,\,\Delta_2.%
         \end{array}\right.$$
 Then, by applying Lemma~\ref{LC00} and the inductive hypothesis
 to our both sequents, we can prove that
\bb{1}
\item multiset~$\Gamma_1$ must be empty,
\item multiset~$\Gamma_2$ must be a multiset of the form
$$ \Gamma_2 =  D_\good{q_1}, D_\good{q_2}, \ldots, D_\good{q_n},$$
\item both~$\bang{\Delta_1}$ and~$\bang{\Delta_2}$ must be
      {\em degenerate}, which results in the desired
      {\em degeneracy} of the whole~$\bang{\Delta}$.
\ee
 Hence, the whole~$\Gamma$ is of the required form
         $$ \Gamma = \Gamma_2.$$

\CASE{1.2} The {\em principal\/} formula~$B$ is of the form
            $$ B = (D_\good{q_1} \otimes D_\good{Z'}) $$
           where
$$ D_\good{Z'} = (D_\good{q_2}\otimes\cdots\otimes D_\good{q_n}),$$
 and, according to rule~{\rm {\bf R$\otimes$}}, our sequent is
 derived from two sequents of the form
$$ \seq{\Gamma_1,\,\bang{\Delta_1}}{D_\good{q_1}} $$
 and
$$ \seq{\Gamma_2,\,\bang{\Delta_2}}{D_\good{Z'}} $$
 where
$$\left\{\begin{array}{lcl}%
    \Gamma  & = & \Gamma_1,\,\Gamma_2,\\%
    \Delta  & = & \Delta_1,\,\Delta_2.%
         \end{array}\right.$$
 Then, by applying Lemma~\ref{LDq} and the inductive hypothesis
 to our both sequents, we can prove that
\bb{1}
\item multiset~$\Gamma_1$ must be a singleton of the form~\footnote%
      {Take into account that $\Gamma$~does not contain any~$H_1$.}
      $$ \Gamma = D_\good{q_1},$$
\item multiset~$\Gamma_2$ must be a multiset of the form
      $$ \Gamma_2 = D_\good{q_2}, \ldots, D_\good{q_n},$$
\item both~$\bang{\Delta_1}$ and~$\bang{\Delta_2}$ must be
      {\em degenerate}, which results in the desired
      {\em degeneracy} of the whole~$\bang{\Delta}$.
\ee
 Hence, the whole~$\Gamma$ is of the required form
$$ \Gamma =  D_\good{q_1}, D_\good{q_2}, \ldots, D_\good{q_n}.$$

\CASE{1.3} The {\em principal\/} formula~$B$ is of the form
$$D_\good{q} = \limply{\limply{H_1}{\bot^\good{a}}}{\bot^\good{a}},$$
 and, according to rule~{\rm {\bf R$\llto$}}, our sequent is
 derived from the sequent
 $$ \seq{\limply{H_1}{\bot^\good{a}},\,\Gamma,\,\bang{\Delta}}
        {\bot^\good{a}}.$$
 By Lemma~\ref{LDq}, $\Gamma$~must be a singleton of the form
 $$\Gamma = D_\good{q} =
          \limply{\limply{H_1}{\bot^\good{a}}}{\bot^\good{a}},$$
 and $\bang{\Delta}$~can be produced by rules~{\rm {\bf W\bang{}}}
 and~{\rm {\bf C\bang{}}} only.

\CASE{2} Assume that the {\em principal\/} formula belongs
         to~$\Gamma$, and it is of the form~$F_\good{A}$
         \mbox{(or~$F_\good{Y}$).}

 The following subcases are to be considered.

\CASE{2.0} The {\em principal\/} formula is of the
           form~\mbox{$(F_\good{A_1} \& F_\good{A_2})$,}
 and, by rule~{\rm {\bf L$\&$}}, our sequent is derived either
 from a sequent of the form
$$ \seq{K,\,\Gamma',\, F_\good{A_1},\,\bang{\Delta}}{B} $$
  or from a sequent of the form
$$ \seq{K,\,\Gamma',\, F_\good{A_2},\,\bang{\Delta}}{B} $$
 Then we get a contradiction because, according to the inductive
 hypothesis, either the multiset
      $$ \Gamma', \ F_\good{A_1} $$
 or the multiset
      $$ \Gamma', \ F_\good{A_2} $$
 is in the {\em wrong} form.

\CASE{2.1} Assume that the {\em principal\/} formula is of the
             form~\limply{E_\good{X}}{E_\good{Y}}, and,
             according to rule~{\rm {\bf L$\llto$}}, our sequent
             is derived from two sequents of the form
$$\seq{C_{00}^\good{k_1},\,\Gamma_1,\,\bang{\Delta_1}}{E_\good{X}} $$
 and
$$\seq{C_{00}^\good{k_2},\,E_\good{Y},\,\Gamma_2,\,\bang{\Delta_2}}
      {B} $$
 where
$$\left\{\begin{array}{lcl}%
       k  & = & k_1 + k_2,\\%
   \Gamma &\supset & \Gamma_1,\,\Gamma_2,\\%
  \Delta  & = & \Delta_1,\,\Delta_2.%
\end{array}\right.$$
 Then Lemma~\ref{L1} and Lemma~\ref{lcost} yield:
$$\left\{\begin{array}{lcl}%
       -2Nk_1      & = & 6N  \pmod{9N},\\%
       -2Nk_2 + 6N & = & -2Nm \pmod{9N},\\%
\end{array}\right.$$
 which is a contradiction.

\CASE{2.2} Assume that the {\em principal\/} formula is of the
             form~\limply{E_\good{X}}{E_\good{Y}}, and our sequent
             is derived from two sequents of the form
$$\seq{C_{00}^\good{k_1},\,\Gamma_1,\,\bang{\Delta_1}}
      {E_\good{X}, B} $$
 and
$$\seq{C_{00}^\good{k_2},\,E_\good{Y},\,\Gamma_2,\,\bang{\Delta_2}}
      { } $$
 Then Lemma~\ref{L1} and Lemma~\ref{lcost} yield:
$$\left\{\begin{array}{lcl}%
       -2Nk_1      & = & 6N -2Nm - 1  \pmod{9N},\\%
       -2Nk_2 + 6N & = & 1 \pmod{9N},\\%
\end{array}\right.$$
 which is also a contradiction.

\CASE{2.3} Assume that the {\em principal\/} formula is of the
             form~\limply{F_\good{A}}{F_\good{Y}}, and,
             according to rule~{\rm {\bf L$\llto$}}, our sequent
             is derived from two sequents of the form
$$ \seq{C_{00}^\good{k_1},\,\Gamma_1,\,\bang{\Delta_1}}
       {F_\good{A}} $$
 and
$$\seq{C_{00}^\good{k_2},\,F_\good{Y},\,\Gamma_2,\,\bang{\Delta_2}}
      {B} $$
 where
$$\left\{\begin{array}{lcl}%
       k  & = & k_1 + k_2,\\%
   \Gamma &\supset & \Gamma_1,\,\Gamma_2,\\%
  \Delta  & = & \Delta_1,\,\Delta_2.%
\end{array}\right.$$
 Then, by Lemma~\ref{L1} and Lemma~\ref{lcost}, we have:
$$\left\{\begin{array}{lcl}%
       -2Nk_1      & = & 0  \pmod{9N},\\%
       -2Nk_2      & = & -2Nm \pmod{9N}.\\%
\end{array}\right.$$
 Hence,
         $$ k_2 = m,$$
 and we get a contradiction because, according to the inductive
 hypothesis, the multiset
       $$ F_\good{Y},\ \Gamma_2 $$
 is in the {\em wrong} form.

\CASE{2.4} Assume that the {\em principal\/} formula is of the
             form~\limply{F_\good{A}}{F_\good{Y}}, and our sequent
             is derived from two sequents of the form
$$ \seq{C_{00}^\good{k_1},\,\Gamma_1,\,\bang{\Delta_1}}
       {F_\good{A}, B} $$
 and
$$\seq{C_{00}^\good{k_2},\,F_\good{Y},\,\Gamma_2,\,\bang{\Delta_2}}
      { } $$
 Then Lemma~\ref{L1} and Lemma~\ref{lcost} show a contradiction
 as well:
$$\left\{\begin{array}{lcl}%
       -2Nk_1      & = & -2Nm - 1 \pmod{9N},\\%
       -2Nk_2      & = & 1 \pmod{9N}.\\%
\end{array}\right.$$

\CASE{2.5} Case of the {\em principal\/} formula of the
 form~\lvariant{E_\good{X}}{E_\good{Y_1}}{E_\good{Y_2}}
 is handled similarly to~\mbox{{\bf Cases 2.1.}}
 and~\mbox{{\bf Cases 2.2.}}

\CASE{3} Assume that the {\em principal\/} formula belongs
         to~$\Gamma$, and it is of the form
 $$ D_\good{q} =
    \limply{\limply{H_1}{\bot^\good{b}}}{\bot^\good{b}} $$
 where
          $$ 4 \ \leq\  b \ \leq\  N-3.$$

\CASE{3.1} According to rule~{\rm {\bf L$\llto$}}, let our sequent
            be derived from two sequents of the form
$$ \seq{C_{00}^\good{k_1},\,\Gamma_1,\,\bang{\Delta_1}}
       {\limply{H_1}{\bot^\good{b}}} $$
 and
$$ \seq{C_{00}^\good{k_2},\,\bot^\good{b},\,
                   \Gamma_2,\,\bang{\Delta_2}}{B} $$
 where
$$\left\{\begin{array}{lcl}%
       k  & = & k_1 + k_2,\\%
   \Gamma &\supset & \Gamma_1,\,\Gamma_2,\\%
  \Delta  & = & \Delta_1,\,\Delta_2.%
\end{array}\right.$$
 Then Lemma~\ref{L1} and Lemma~\ref{lcost} yield a contradiction:
$$\left\{\begin{array}{lcl}%
       -2Nk_1     & = & b  \pmod{9N},\\%
       -2Nk_2 + b & = & -2Nm \pmod{9N}.\\%
\end{array}\right.$$

\CASE{3.2} Now let our sequent of be derived from two
 sequents of the form
$$ \seq{C_{00}^\good{k_1},\,\Gamma_1,\,\bang{\Delta_1}}
       {\limply{H_1}{\bot^\good{b}}, B} $$
 and
$$ \seq{C_{00}^\good{k_2},\,\bot^\good{b},\,
                   \Gamma_2,\,\bang{\Delta_2}}{ } $$
 Then, by Lemma~\ref{L1} and Lemma~\ref{lcost}, we have:
$$\left\{\begin{array}{lcl}%
       -2Nk_1     & = & b -2Nm - 1 \pmod{9N},\\%
       -2Nk_2 + b & = & 1 \pmod{9N},\\%
\end{array}\right.$$
 which is a contradiction as well.

\CASE{4} Finally, let the left-hand {\em principal\/} formula
         belong neither to~$\Gamma$ nor to~$\bang{\Delta}$.
         Hence, it is of the form
 $$ C_{00} = \limply{\limply{H_{00}^\good{2}}{\bot^\good{3}}}
                    {\bot^\good{3}},$$
 and, according to rule~{\rm {\bf L$\llto$}}, our sequent is
 derived from two sequents of the form
$$ \seq{C_{00}^\good{k_1},\,\Gamma_1,\,\bang{\Delta_1}}
       {\limply{H_{00}^\good{2}}{\bot^\good{3}}} $$
 and
$$ \seq{C_{00}^\good{k_2},\,\bot^\good{3},\,
                  \Gamma_2,\,\bang{\Delta_2}}{B} $$
 where
$$\left\{\begin{array}{lcl}%
     k-1  & = & k_1 + k_2,\\%
  \Gamma  & = & \Gamma_1,\,\Gamma_2,\\%
  \Delta  & = & \Delta_1,\,\Delta_2.%
\end{array}\right.$$
 Then, by Lemma~\ref{L1} and Lemma~\ref{lcost}, the following
 contradiction is immediate:
$$\left\{\begin{array}{lcl}%
       -2Nk_1     & = &  2N + 3 \pmod{9N},\\%
       -2Nk_2 + 3 & = & -2Nm \pmod{9N}.\\%
\end{array}\right.$$
 If our sequent were derived from two sequents of the form
$$ \seq{C_{00}^\good{k_1},\,\Gamma_1,\,\bang{\Delta_1}}
       {\limply{H_{00}^\good{2}}{\bot^\good{3}}, B} $$
 and
$$ \seq{C_{00}^\good{k_2},\,\bot^\good{3},\,
                  \Gamma_2,\,\bang{\Delta_2}}{ } $$
 then we had a contradiction as well:
$$\left\{\begin{array}{lcl}%
       -2Nk_1     & = & 2N + 2 -2Nm \pmod{9N},\\%
       -2Nk_2 + 3 & = & 1 \pmod{9N}.\\%
\end{array}\right.$$
 Now, bringing together all the cases considered, we can
 complete the proof of Lemma~\ref{Laxiom}.
\QED }






\end{document}

%% file: myth.tex
\newtheorem{lemma}{Lemma}[section] 
\newtheorem{corollary}{Corollary}[section] 

\newtheorem{Defi}{Definition}[section]
\newenvironment{definition}{\begin{Defi}\em }{\end{Defi}}
\newtheorem{Exa}{Example}[section]

\newcommand{\proof}{\noindent {\bf Proof.}\hspace{2ex}}
\newcommand{\CASE}[1]{\noindent\mbox{{\bf Case #1}}\hspace{2ex}}
\newcommand{\QED }{\hspace*{\fill}\vrule height6pt width6pt depth0pt}

\newtheorem{The}{Theorem}[section]
\newenvironment{theorem}{\begin{The}\setcounter{TLem}{0}}{\end{The}}
\newtheorem{TLem}{Lemma}